\documentclass[trackchanges,twocolumn]{aastex62}
\usepackage{graphicx} 
\usepackage{txfonts} 
\usepackage{natbib}
\usepackage{booktabs}
\graphicspath{{imgArxiv/}}

\submitjournal{\apj}
\accepted{21/11/2019}
\shorttitle{Kinematics of multiple populations in globular clusters} 
\shortauthors{G.\,Cordoni et al.} 
\begin{document} 

\title{Three-Component Kinematics of Multiple Stellar Populations in Globular Clusters with Gaia and VLT.}  
\author{G.\,Cordoni }
\affiliation{Dipartimento di Fisica e Astronomia ``Galileo Galilei'' -
  Univ. di Padova, Vicolo dell'Osservatorio 3, Padova, IT-35122}
\author{A.\ P.\,Milone}
\affiliation{Dipartimento di Fisica e Astronomia ``Galileo Galilei'' -
 Univ. di Padova, Vicolo dell'Osservatorio 3, Padova, IT-35122}
\author{A.\,Mastrobuono-Battisti}
\affiliation{Max-Planck Institut f{\"u}r Astronomie, K{\"o}nigstuhl 17, D-69117 Heidelberg, Germany}
\author{A.\ F.\,Marino} 
\affiliation{Dipartimento di Fisica e Astronomia ``Galileo Galilei'' - Univ. di Padova, Vicolo dell'Osservatorio 3, Padova, IT-35122}
\affiliation{Centro di Ateneo di Studi e Attivit\`a Spaziali ``Giuseppe Colombo'' - CISAS, Via Venezia 15, Padova, IT-35131}
\author{E.\ P.\,Lagioia }
\affiliation{Dipartimento di Fisica e Astronomia ``Galileo Galilei'' -
 Univ. di Padova, Vicolo dell'Osservatorio 3, Padova, IT-35122}
\author{M.\,Tailo}
\affiliation{Dipartimento di Fisica e Astronomia ``Galileo Galilei'' -
 Univ. di Padova, Vicolo dell'Osservatorio 3, Padova, IT-35122}%
 \author{H.\,Baumgardt}
\affiliation{School of Mathematics and Physics, The University of Queensland, 
St. Lucia, QLD 4072, Australia}
 \author{M.\,Hilker}
\affiliation{European Southern Observatory, Karl-Schwarzschild-Str 2, 
D-85748 Garching, Germany}%

\correspondingauthor{G.\ Cordoni}
\email{giacomo.cordoni@phd.unipd.it} 
\begin{abstract}
The internal dynamics of multiple stellar populations in Globular Clusters (GCs) provides unique constraints on the physical processes responsible for their formation. Specifically, the present-day kinematics of cluster stars, such as rotation and velocity-dispersion, could be related to the initial configuration of the system. 
In recent work, we provided the first study of the kinematics of different stellar populations in NGC$\,$0104 over a large field of view in the plane of the sky, exploiting Gaia Data Release 2 (DR2) proper motions combined with multi-band ground-based photometry. \\ In this paper, we combine Gaia DR2 proper motions with Very Large Telescope radial velocities to investigate the kinematics along the line of sight and in the plane of the sky of multiple populations in seven GCs, namely NGC$\,$0104, NGC$\,$0288, NGC$\,$5904, NGC$\,$6121, NGC$\,$6254, NGC$\,$6752 and NGC$\,$6838. Among the analyzed clusters only NGC$\,$0104 and NGC$\,$5904 show significant rotation.\\
Separating our sample into two groups of first- and second-population stars (1P and 2P) we find that overall these two populations exhibit a similar rotation pattern in NGC$\,$0104. However, some hints of different rotation are observed in the external regions of this cluster. Interestingly, 1P and 2P stars in NGC$\,$5904 show different rotation curves, with distinct phases and such difference is significant at the $\sim$2.5-$\sigma$ level. The analysis of the velocity-dispersion profiles of multiple populations confirms that 2P stars of NGC$\,$0104 show stronger anisotropy than the 1P.
\end{abstract} 
 
\keywords{
  globular clusters: general, stars: population II, stars: abundances, dynamics, techniques: photometry.} 

\section{Introduction}\label{sec:intro}
Studies based on {\it Hubble Space Telescope} ({\it HST}) images revealed that the photometric diagrams of nearly all GCs are composed of two main groups of first- and second-population stars \citep[1P, 2P, e.g.][]{milone2017} with different chemical compositions \citep[e.g.][]{marino2019}. Many efforts have been made to understand their origin, but, so far, none of the proposed scenarios have been able to reach a satisfactory agreement with observations \citep[e.g.][]{renzini2015}. \\
According to many of these scenarios, 2P stars formed out of the ejecta of 1P more-massive stars \citep[e.g.][]{ventura2001, decressin2007, dercole2010, denissenkov2014} after the segregation of the gas in the cluster center. As a consequence, 2P stars may have formed in a more centrally-concentrated environment. \\
As an alternative hypothesis, GCs host a single stellar generation and stars
with different chemical composition are the product of exotic physical phenomena specific of proto-GCs \citep[e.g.][]{demink2009, bastian2013, gieles2018}. \\
An important signature of the physical processes responsible for the formation of multiple populations is the kinematics of cluster stars. Specifically, $N$-body simulations suggest that the dynamical evolution of more centrally-concentrated 2P stars should be significantly different from that of 1P stars, and such difference could still be observable in present-day GC kinematics \citep[e.\,g.\,][]{vesperini2013, mastrobuono2013, mastrobuono2016, henault2015, tiongco2019}. \\
In the past decade, nearly all works on the internal kinematics of GCs were based on radial velocities of a relatively-small sample of stars \citep[e.\,g.\,][]{norris1997, bellazzini2012, marino2014, cordero2014} with the study of 650 stars of NGC\,5139 ($\omega$\,Centauri) by \cite{pancino2007}, being a remarkable exception. 

More recently, {\it HST} provided high-precision relative proper motions of a small but increasing number of clusters, namely NGC\,0104 (47\,Tucanae), NGC\,0362, NGC\,2808, NGC\,5139 and NGC\,6352 that allowed the investigation of the kinematics of multiple populations in the plane of the sky \citep{richer2013, bellini2015, bellini2018, libralato2018, libralato2019}. In all the studies the authors concluded that 2P stars show a more-radially anisotropic velocity distribution. 
While these works are based on high-precision relative proper motions of thousands of stars, the small field of view of {\it HST} does not allow the study of the entire cluster. 

To overcome this shortcoming and study the kinematics of multiple stellar populations over a large field of view, we started a project based on Gaia Data Release 2 \citep[DR2\footnote{https://gea.esac.esa.int/archive/},][]{gaia2018a} accurate proper motions and multi-band wide-field ground-based photometry.
In the pilot paper of this project, we investigated for the first time the kinematics of 1P and 2P stars of NGC\,0104 over a wide field of view, up to $\sim$18 arcmin from the cluster center \citep[corresponding to $\sim 22$\,pc,][]{milone2018}.
In the present work, we analyse the spatial distributions and the 3D kinematics of NGC\,0104 and other six GCs, namely NGC\,0288, NGC\,5904 (M\,5), NGC\,6121 (M\,4), NGC\,6254 (M\,10), NGC\,6752 and NGC\,6838 (M\,71), whose physical parameters are listed in Table~\ref{tab:parameters}.

The paper is organized as follows: in Section~\ref{sec:data} we introduce the dataset and present the photometric diagrams of the analyzed clusters. In Section~\ref{sec: spatial} we analyze the spatial distribution of multiple stellar populations.
The 3D rotation of 1P and 2P stars and their velocity dispersion are investigated in Sections~\ref{sec:rot} and~\ref{sec:vprofile}, respectively.
 Finally, Section~\ref{sec:summary} provides the summary and the discussion of the results.
 
\section{Data and data analysis} \label{sec:data}
To investigate the internal kinematics of multiple stellar populations in each GC, we combined ground-based wide-field photometry, proper motions from Gaia DR2 and high-precision radial velocities provided by \citet{baumgardt2018} and derived from archival ESO/VLT and Keck spectra together with published radial velocities from the literature.
 
Photometry in $U$, $B$, $V$, $I$ bands has been derived by Peter Stetson from images collected with various facilities and by using the methods and the computer programs by \cite{stetson2005} and \cite{stetson2019}. Photometry has been calibrated on the reference system by \cite{landolt1992}. Details on the dataset and on the data reduction are provided by \cite{monelli2013} and \cite{stetson2019}.  
The photometric catalogs by Stetson and collaborators have been widely used to investigate multiple populations in GCs \citep[e.,g.\,][]{monelli2013, marino2016, marino2017, milone2012b, milone2018, stetson2019}. Most of these works are based on the pseudo color $C_{\rm U,B,I}=(U-B)-(B-I)$, which is an efficient tool to identify stellar populations with different light-element abundance along the RGB and will be used in the following to identify 1P and 2P stars.  
We corrected proper motions for the perspective expansion/contraction due to the bulk motion of the GC by means of Equation~6 in \cite{vandenven2006}. 

As well established in the literature, Gaia DR2 proper motions suffer from systematic errors that mostly depend on stellar colors and positions \citep[e.g.\,][]{gaia2018a, bianchini2018, lindegren2018, sollima2019,vasiliev2019, vasiliev2019b}. 
In this work we are interested in the relative motions of groups of 1P and 2P stars that are almost indistinguishable to the eye of Gaia, as they have similar colors and magnitudes and, to a first approximation, share the same spatial distributions. As a consequence, the systematic errors associated with the motions of both populations are similar and their effects on the relative motions of 1P and 2P stars may be cancelled out when infinite amounts of 1P and 2P stars are available. However, in the case of finite numbers of stars the effect of systematics on the relative proper motions of the two populations may not entirely cancel out.
In this work, we used a conservative approach and followed the recipe described in \citet[][]{vasiliev2019b}\footnote{code publicly available at \url{https://github.com/GalacticDynamics-Oxford/GaiaTools}} to entirely account for systematic errors in our analysis of 1P and 2P stars. 
Since we are considering relative motions, our error estimates would overestimate the true uncertainties.

\subsection{Selection of cluster members}
\begin{figure*}
  \centering
\includegraphics[width=14cm,trim={0cm 0cm 0cm 2cm},clip]{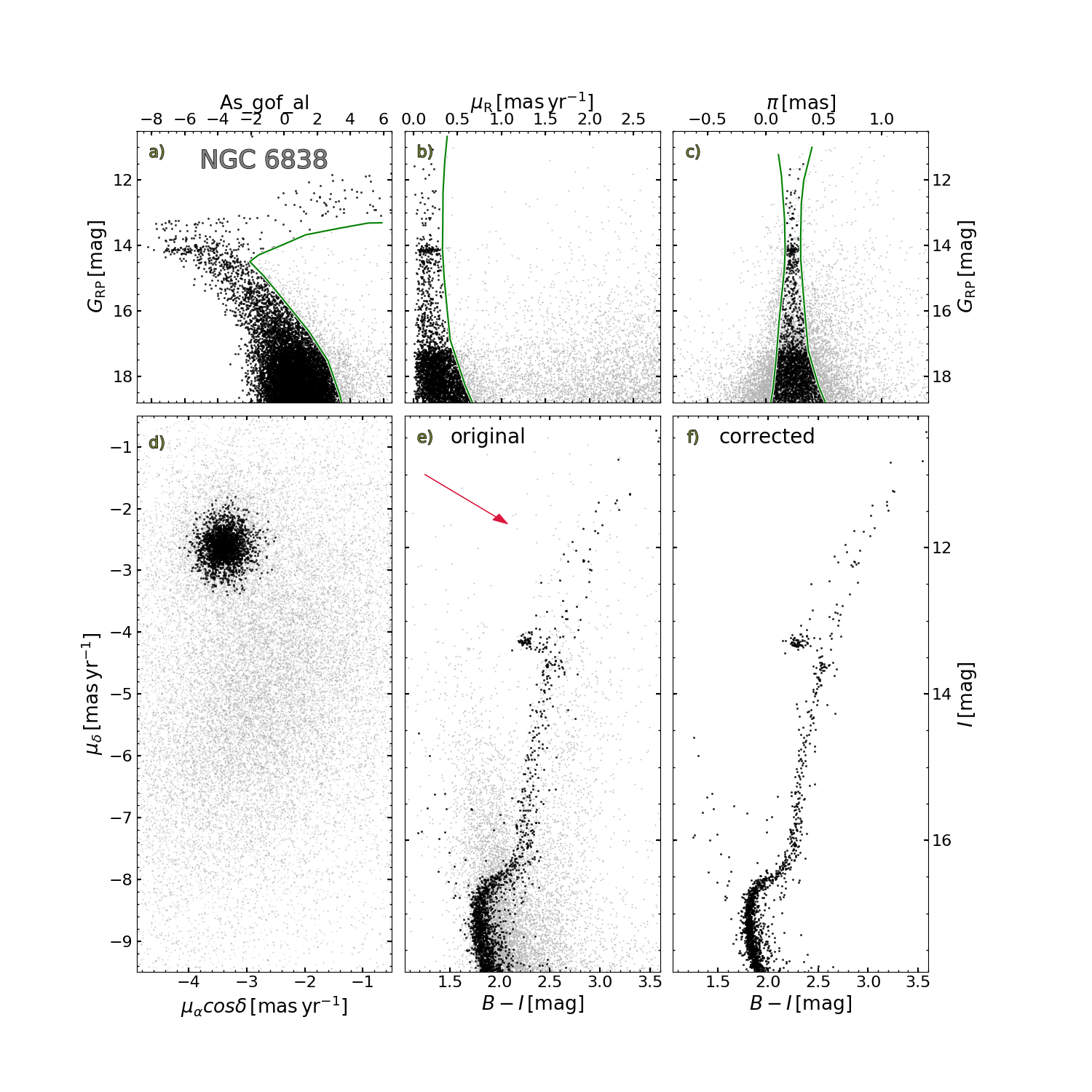}
  \caption{Illustration of the procedure to select stars with high-quality proper motions and to determine the {\it bona-fide} cluster 
  members of NGC\,6838. 
  Panels a, b, and c show the $G_{\rm RP}$ magnitude from Gaia DR2 against the \texttt{As\_gof\_al} parameter, stellar proper motions relative to the cluster mean motion, $\mu_{\rm R}$, and parallax, $\pi$, respectively. The green lines separate cluster members (black points) from field stars (gray points). The proper motion vector-point diagram  is plotted in panel d. Panels e and f compare the original $I$ vs.\,$B-I$ CMD of cluster members with the CMD corrected for differential reddening. The red arrow in panel e represents the reddening vector and corresponds to a reddening variation of E($B-V$)=0.3. See text for details. }
  \label{fig:cluster}
\end{figure*}

To  study the kinematics of stellar populations in GCs we need accurate stellar proper motions. To identify a sample of RGB stars with high-quality astrometric measurements we exploited the method used by \cite{milone2018} and \cite{cordoni2018}, which is illustrated in Figure \ref{fig:cluster} for NGC\,6838, and exploits the parameters provided by the Gaia DR2. 

In a nutshell, we first selected a sample of stars with high-accuracy proper motions, by using the \texttt{astrometric\_gof\_al} (\texttt{As\_gof\_al}) parameter, indicative of the goodness of fit statistics of the astrometric solution for the source in the along-scan direction \citep[see][for details]{gaia2018a}, and the renormalized unit weight error \citep{lindegren2018}.
To do this, we divided the $G_{\rm RP}$-magnitude range between 11.0 and 18.5 into bins of 0.5 mag. We calculated the median magnitude ($G_{\rm RP, i})$, the median \texttt{As\_gof\_al} value (\texttt{As\_gof\_al$_{\rm i}$}) and the corresponding random mean scatter ($\sigma_{\rm i}$) for stars in each magnitude bin (i). We associated the values of $G_{\rm RP, i}$ and \texttt{As\_gof\_al$_{\rm i}$} + 4 $\sigma_{\rm i}$ and linearly interpolated these points to derive the green line of Figure~\ref{fig:cluster}a. We considered those stars that lie on the left side of the green line as well measured. 
Moreover, only stars with proper motion uncertainties smaller than 0.35\,mas/yr have been included in our analysis.

We determined cluster membership of each star using the same procedure described in \citet[][see their Section~2]{cordoni2018}. Briefly, we analyzed the proper motion vector-point diagram (VPD) shown in panel d of Figure~\ref{fig:cluster}, and derived by eye a circle enclosing most cluster stars. Then,  we calculated  the proper motion of each star relative to the cluster mean motion ($\mu_{\rm R}$). We plotted $\mu_{\rm R}$ against   
the $G_{\rm RP}$-magnitude and selected only stars with dispersion lower than $4\sigma$ from the mean relation. We then repeated the same procedure for the parallax, $\pi$. 
This procedure has been iterated three times.  We verified that the sample of cluster stars identified from the criteria described above is nearly coincident with that obtained by following the method by \citet{vasiliev2019}, which is based on Gaussian mixture models. When we adopt the latter stellar sample the conclusions of the paper remain unchanged.\\
As a final step, the $U$, $B$, $V$, $I$ photometry of cluster members has been corrected for differential reddening using the method described in \cite[][see their Section~3.1]{milone2012}.  In a nutshell, we first derived the fiducial line of MS and SGB stars in the $I$ vs.\,($B-I$) plane, where 1P/2P stars are almost indistinguishable, and we calculated the residuals from this line. Then we selected 35 neighbors MS and SGB bright cluster members and computed the median of the color-residuals, calculated along the reddening direction, as our differential-reddening estimate. 
In panels e and f of Figure~\ref{fig:cluster} we compare the original $I$ vs.\,($B-I$) CMD of NGC\,6838 members and the corresponding CMD corrected for differential reddening. Clearly, the comparison between the original and the differential-reddening free CMD suggests that our correction provides much narrower photometric sequences, demonstrating the goodness of our procedure. 

\subsection{Multiple populations along the color-magnitude diagrams}\label{subsec:cmds}

To distinguish 2P from 1P stars we exploit photometric diagrams based on the $C_{\rm U,B,I}$ index. Indeed, a visual inspection of our $V$ vs. $C_{\rm U,B,I}$ diagrams of cluster members, reveals that 1P and 2P stars of the analyzed GCs define two distinct RGBs \citep[see also\,][]{monelli2013, marino2016, marino2017, milone2012b, milone2018}. 
 
\begin{figure*}
  \centering
  \includegraphics[width=0.95\textwidth,height=0.25\textheight]{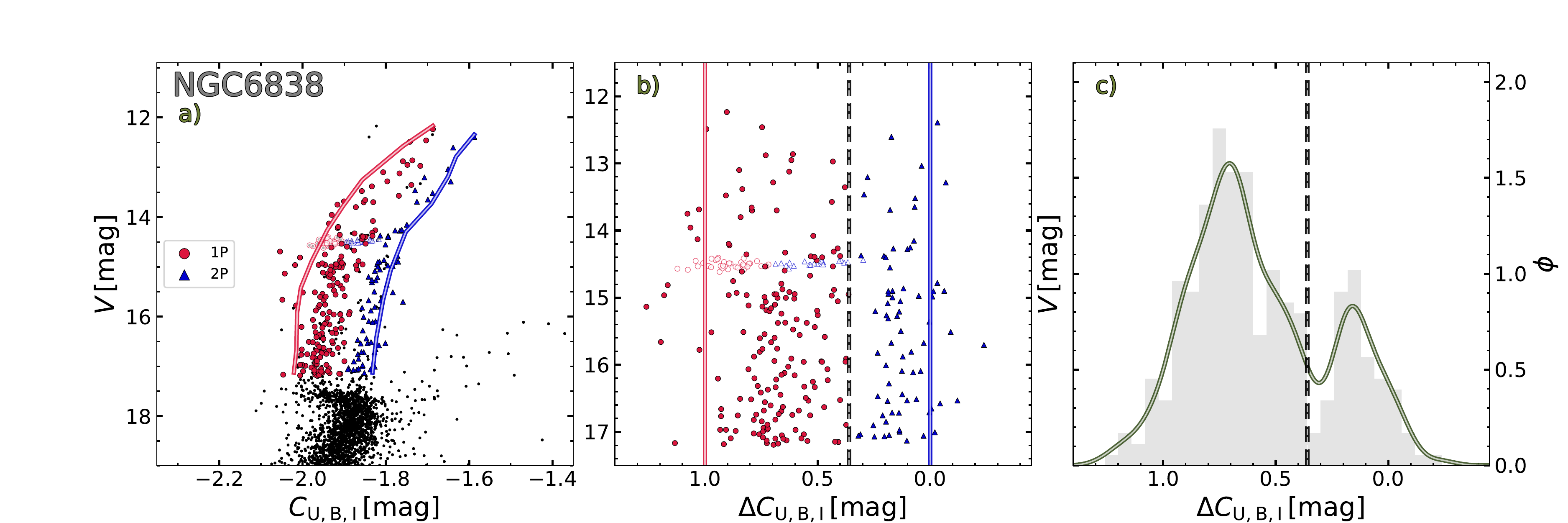}
  \caption{This figure illustrates the procedure to select 1P and 2P stars along the RGB of NGC\,6838.  
   Panel a shows the $V$ vs.\,$C_{\rm U, B, I}$ diagram for cluster members, while the verticalized $V$ vs.\,$\Delta C_{\rm U,B,I}$ diagram for RGB stars and the corresponding $\Delta C_{\rm U,B,I}$ histogram distribution are plotted in panel b and c, respectively. The red and blue continuous lines mark the boundaries of the RGB, while the dashed gray vertical line is used to separate 1P (red circles) stars from 2P stars (blue triangles). HB stars are marked with empty symbols. The continuous line superimposed on the histogram represents the $\Delta C_{\rm U,B,I}$ kernel-density distribution of RGB stars. See text for details.}
  \label{fig:popsel}
\end{figure*}

The procedure that we used to identify the sample of 1P and 2P stars is illustrated in Figure~\ref{fig:popsel} for NGC\,6838 and is based on the $V$ vs. $C_{\rm U,B,I}$ diagram plotted in panel a. 
The red and blue lines superimposed on the diagram correspond to the RGB boundaries and are derived as in \citet[][see their Section~3]{milone2017}. 
In the case of NGC\,6838 we only used stars in the magnitude interval between $V$=12.0 and $V$=17.5, where the RGB split is clearly visible.
In a nutshell, we first divided the magnitude interval between $V$=14.0 and $V$=17.5 into a series of bins of size $dV=0.9$ 
mag. The bins are defined over a grid of points separated by 0.3 mag. 
For each bin we calculated the average $V$ magnitude and associate its value to the $4^{\rm th}$ and the $96^{\rm th}$ percentile of the $C_{\rm U,B,I}$ distribution of RGB stars.
We smoothed these points by using boxcar averaging, where we substituted each point with the average of its three adjacent points.
Due to the small number of stars brighter than $V=14.0$, the fiducial points of the portion of the RGB with $12.0\lesssim V \lesssim 14.0$ are drawn by eye. 

The fiducial lines are verticalized as in \citet[][see their Section~3.1]{milone2015} to derive the $V$ vs. $\Delta C_{\rm U,B,I}$ diagram plotted in panel b. 
Panel c of Figure~\ref{fig:popsel} shows the histogram and the kernel-density distribution of the $\Delta C_{\rm U,B,I}$ for RGB stars with $12.0<V<17.5$. Clearly, the $\Delta C_{\rm U,B,I}$ distribution represented in panels b and c allows us to distinguish 1P stars (represented with red circles) from 2P stars (blue triangles), based on the vertical dashed line.

\begin{figure*}
  \centering
  \includegraphics[width=8.8cm,height=5cm,trim={0cm 0.8cm 0.2cm 0.85cm},clip]{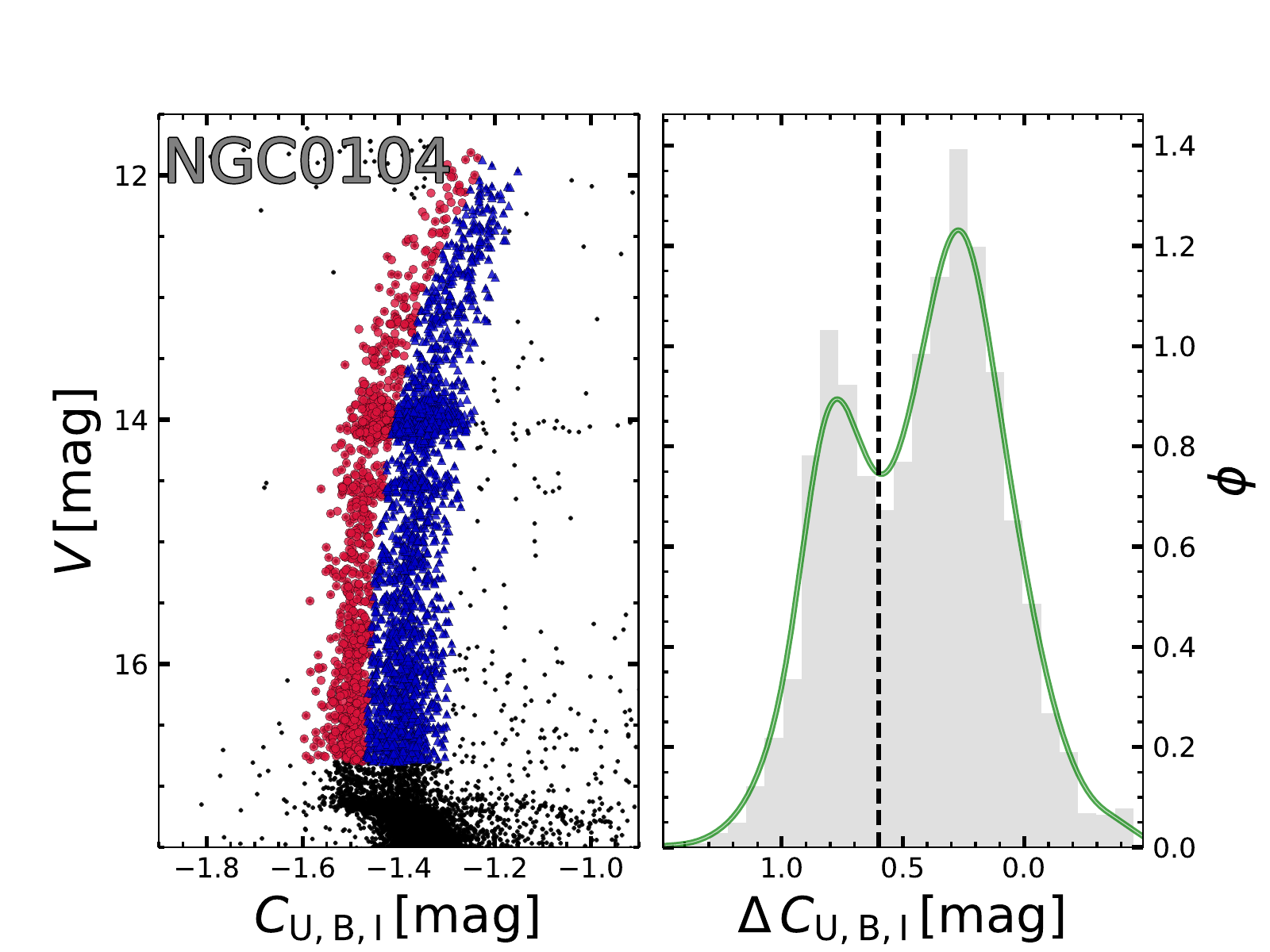}
  \includegraphics[width=8.8cm,height=5cm,trim={0.3cm 0.8cm 0cm 0.85cm},clip]{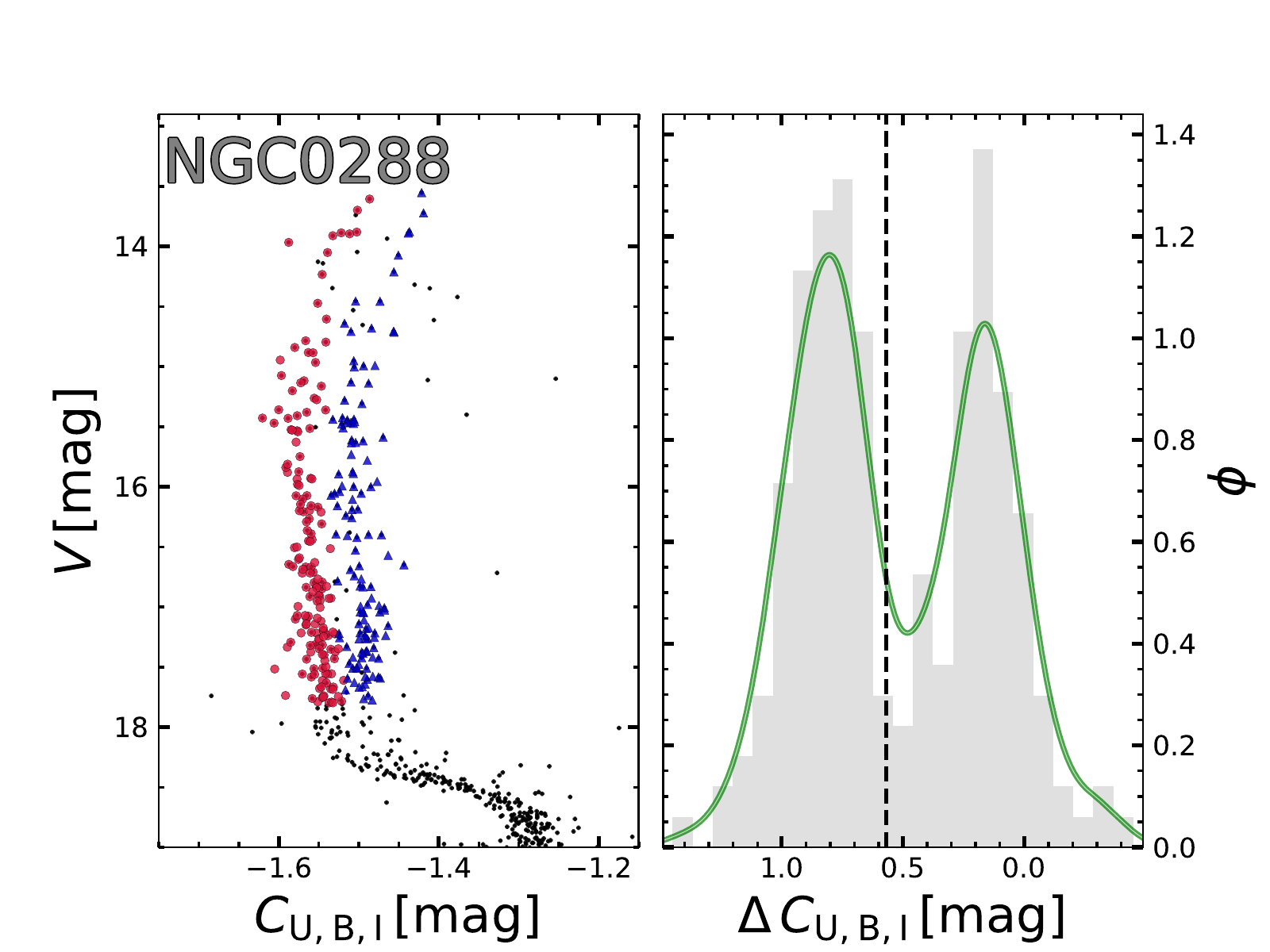}
  \includegraphics[width=8.8cm,height=5cm,trim={0cm 0.8cm 0.2cm 0.85cm},clip]{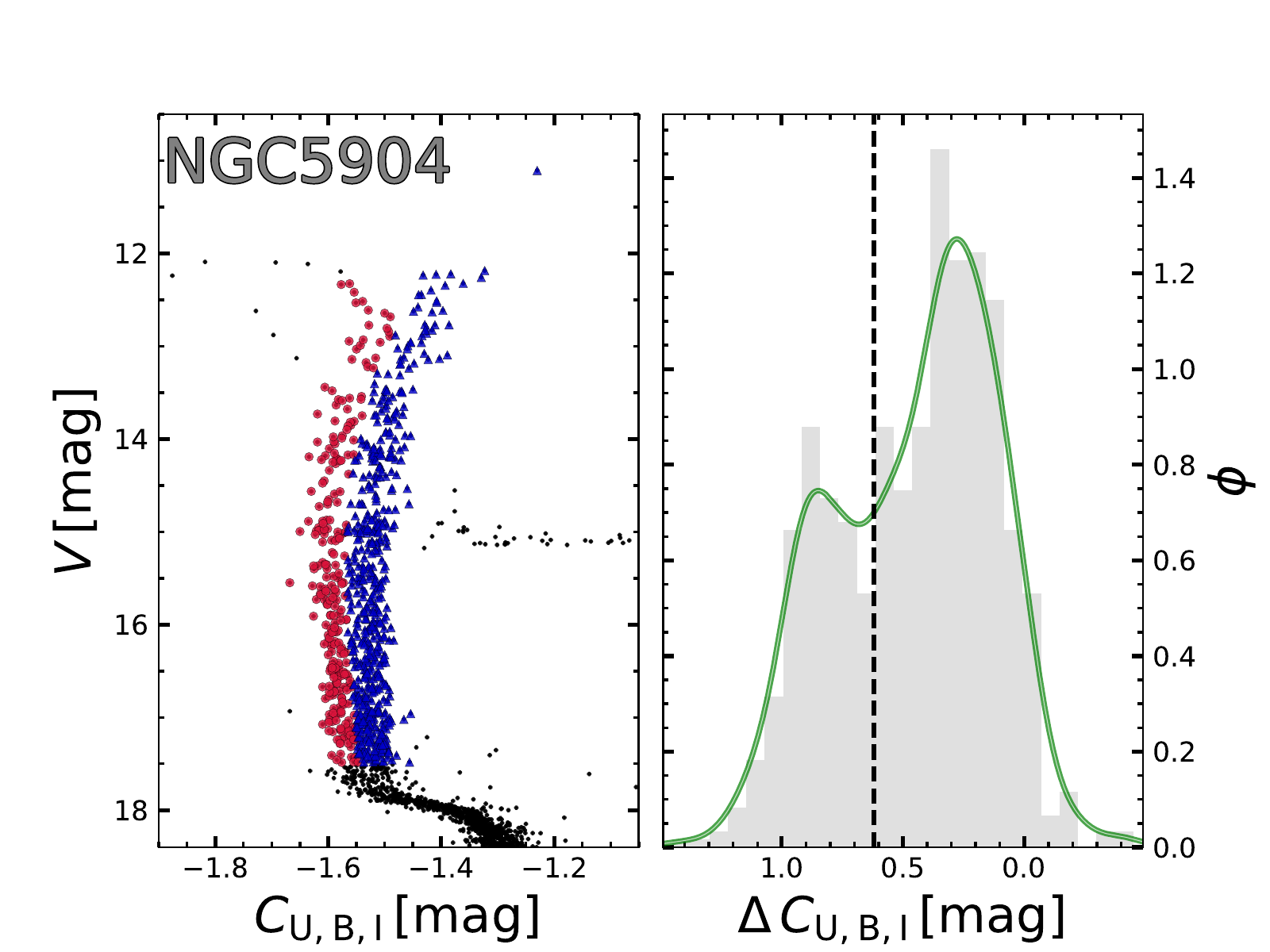}
  \includegraphics[width=8.8cm,height=5cm,trim={0.3cm 0.8cm 0cm 0.85cm},clip]{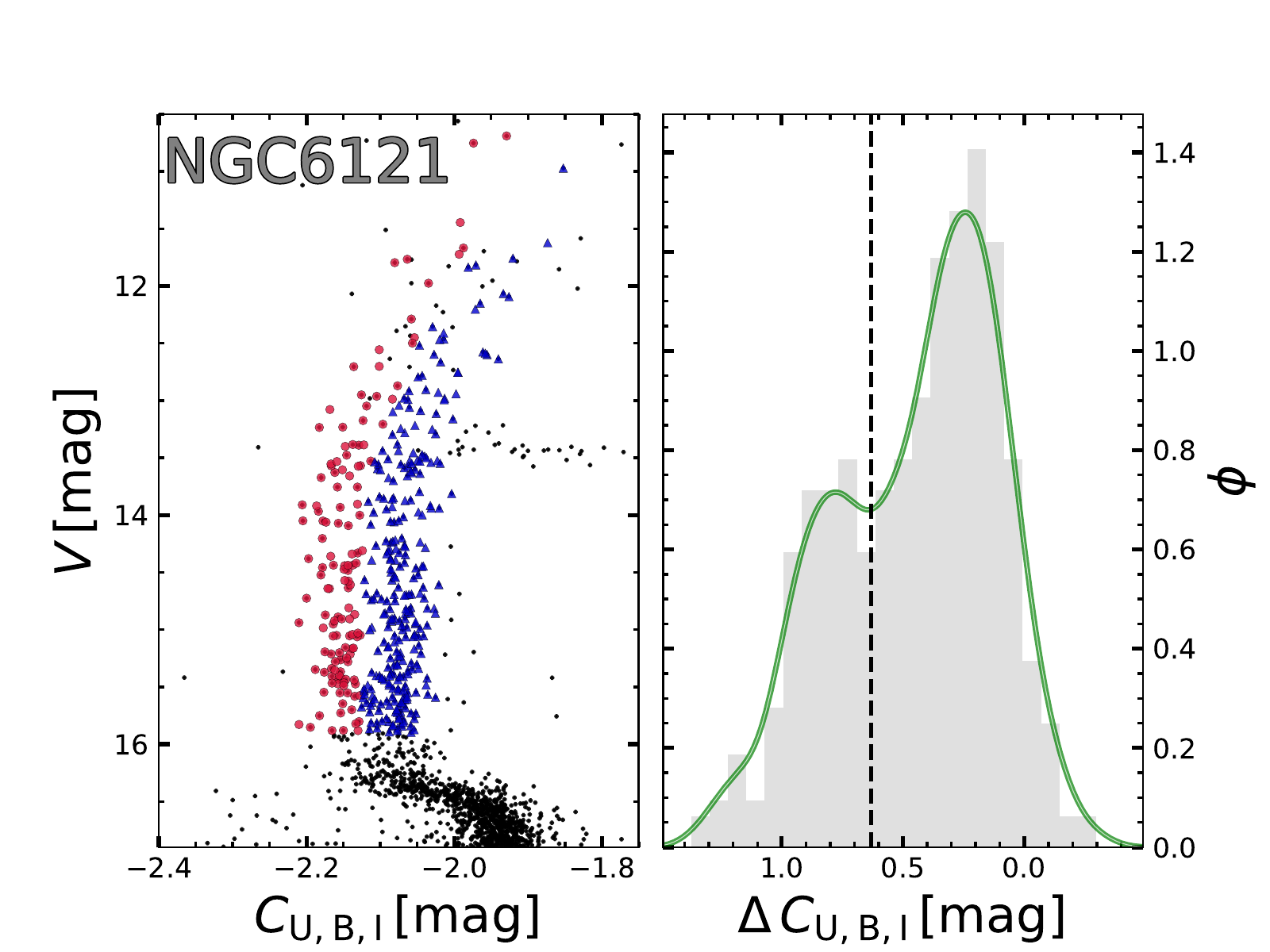}
  \includegraphics[width=8.8cm,height=5cm,trim={0cm 0.0cm 0.2cm 0.85cm},clip]{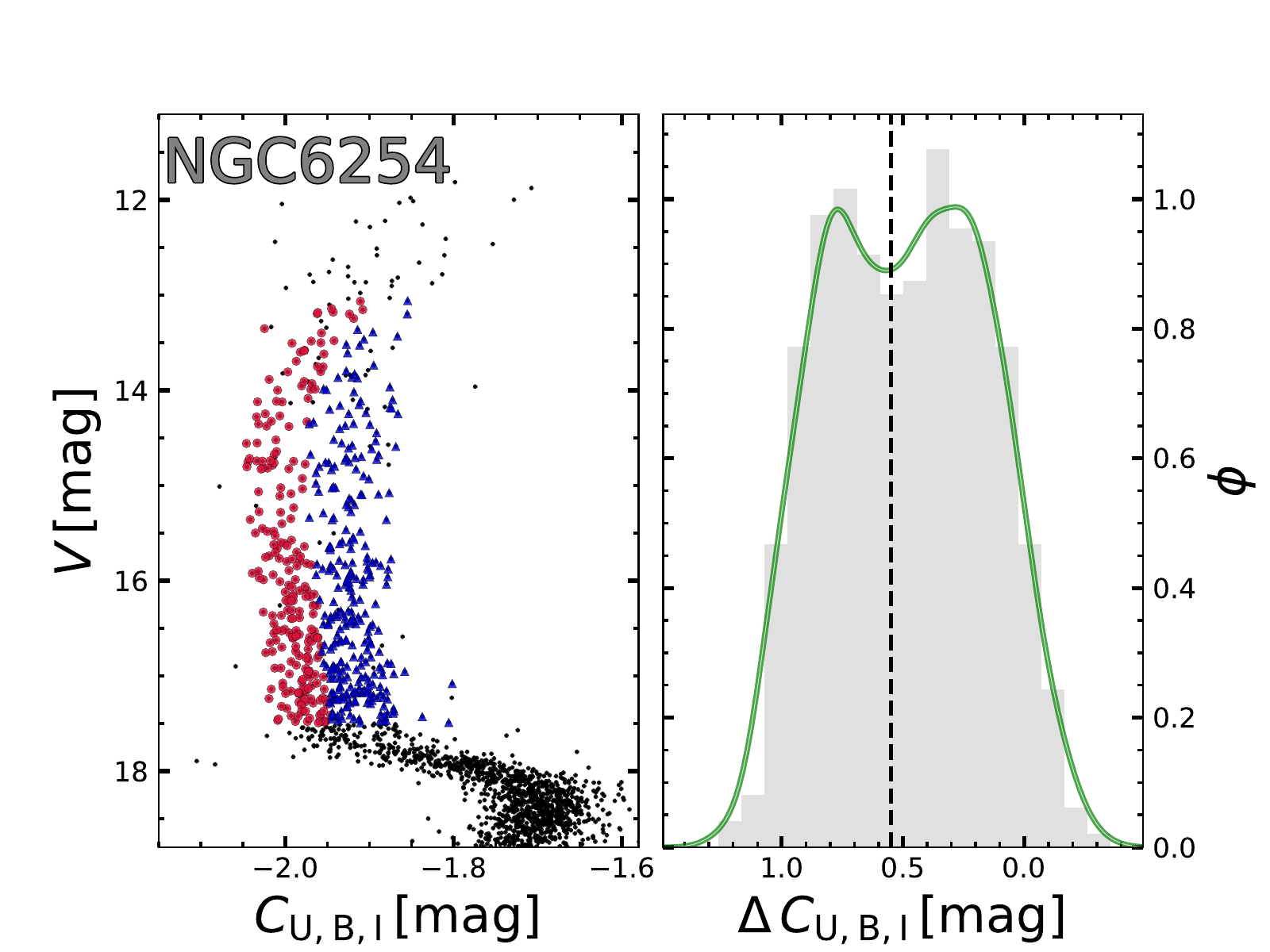}
  \includegraphics[width=8.8cm,height=5cm,trim={0.3cm 0.0cm 0cm 0.85cm},clip]{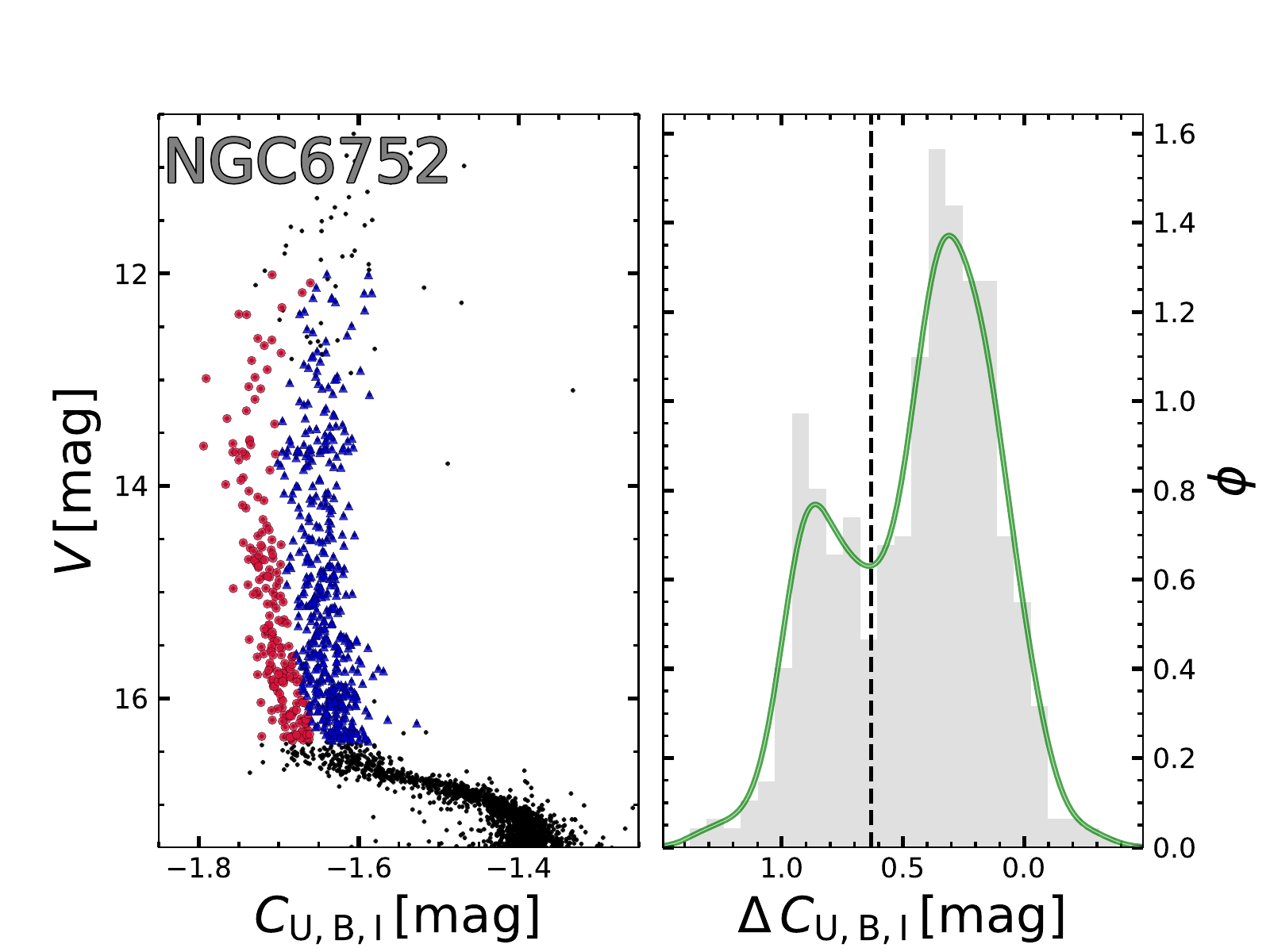}

  \caption{$V$ vs.\,$C_{\rm U,B,I}$ diagrams for the selected cluster members of NGC\,0104, NGC\,0288, NGC\,5904, NGC\,6121, NGC\,6254 and NGC\,6752 (left panels). The panels on the right show the histogram and the kernel-density $\Delta C_{\rm U,B,I}$ distributions for the RGB stars that we used to investigate the internal kinematics of stellar populations. The vertical dashed lines separate the selected 1P and 2P stars that are colored red and blue, respectively, in the left-panel diagrams.}
  \label{fig:cmds}
\end{figure*}

The same procedure illustrated for NGC\,6838 has been applied to the other six clusters that we have analyzed. 
Figure~\ref{fig:cmds} shows the $V$ vs.\,$C_{\rm U,B,I}$ diagrams and the corresponding $\Delta C_{\rm U,B,I}$ histograms and kernel-density  distributions of RGB stars for NGC\,0104, NGC\,0288, NGC\,5904, NGC\,6121, NGC\,6254 and NGC\,6752. The RGB of each cluster defines two distinct sequences  and allows us to select the groups of 1P (red dots) and 2P stars (blue triangles). Only the selected 1P and 2P RGB stars will be used to explore the kinematics of multiple populations in each GC.
In  NGC\,0104 and NGC\,6838, we included in the analysis 1P and 2P HB stars that we selected from the $U-B$ vs.\,$B-I$ two-color diagram as in \citet{milone2012b}.

\begin{table*}
\centering
\begin{tabular}{l|llllllllll}
\toprule
\toprule
ID & RA (J2000)\footnote{from \citet[][updated as in 2010]{harris1996}} & DEC (J2000)$^{\rm a}$ & mass$^{\rm b}$ & $ d_{\rm sun}$\footnote{from \cite{baumgardt2018}} & $R_{\rm Gal}^{\rm a}$ & $R_{\rm c}^{\rm b}$ & $R_{\rm h }^{\rm b}$ & $\bar{V}_{\rm LoS}^{\rm b}$ & $\log\tau_{\rm RH}^{\rm b}$ \\
          &               &             & [$M_{\rm \odot}$]   & [kpc] & [kpc] & [arcmin] & [arcmin] & [km/s] & yr \\ 
\midrule
NGC\,0104 &   00 24 05.67 & $-$72 04 52.6  & $7.79\times 10^5$ & 4.41 & 7.40 &  0.38  &  2.78  & -17.20  & 9.58 \\ 
NGC\,0288 &   00 52 45.24 & $-$26 34 57.4  & $1.16\times 10^5$ & 9.80 & 12.0 &  1.67  &  2.45  & -44.83  & 9.58 \\ 
NGC\,5904 &   15 18 33.22 & $+$02 04 51.7  & $3.72\times 10^5$ & 7.50 & 6.20 &  0.55  &  1.65  &  53.70  & 9.45 \\ 
NGC\,6121 &   16 23 35.22 & $-$26 31 32.7  & $9.69\times 10^4$ & 2.14 & 5.90 &  1.06  &  4.53  &  71.05  & 8.99 \\ 
NGC\,6254 &   16 57 09.05 & $-$04 06 01.1  & $1.84\times 10^5$ & 4.71 & 4.60 &  0.59  &  2.03  &  74.02  & 9.15 \\ 
NGC\,6752 &   19 10 52.11 & $-$59 59 04.4  & $2.39\times 10^5$ & 4.30 & 5.20 &  0.15  &  1.92  & -26.28  & 9.16 \\ 
NGC\,6838 &   19 53 46.49 & $+$18 46 45.1  & $4.91\times 10^4$ & 3.86 & 6.70 &  0.46  &  2.63  & -22.27  & 8.90 \\ 
           
\bottomrule
\bottomrule
\end{tabular}
\caption{Identification, positional data and adopted structural parameters for the analyzed clusters. For each cluster we list position (RA, DEC), distance from the Sun, galactocentric radius ($R_{\rm Gal}$), mass, core radius ($R_{\rm c}$), half-light radius ($R_{\rm h}$), line-of-sight mean velocity ($\bar{V}_{\rm LoS}$) and half-mass relaxation time ($\log\tau_{\rm RH}$).}

\label{tab:parameters}
\end{table*}

\section{Spatial distribution}
\label{sec: spatial}
\begin{figure*}
    \centering
    \includegraphics[width=8cm, trim={3.2cm 1.3cm 2.5cm 0.5cm},clip]{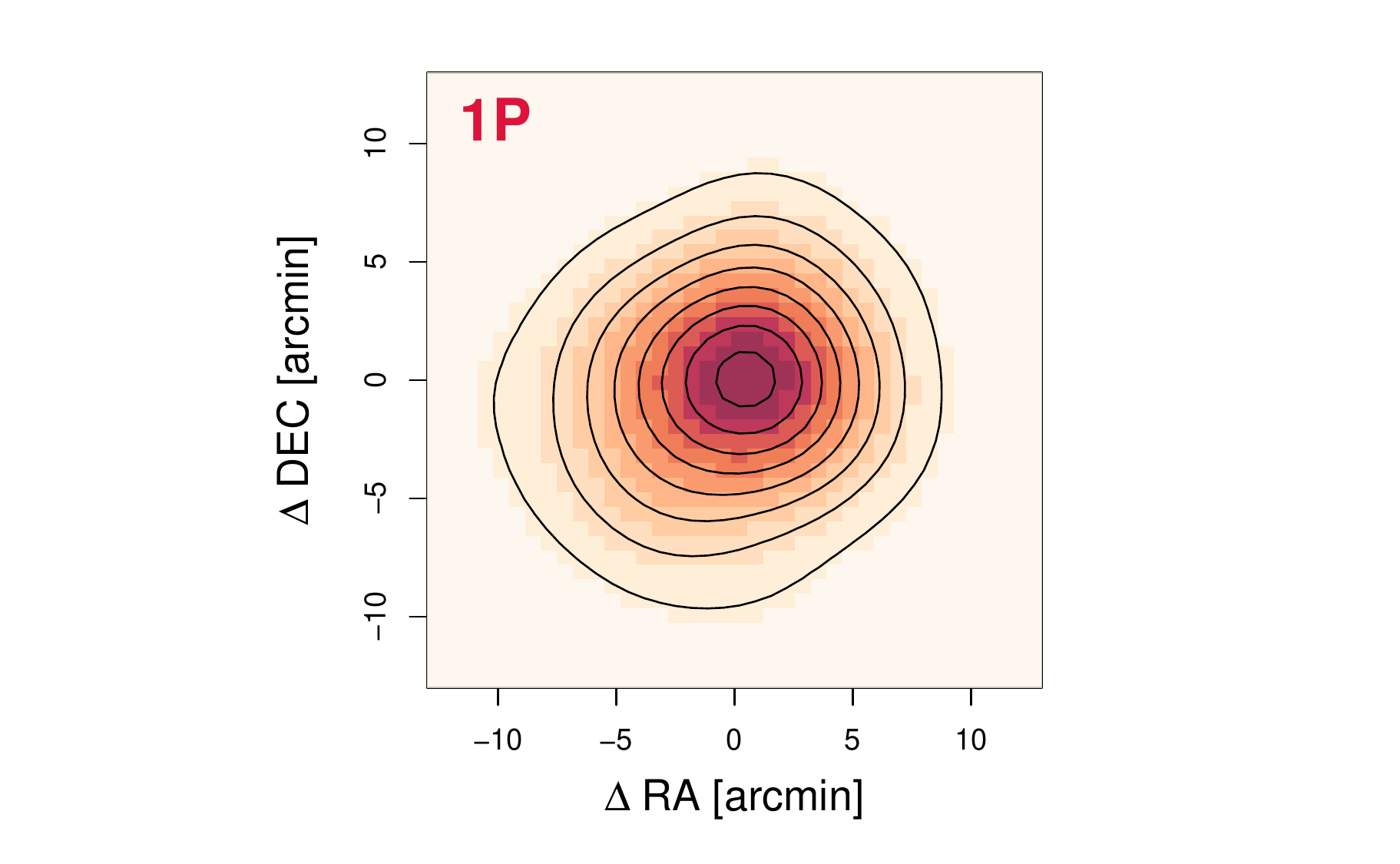}
    \includegraphics[width=8cm, trim={3.2cm 1.3cm 2.5cm 0.5cm},clip]{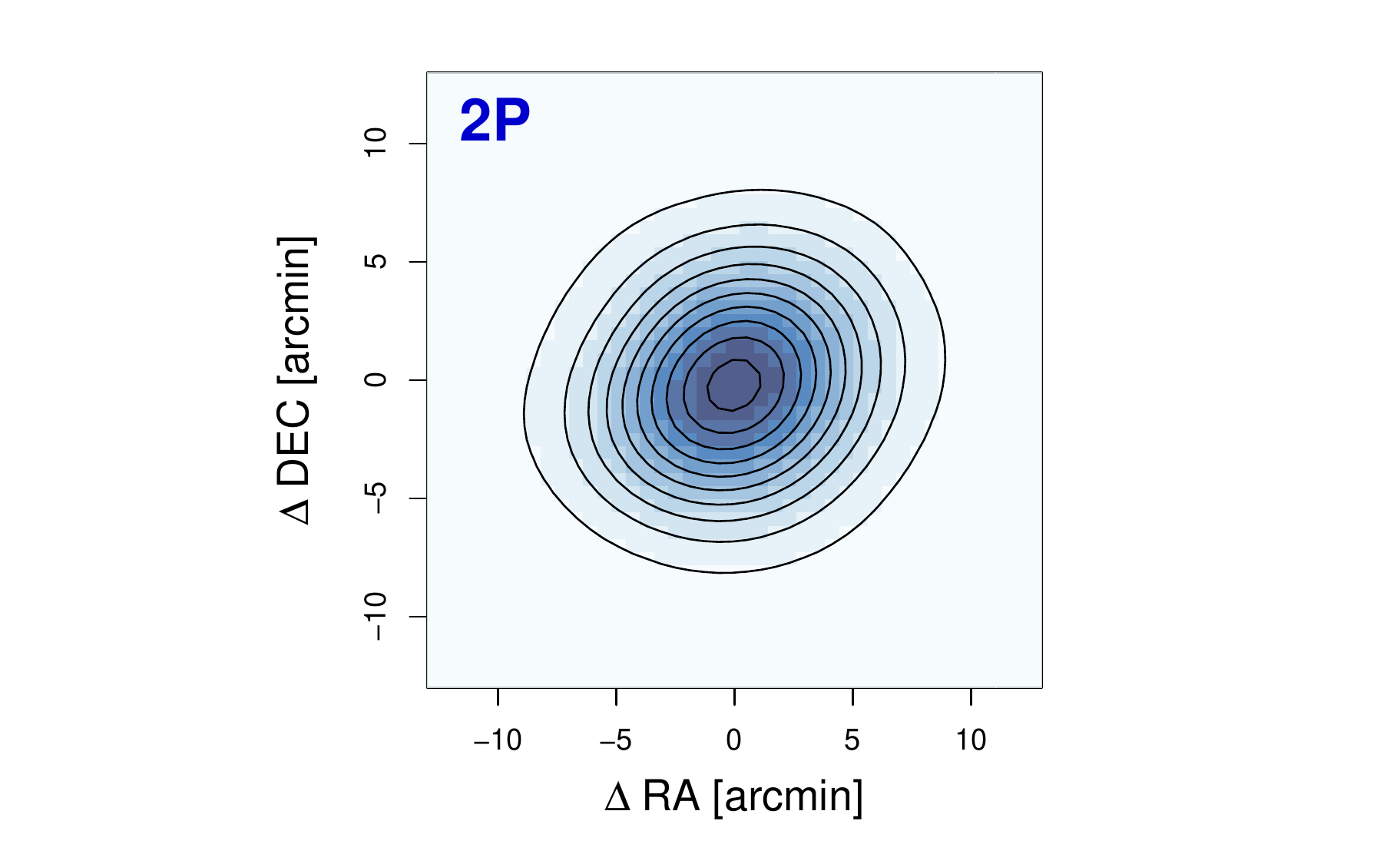}
    \includegraphics[width=8cm, trim={3.2cm 0.0cm 2.5cm 0.6cm},clip]{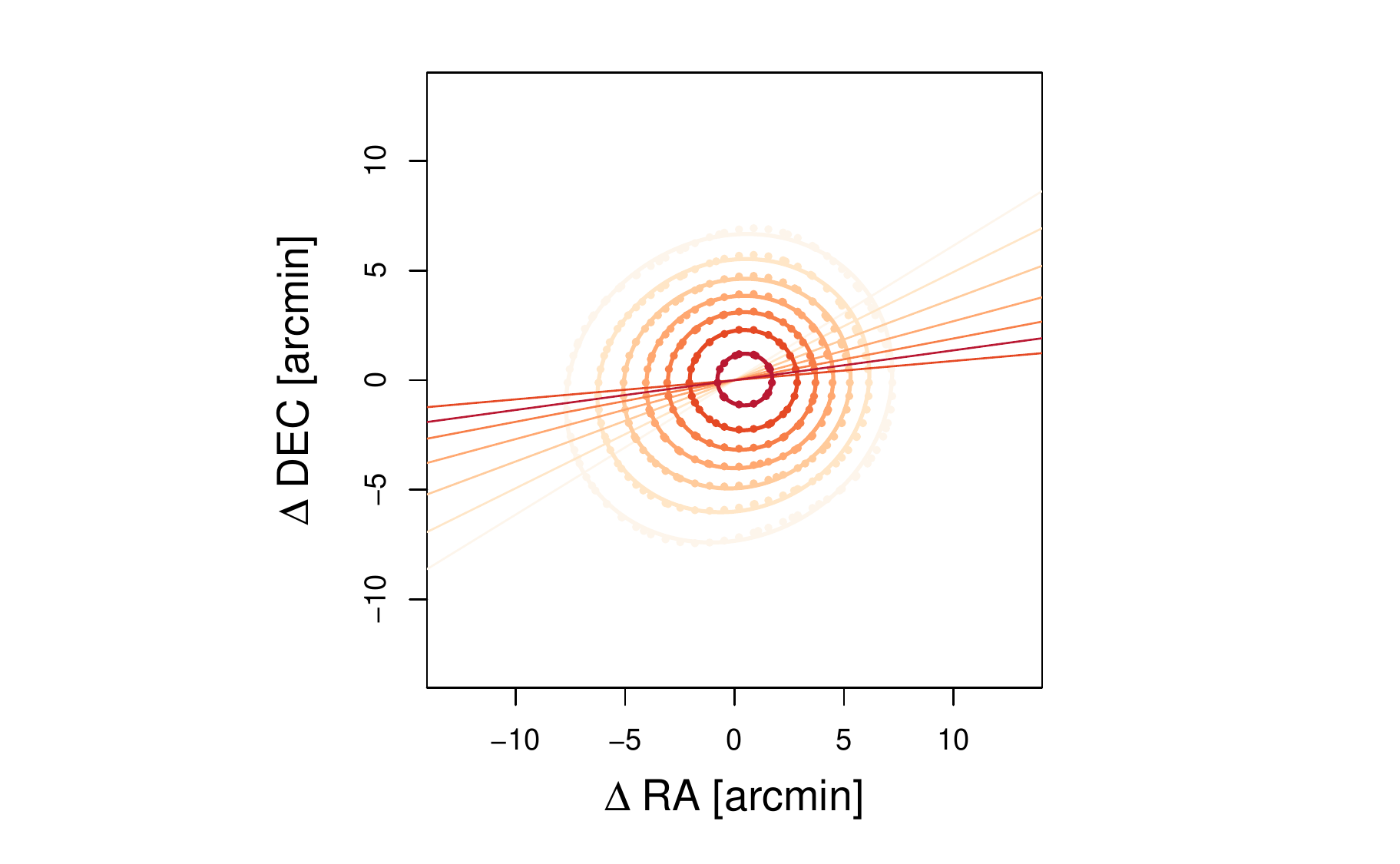}  
    \includegraphics[width=8cm, trim={3.2cm 0.0cm 2.5cm 0.6cm},clip]{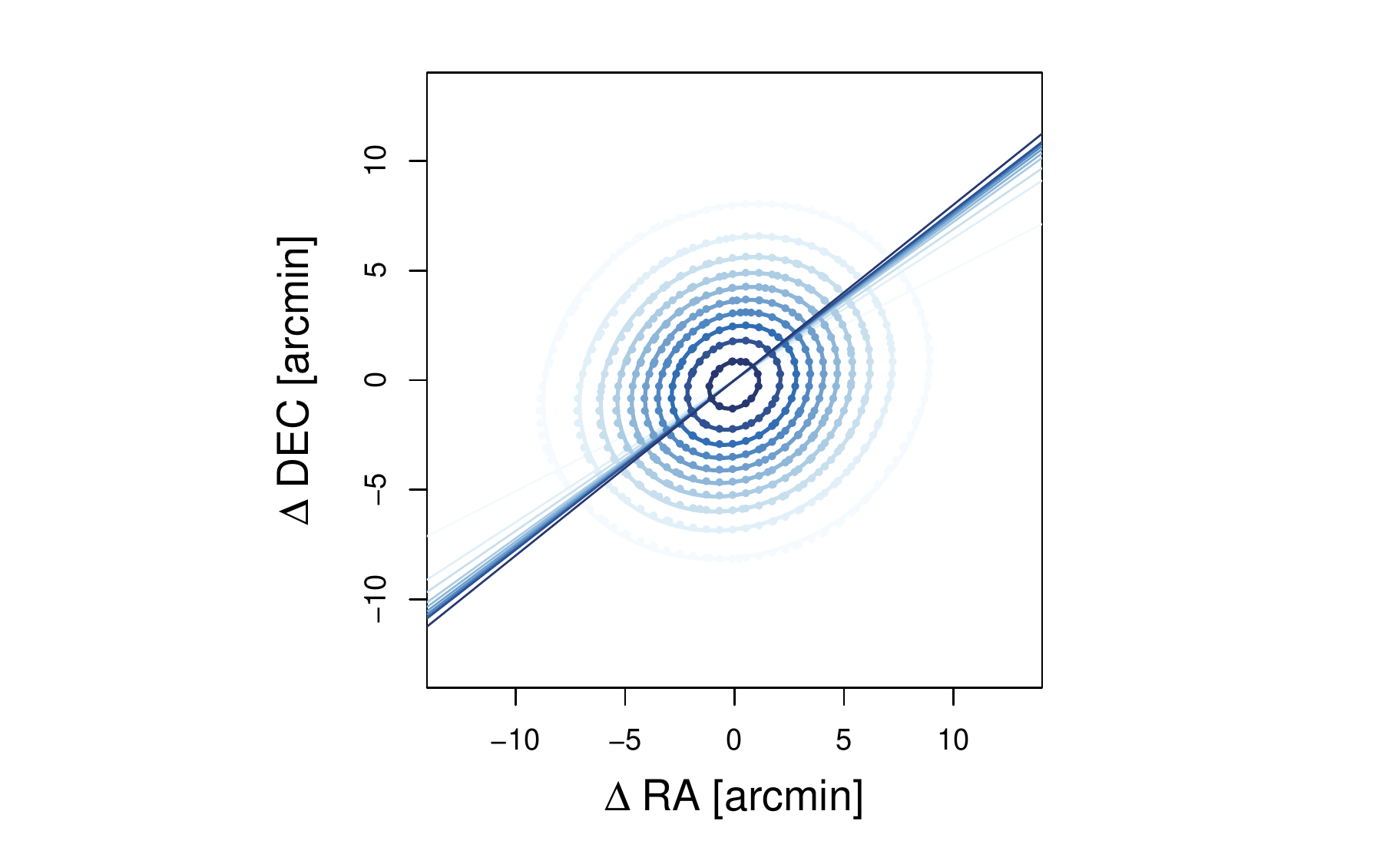}    
    \caption{Spatial distribution of multiple stellar populations in NGC\,5904. \textit{Top panels.} 2D Binned Kernel Density Estimate \citep{wand2015} with iso-density contour lines. \textit{Bottom panels.} Least squares fit ellipses to the iso-density contours. The ellipses have been interpolated using the algorithm described in \citet{Halir98}. }
    \label{fig: spatial dist}
\end{figure*}

In the following, we analyze the spatial distribution of the two groups of 1P and 2P stars that we identified in the previous section for the seven analyzed clusters. 
To do this, we used a procedure, which is based on the 2D Binned Kernel Density Estimate \citep{wand2015}, illustrated in the left and right panels of Figure~\ref{fig: spatial dist} for the first and second population of NGC\,5904, respectively. 
The levels of red and blue in the upper panels are indicative of the density of 1P and 2P stars  in a reference frame where the origin corresponds to the cluster center and the X and Y axes point towards the directions of increasing RA and DEC. 

\begin{figure*}
    \centering
    \includegraphics[width=17cm, trim={0.0cm 0.0cm 0.0cm 0.0cm},clip]{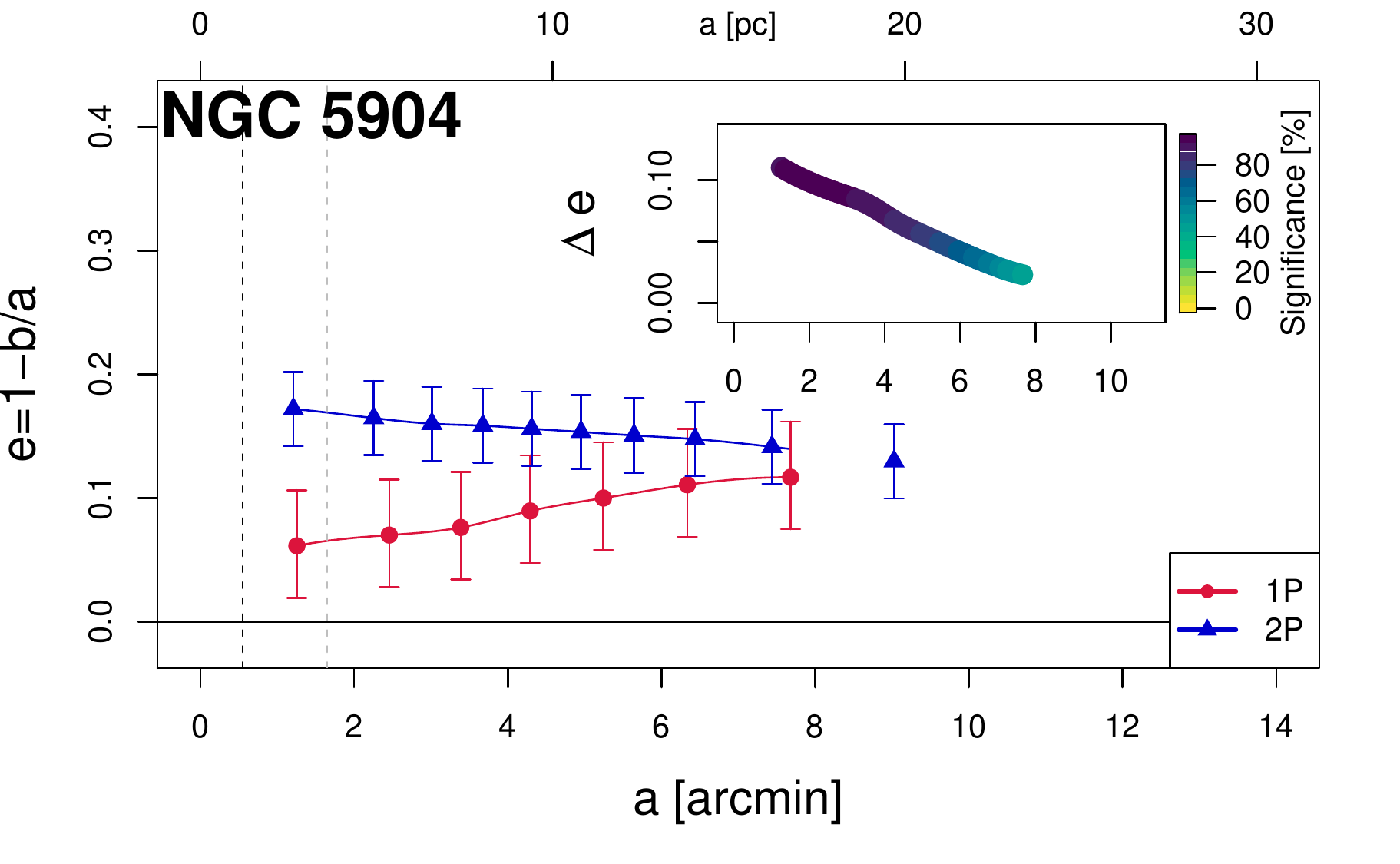}
    \caption{Ellipticity, e, of 1P and 2P (color coded in red and blue respectively) as a function of the semi-major axis, a. Black and gray dashed lines indicate the core radius $(R_{\rm c})$ and the half-light radius $(R_{\rm h})$, respectively. The inset shows the ellipticity difference between 2P and 1P stars, $\Delta {\rm e}$ against a. The colors indicate the significance of such difference, as indicated in the colorbar on the right. See text for details.}
    \label{fig:ell significance ngc5904}
\end{figure*}

We calculated six and nine iso-density contour lines for 1P and 2P respectively, spaced by 0.001 in normalized density unit (black lines in the upper panel).
We used the algorithm by \cite{ellfitting} to fit each contour line with an ellipse by means of least-squares and plot the best-fit ellipses in the bottom-left panel of Figure~\ref{fig: spatial dist}, where we also show the corresponding directions of the major axes.
A visual inspection of Figure~\ref{fig: spatial dist} suggests that 2P stars exhibit more-elongated distributions than the 1P. 
To quantify this fact, we define the ellipticity as $e=1-b/a$ where $a,b$ are respectively the semi-major and semi-minor axis of the interpolated ellipses. \\
The ellipticity radial profile is presented in  Figure~\ref{fig:ell significance ngc5904}.
The uncertainty associated with each ellipticity measurement is determined by means of bootstrapping 1,000 times with replacement.
Clearly, 2P stars exhibit larger values of $e$ than the 1P, as previously found in \citet{lee2017}. The ellipticity difference between 2P and 1P decreases from $\Delta(e) \sim 0.1$, at a radial distance of about 1 arcmin ($\sim$0.61 $R_{\rm h}$, $\sim$2 pc) from the cluster center, to $\sim$0.02 for $a \sim 8$ arcmin ($\sim$4.7 $R_{\rm h}$, $\sim$17 pc). 

To estimate the statistical significance of $\Delta(e)$ we sampled the observed radial profile of the cluster to
create a catalog of 100,000 stars with a radial distribution similar to the observed ones and with ellipticity $e=0$.
We selected two stellar groups with the same number of stars as observed for the 1P and the 2P, derived their ellipticity at different radial distances from the center and calculated $\Delta e_{\rm sim}$ in close analogy with what we did for the observed stars. 
Finally, we computed the ratio between the number of simulations where $\Delta e_{\rm sim} \ge \Delta e$ and the total number of simulations. This quantity corresponds to the probability that the observed ellipticity difference between 2P and 1P stars is not due to observational uncertainties.
As shown in the inset of Figure~\ref{fig:ell significance ngc5904}, where we plot $\Delta e$ against the semi-major axis of the corresponding the best-fit ellipse, the significance of the ellipticity difference between the stellar populations of NGC\,5904 ranges from more than 90\% in the innermost regions, to $\sim$60\% for $a \sim$ 8 arcmin.

\begin{figure*}
    \centering
    \includegraphics[height=4.5cm, trim={0.0cm 1.2cm 0.0cm 0.1cm},clip]{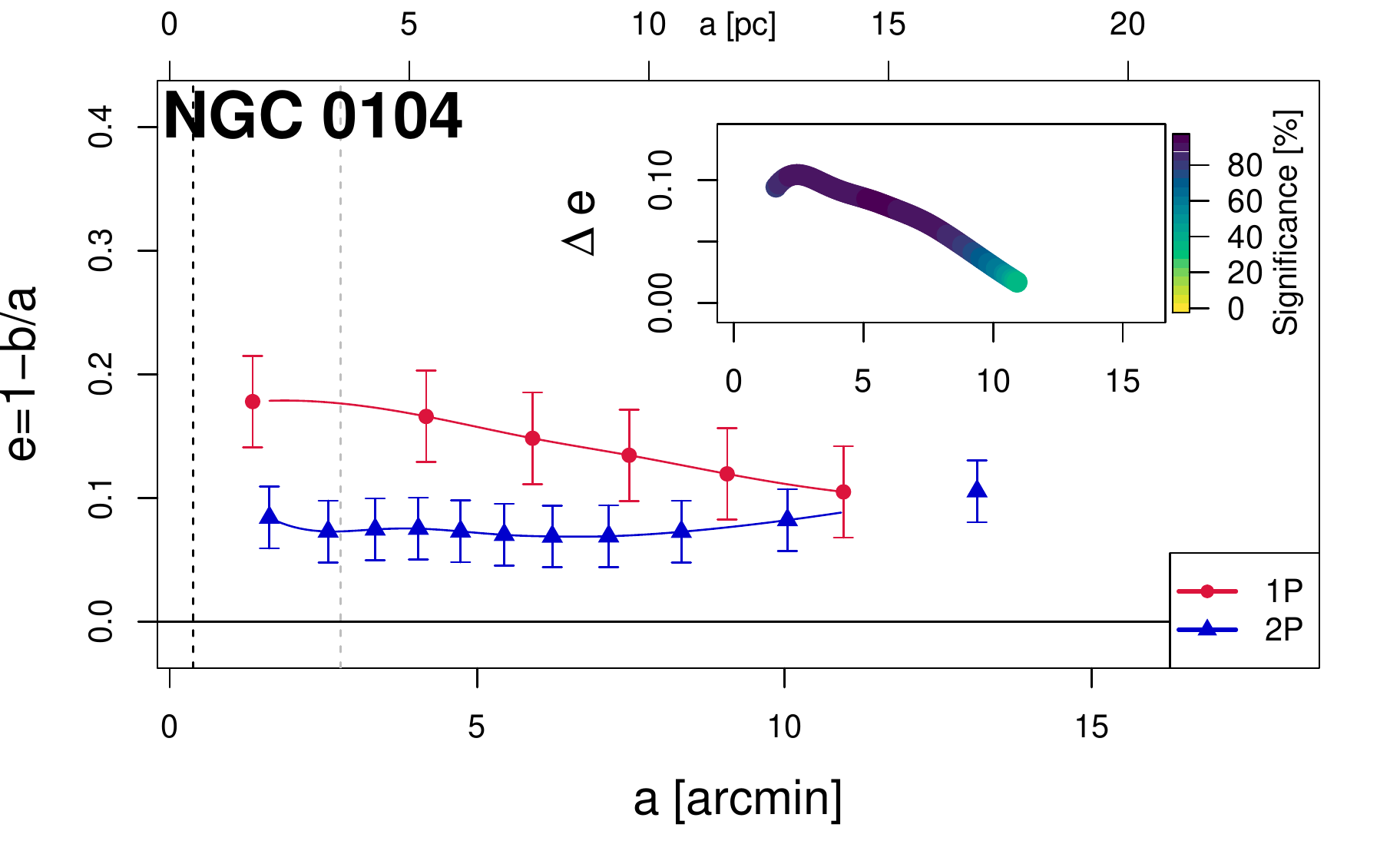}
    \includegraphics[height=4.5cm, trim={0.9cm 1.2cm 0.0cm 0.1cm},clip]{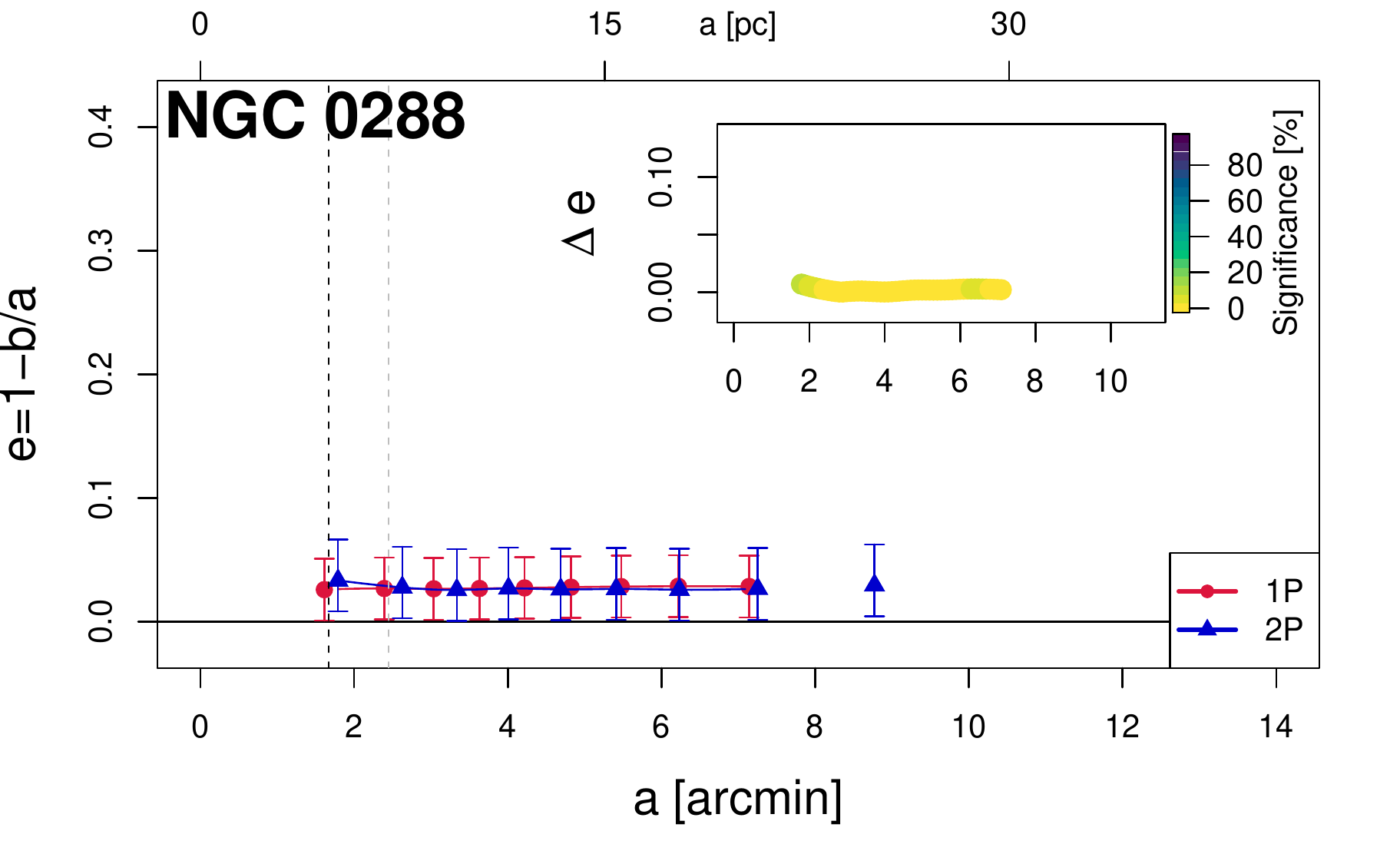}
    \includegraphics[height=4.5cm, trim={0.0cm 1.2cm 0.0cm 0.1cm},clip]{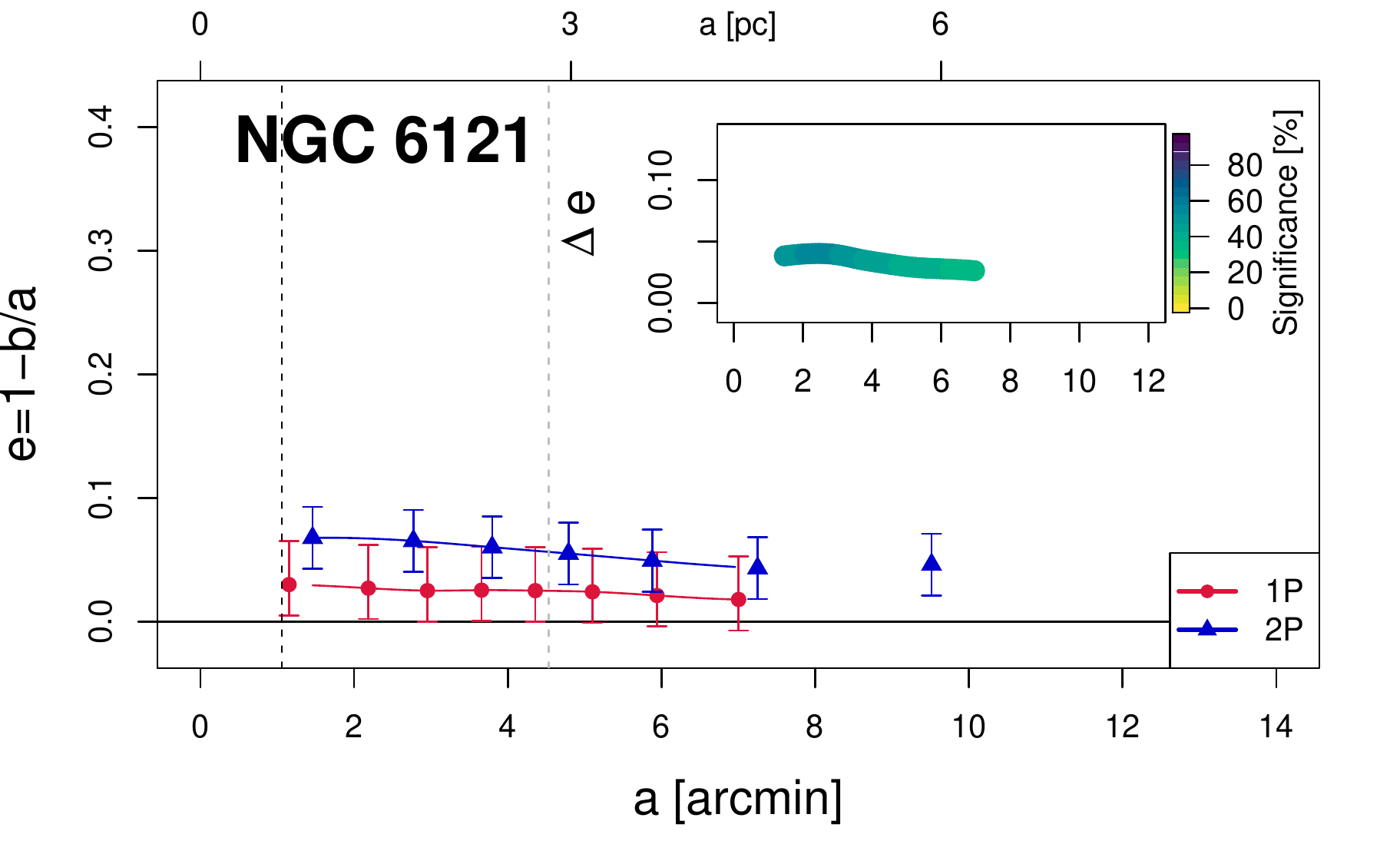}
    \includegraphics[height=4.5cm, trim={0.9cm 1.2cm 0.0cm 0.1cm},clip]{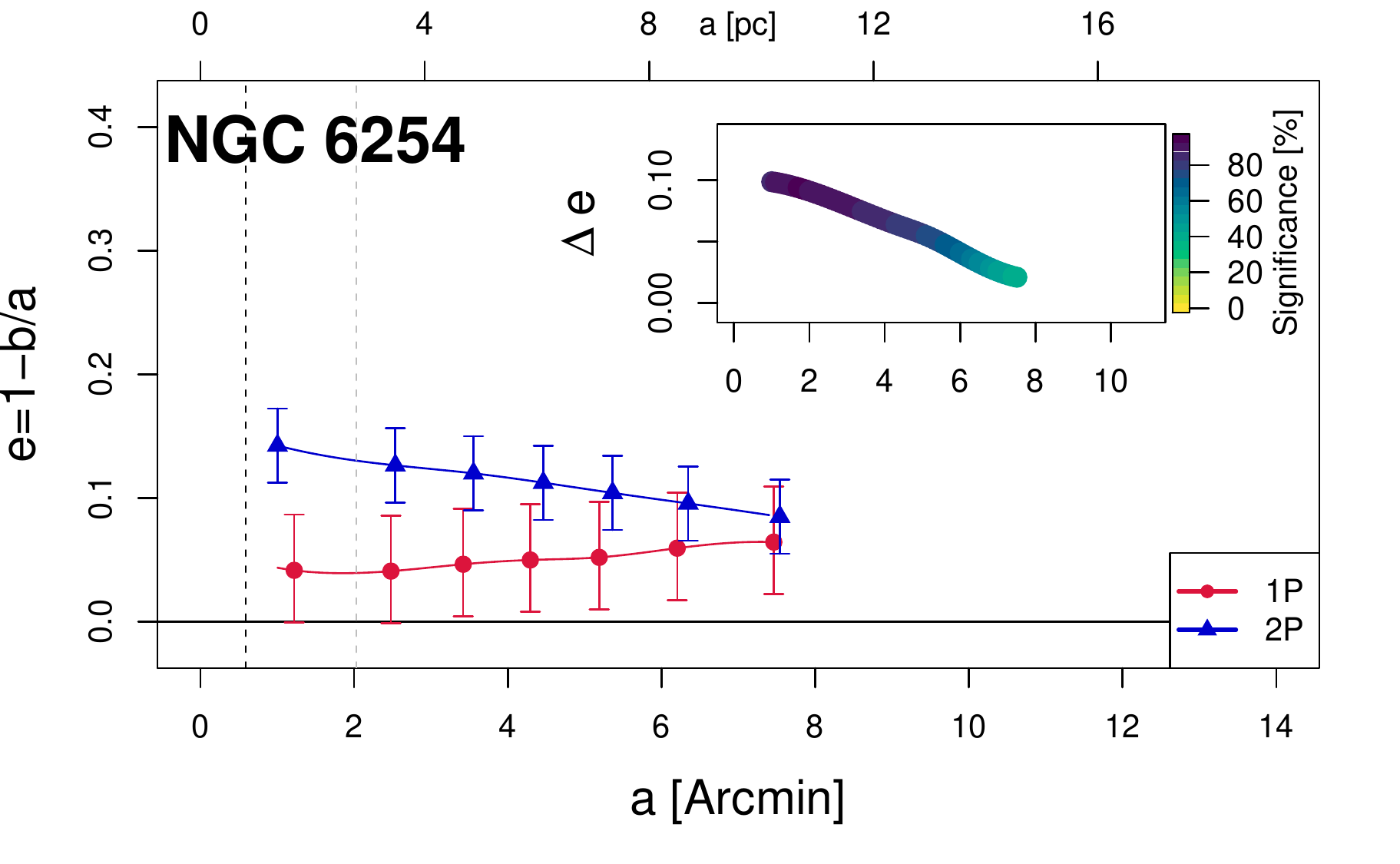}
    \includegraphics[height=5.1cm, trim={0.0cm 0.0cm 0.0cm 0.1cm},clip]{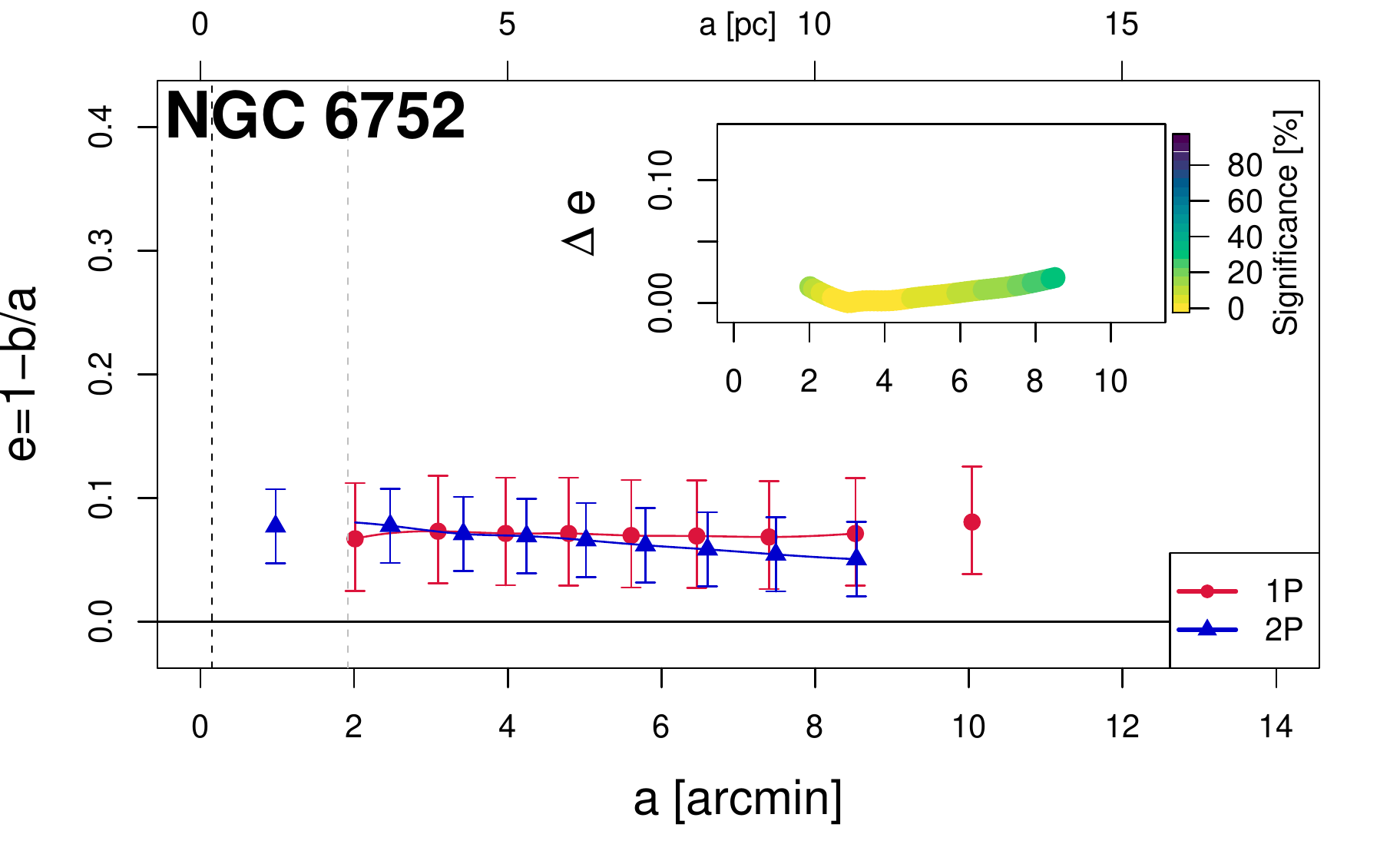}
    \includegraphics[height=5.1cm, trim={0.9cm 0.0cm 0.0cm 0.1cm},clip]{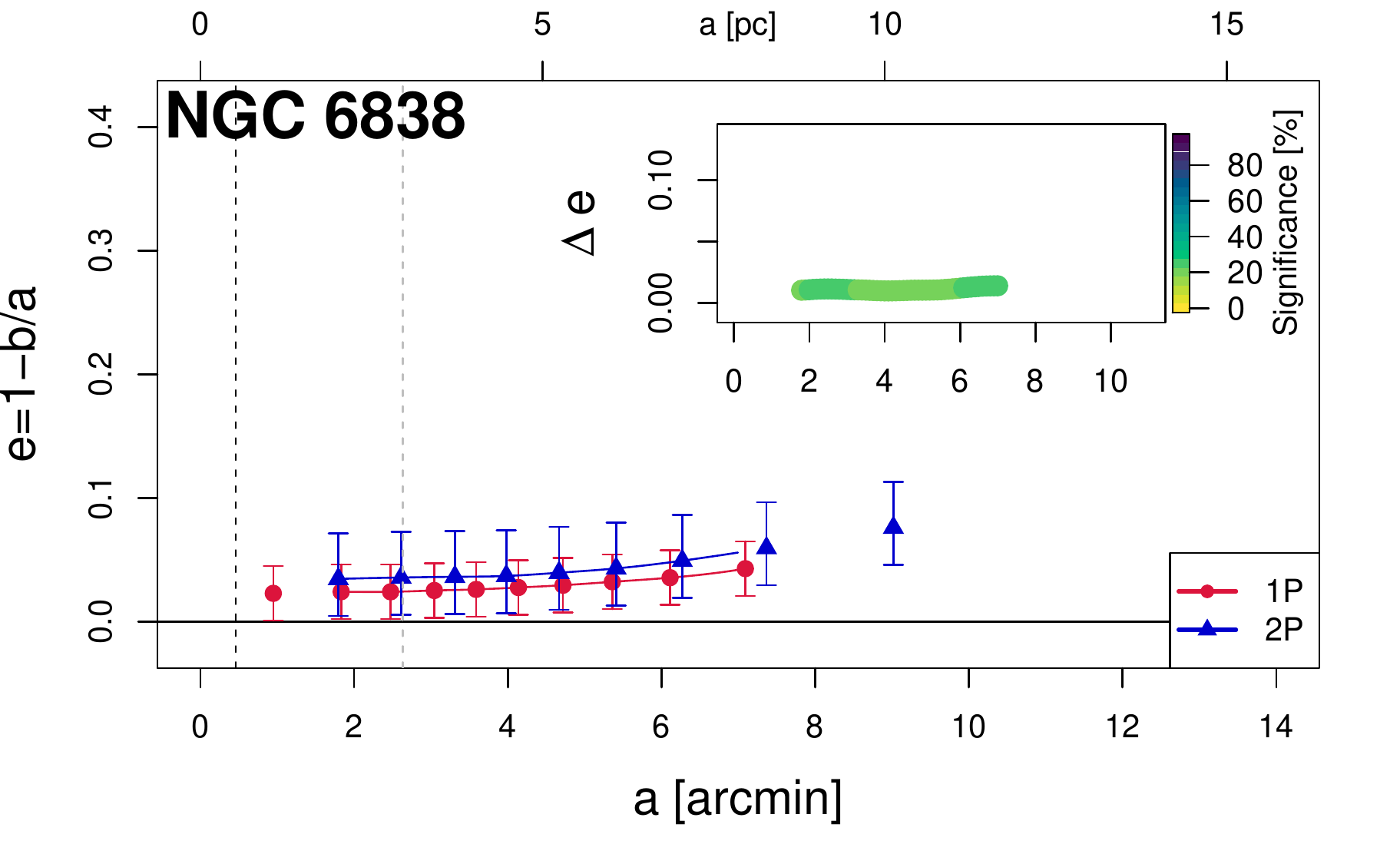}
    \caption{Same as Figure~\ref{fig:ell significance ngc5904} for NGC\,0104, NGC\,0288, NGC\,6121, NGC\,6254, NGC\,6752 and NGC\,6838.}
    \label{fig: ellProfile all}
\end{figure*}

Results for the other clusters are shown in Figure~\ref{fig: ellProfile all}.
We find significant differences in the spatial distribution of 1P and 2P stars in NGC\,0104, NGC\,5904 and NGC\,6254. The remaining clusters do not show hints of different distribution between both populations. It is worth mentioning that while NGC\,5904 and NGC\,6254 are consistent with a more elliptical 2P, NGC\,0104 shows the opposite trend, with a more elliptical 1P. 

\section{Rotation of multiple populations.}
\label{sec:rot}

\begin{figure*}
  \centering

  \includegraphics[width=8.cm, height=8.6cm,trim={0cm 0.1cm 0cm 0cm},clip]{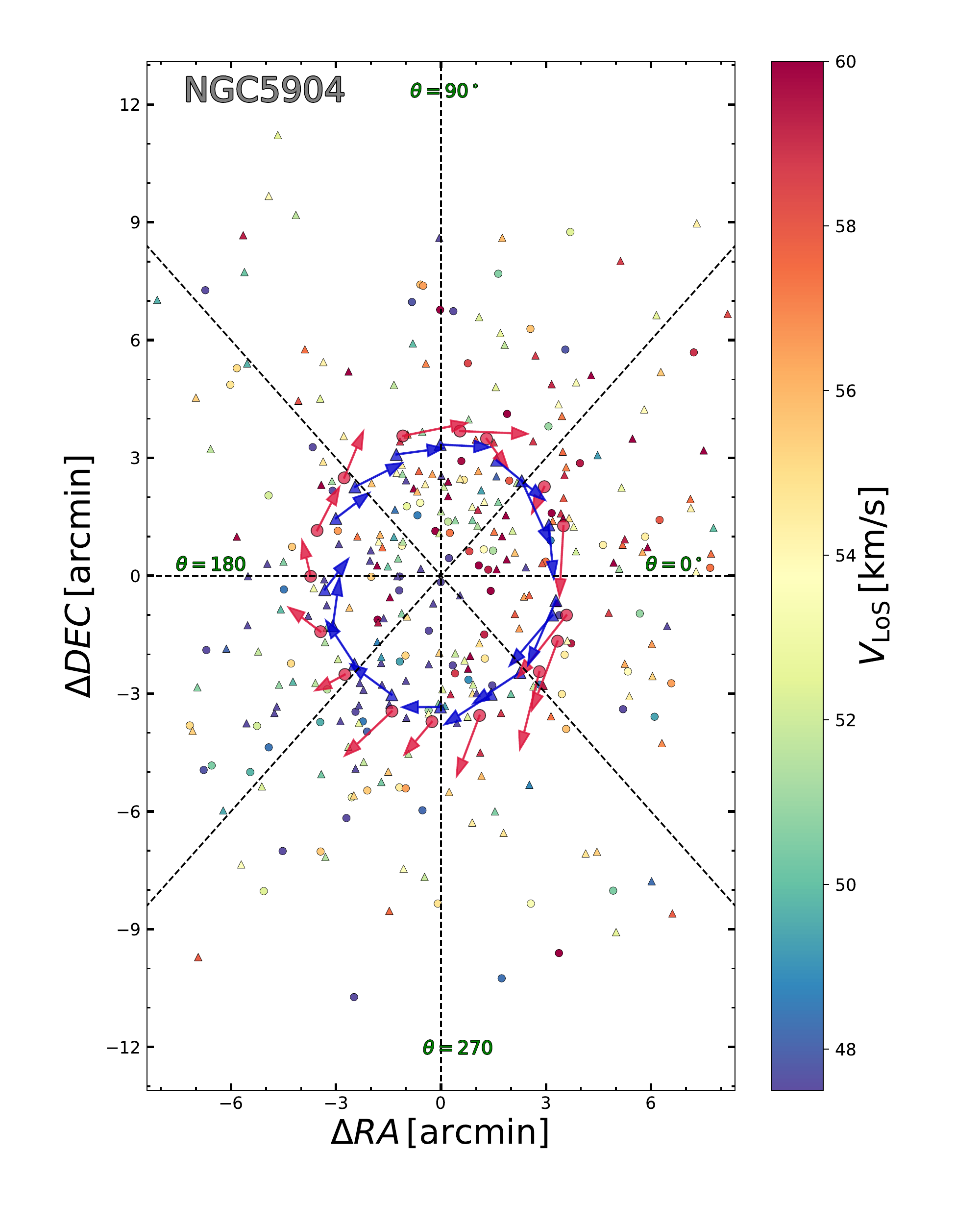}
  \includegraphics[width=9.8cm,trim={0cm 0cm 0cm 0.cm},clip]{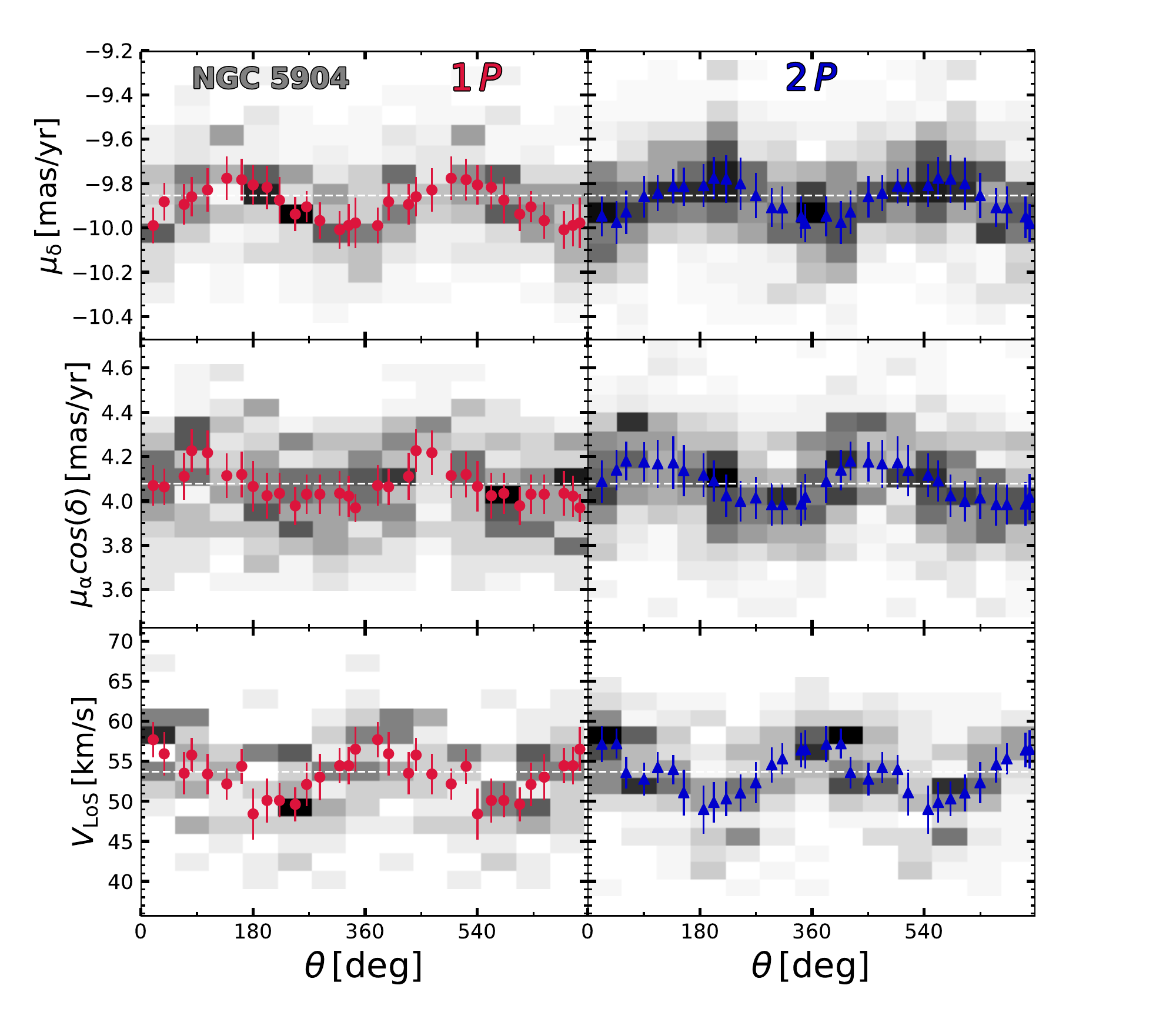} 
  \caption{\textit{Left}.Relative position of the analyzed RGB stars of NGC\,5904 with respect to the cluster center. 1P and 2P stars are shown with circles and triangles, while the color is representative of the LoS velocity as shown in the colorbar. The red and blue arrows indicate the average rotation field, in the plane of the sky, of 1P stars and 2P stars for the 16 analyzed circular sectors. \textit{Right.} $\mu_{\alpha}\cos{\delta}$, $\mu_{\delta}$ and $V_{\rm LoS}$ as a function of the position angle, $\theta$, for 1P and 2P stars of NGC\,5904. The gray levels are indicative of the density of stars. The red dots and the blue triangles represent the average motions of 1P and 2P stars in angular sectors.}

  \label{fig:NGC5904pos}
\end{figure*}

In the following, we investigate the rotation in the plane of the sky  and along the line-of-sight (LoS) for the selected 1P and 2P stars by using the procedure illustrated in Figure~\ref{fig:NGC5904pos} for NGC\,5904. We applied the orthographic projection of the celestial coordinates and converted  proper motions by using Equation~2 from  \citet[][]{gaia2018b}.

In the left panel of Figure~\ref{fig:NGC5904pos} we plotted the positions of the selected 1P and 2P stars relative to the cluster center and defined the angle $\theta$.
In the right panels of the same figure we show the density diagrams of the proper-motion and LoS velocity components
($\mu_\alpha\cos\delta,\mu_\delta, V_{\rm LoS}$) of each population against $\theta$. 
We divided the field of view in sixteen circular sectors with arc length of $45^\circ$  by using a method based on the naive estimator \citep{silverman1986}. Specifically, we defined a series of points separated by arc length of $l=45^\circ$. The circular sectors are defined over a grid  of points that are separated by steps of $l/2$ in arc length.
We calculated the median proper motions and angular positions of stars in each circular sector. 
The median values are superimposed on the density plots in the right panels of Figure~\ref{fig:NGC5904pos}. 

A visual inspection of this figure reveals that the proper motions of both 1P and 2P stars of NGC\,5904 exhibit sinusoidal patterns, thus suggesting that both populations are rotating. 
  
\begin{figure*}
  \centering
  \includegraphics[width=8.25cm,trim={0.3cm 3.05cm 0cm 4.5cm},clip]{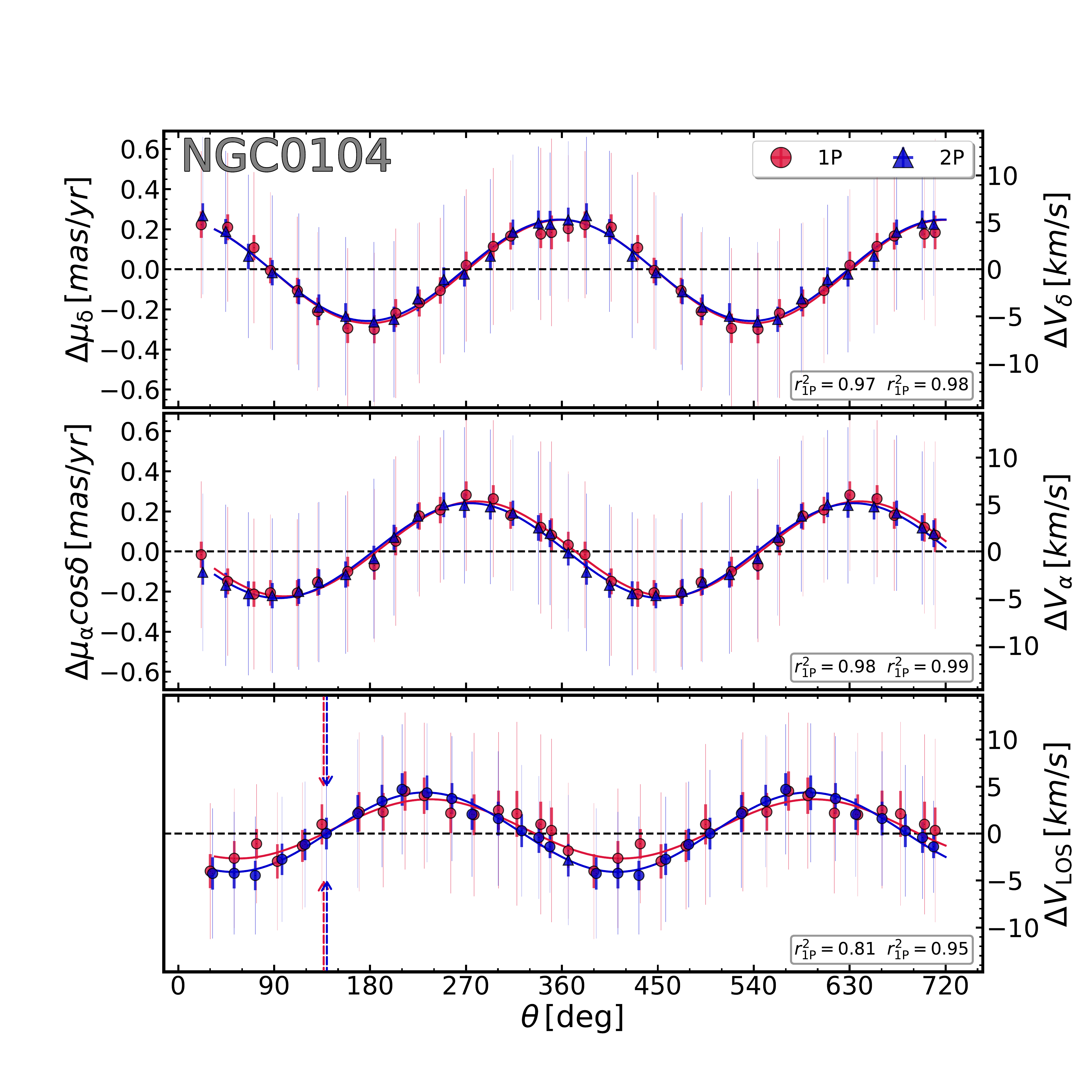}
  \includegraphics[width=8.25cm,trim={0.3cm 3.05cm 0cm 4.5cm},clip]{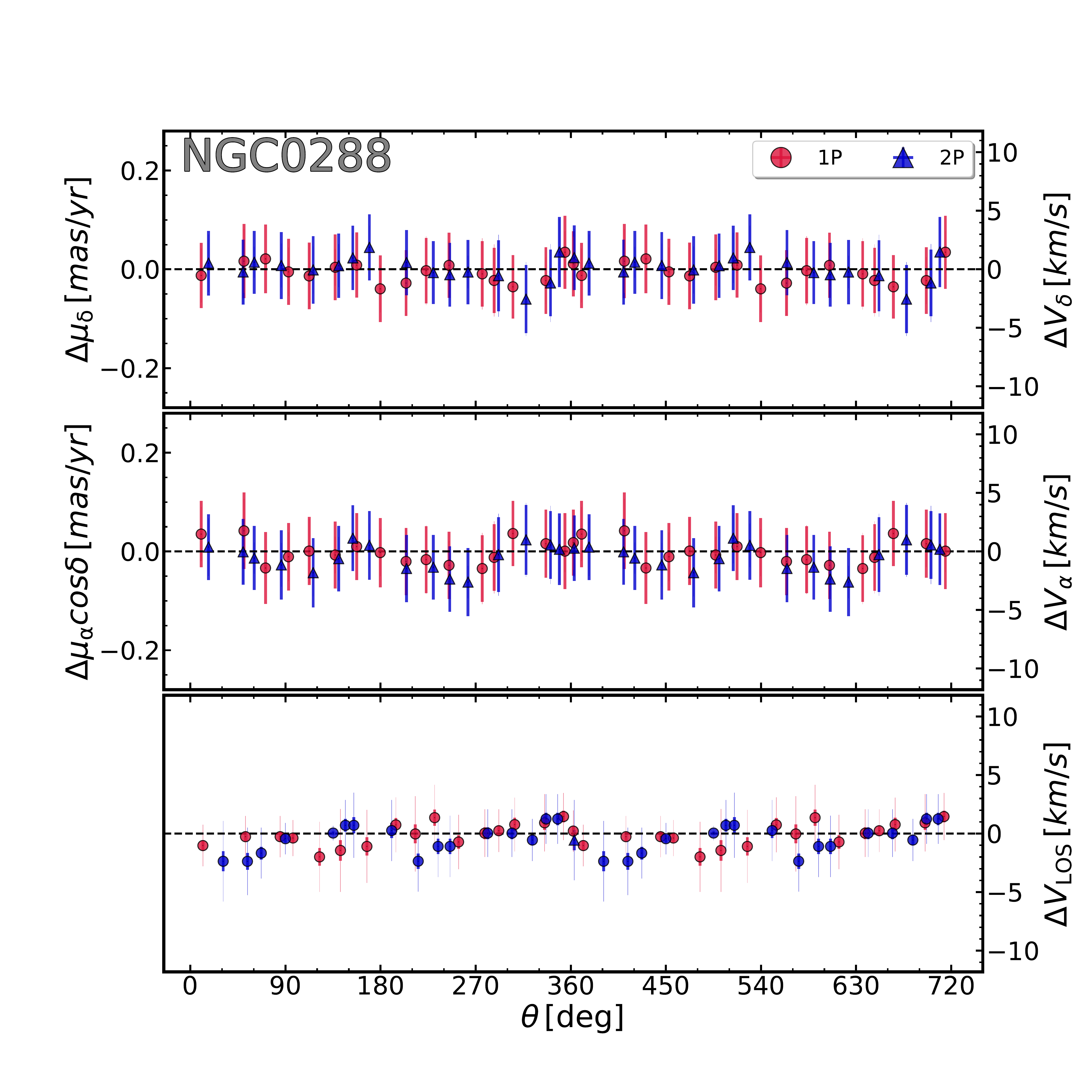}
  \includegraphics[width=8.25cm,trim={0.3cm 3.05cm 0cm 4.5cm},clip]{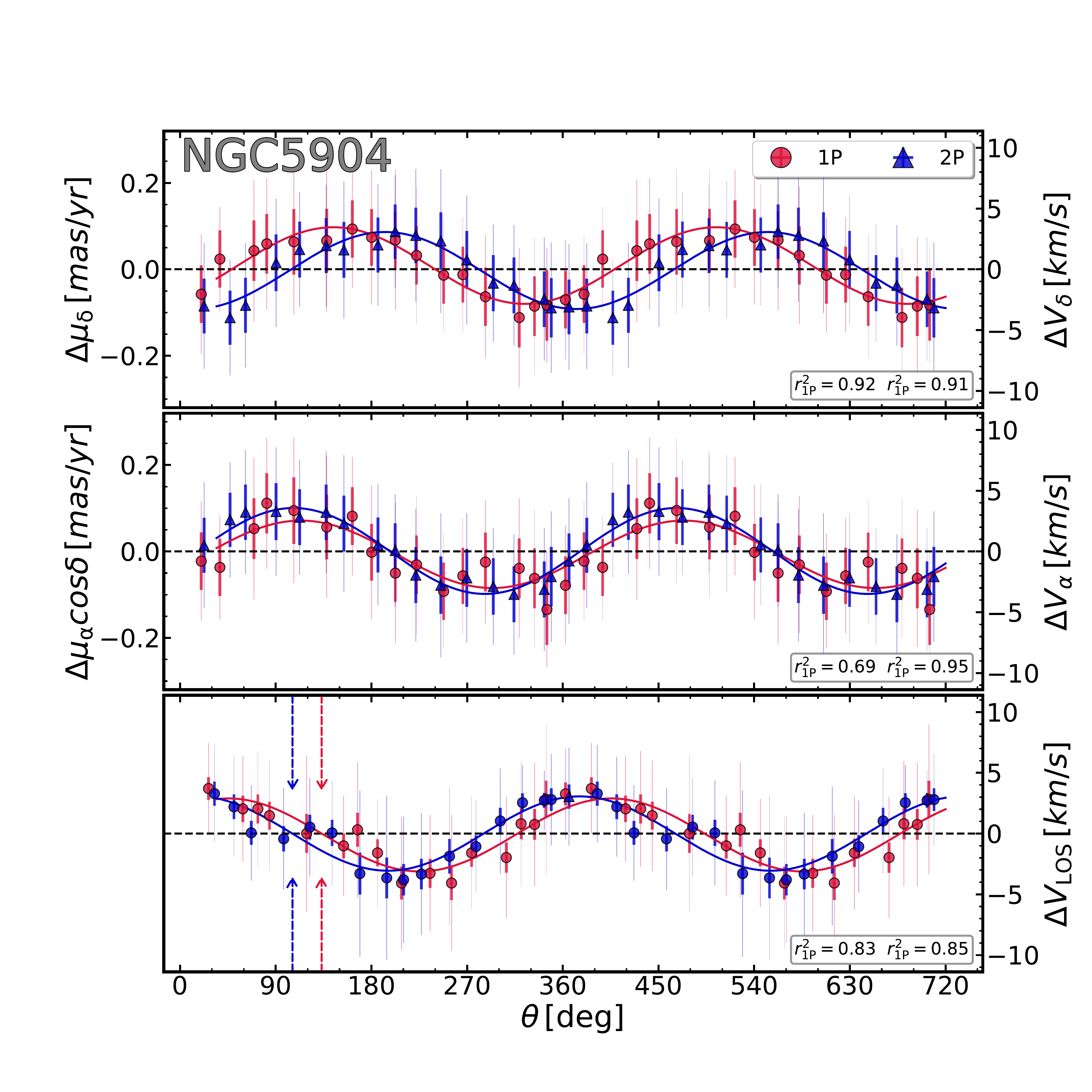}
  \includegraphics[width=8.25cm,trim={0.3cm 3.05cm 0cm 4.5cm},clip]{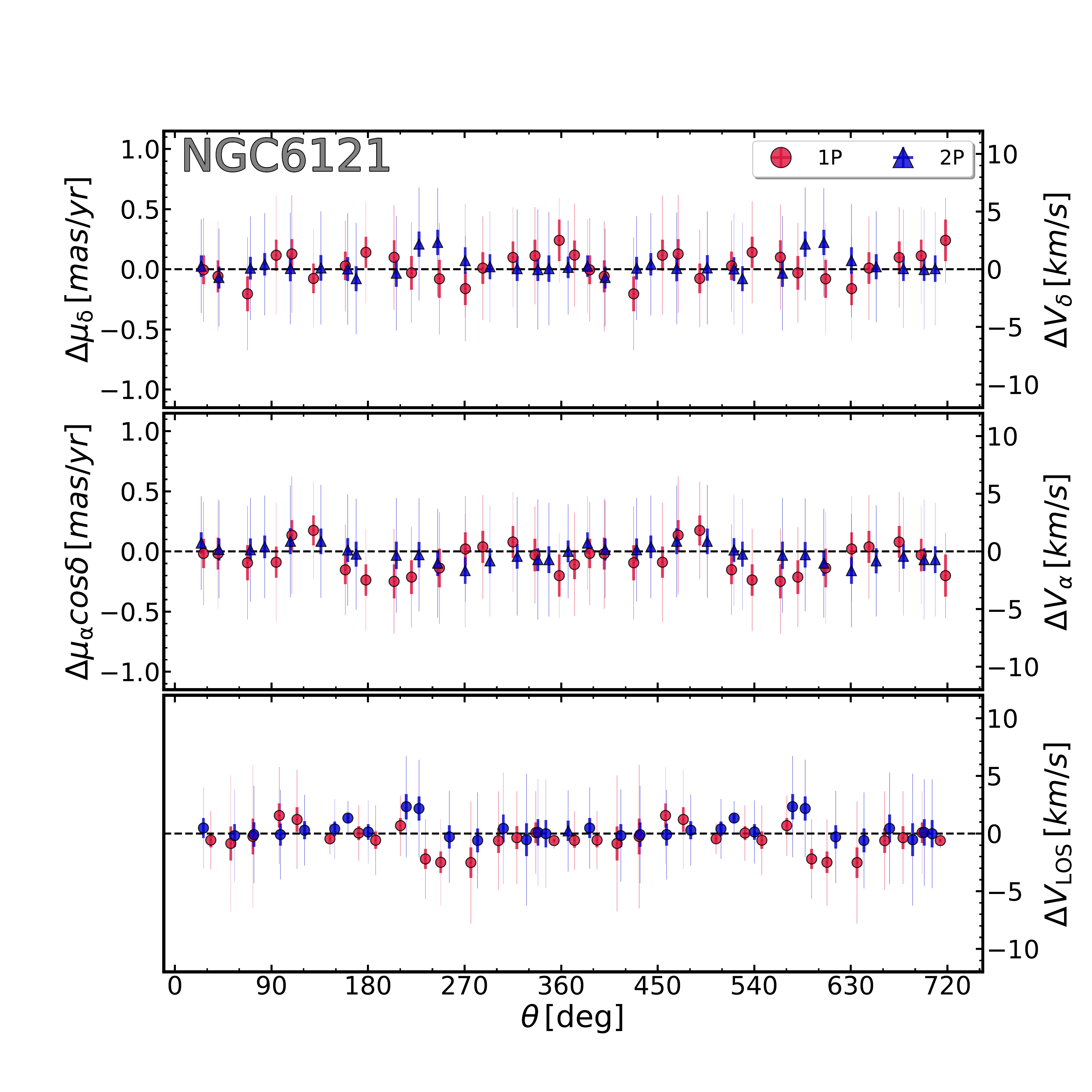}
  \includegraphics[width=8.25cm,trim={0.3cm 1cm 0cm 4.5cm},clip]{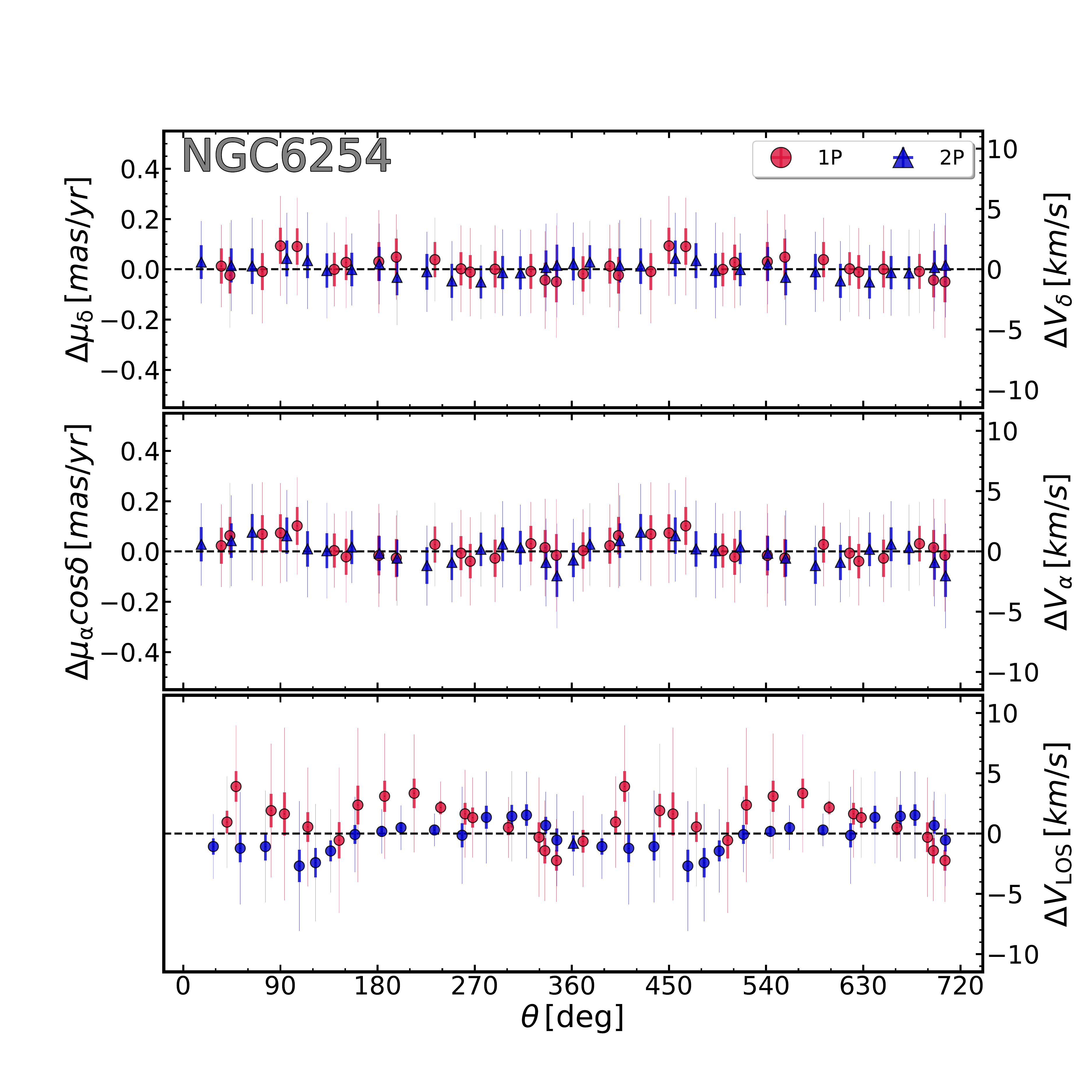}
  \includegraphics[width=8.25cm,trim={0.3cm 1cm 0cm 4.5cm},clip]{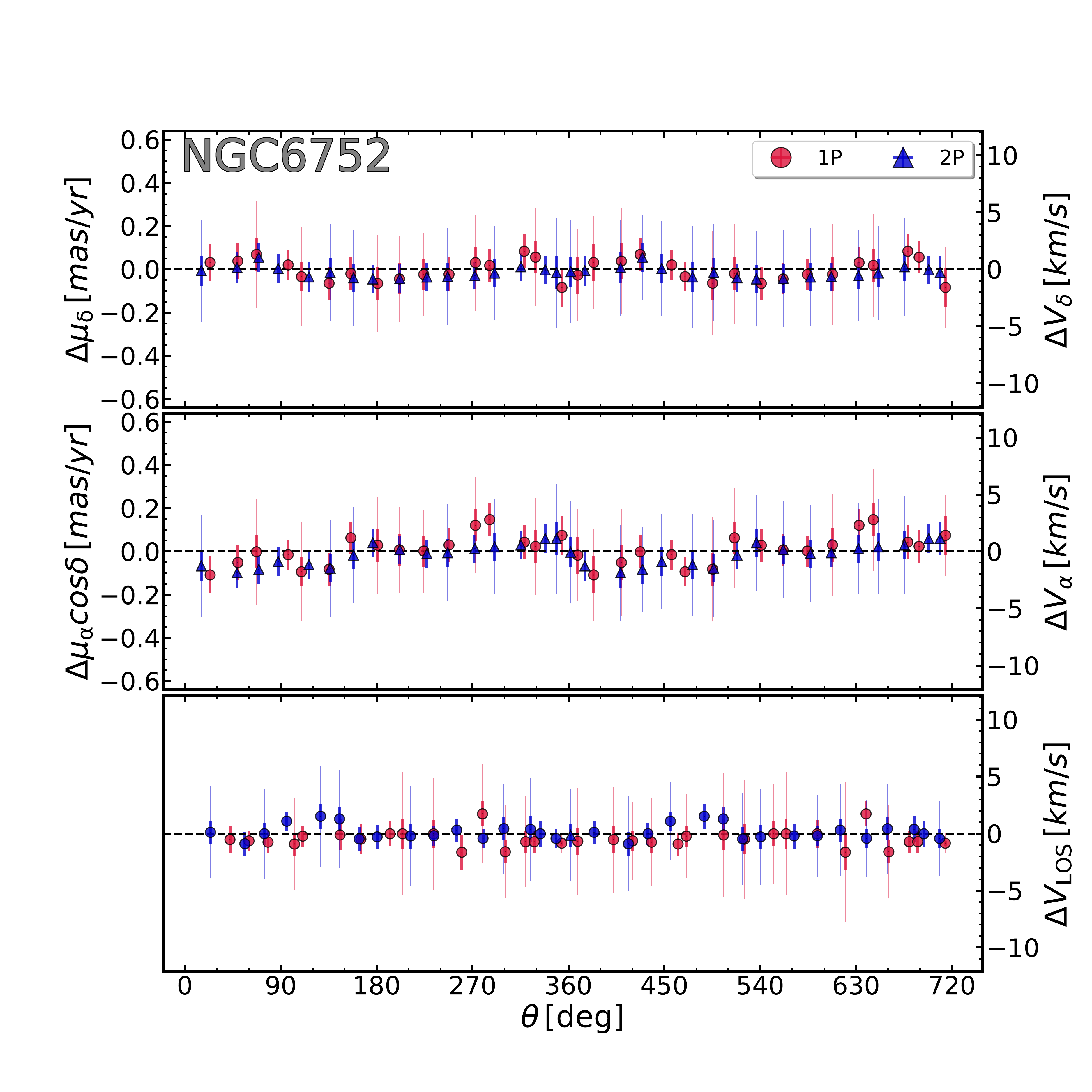}
  \caption{Average proper motions (top and middle panels) and LoS velocities (bottom panels) of 1P (red circles) and 2P stars (blue triangles) as a function of the position angle for 1P and 2P stars of NGC\,0104, NGC\,0288, NCG\,5904, NGC\,6121, NGC\,6254 and NGC\,6752. Thin and thick error bars indicate the uncertainties associated with the average motions and the dispersions, respectively.
    The red and blue lines superimposed on the plots of NGC\,0104 and NGC\,5904 are the least-squares best-fit functions of 1P and 2P stars, respectively. 
    The vertical red/blue arrows plotted in the bottom panels highlight the PA of the rotation axis of 1P and 2P stars, determined as the zero of the rotation curve.}
  \label{fig:AllCLrot}
\end{figure*}

To investigate the rotation of 1P and 2P stars of all the GCs, we calculated the quantities $\Delta \mu_{\alpha} cos{\delta}$, $\Delta \mu_{\delta}$ and $\Delta$V$_{\rm LoS}$   respectively corresponding to the difference between the $\mu_{\alpha} cos{\delta}$, $\mu_{\delta}$ and V$_{\rm LoS}$ of each star, and the cluster median motion. 
Results are shown in Figure~\ref{fig:AllCLrot} where we plot for each cluster the median values of $\Delta \mu_{\alpha} cos{\delta}$, $\Delta \mu_{\delta}$ and $\Delta$V$_{\rm LoS}$ calculated in sixteen circular sectors as a function of $\theta$. 
This analysis suggests that NGC\,0104 and NGC\,5904 are the only two clusters with clear evidence of rotation among both 1P and 2P stars. Remarkably, 1P and 2P stars follow the same random pattern in all the clusters with the possible exception of NGC\,5904\footnote{Work based on N-body simulations \citep[e.g.][]{vesperini2013,mastrobuono2016,tiongco2019} suggest that the force of rotation should vary within the cluster field, as a function of radial distance. 
 Due to the small number of available 1P and 2P stars in each GC, we performed a global analysis that is based on the rotation of stars at different radial distances from the cluster center.
 NGC\,0104 is the only cluster that contains a sufficient number of stars to study rotation in different radial bins, as discussed in Section \ref{subsec:diff rot}.}. 

To quantify the rotation of each population of NGC\,0104 and NGC\,5904 and estimate its amplitude, $A$, and phase, $\phi$, we performed least-squares fitting to all 1P and 2P stars of the function:
\begin{equation}\label{eq:1}
f(\theta)=M+A\cdot\sin(F\cdot\theta+\phi)     
\end{equation}
where $M$ is the zero point of the sine function and $F$ is the frequency.
We exploit the $r^2$ parameter  \citep{glantz} to estimate the statistical significance of the fit:
\begin{equation}\label{eq:2}
r^2=1-\frac{\sum_i  (y_i-f(\theta,i))^2}{\sum_i(f (\theta, i)-\bar{y})^2}
\end{equation}
where $y_{i}$ is the value of $\mu_\alpha\cos\delta (\mu_\delta$) for each star, $i$, $\theta$ is the corresponding position angle, $\bar{y}$ is the average value of $y$, and $f$ is the best-fit function.  
This parameter quantifies the goodness of the fit of a linear function, with the perfect match corresponding to $r^2=1$. We then eye-checked every cluster for consistency between the interpolation and the value of $r^2$. 

The values of $r^2$ for NGC\,0288, NGC\,6121, NGC\,6254, NGC\,6752 and NGC\,6838 are 
 smaller than 0.5 thus demonstrating that the observations are poorly reproduced by Equation~\ref{eq:1}. Hence, there is no evidence for rotation among 1P and 2P stars of these clusters.   

In contrast, NGC\,0104 and NGC\,5904 exhibit a reliable match between the function of Equation~\ref{eq:1} for both populations. 
The obtained $r^{2}$ values for 1P and 2P stars are listed in the bottom right insets of Figure~\ref{fig:AllCLrot} and are larger than 0.6. 
The best-fit functions to all 1P and 2P stars for these two clusters are shown in Figure~\ref{fig:AllCLrot}. 

Once established that 1P and 2P stars of NGC\,0104 and NGC\,5904 rotate, we can further explore the rotation pattern of different stellar populations in these two clusters.
 
\begin{figure}
    \centering
      \includegraphics[width=8.5cm,trim={0.3cm 1cm 0cm 2.5cm},clip]{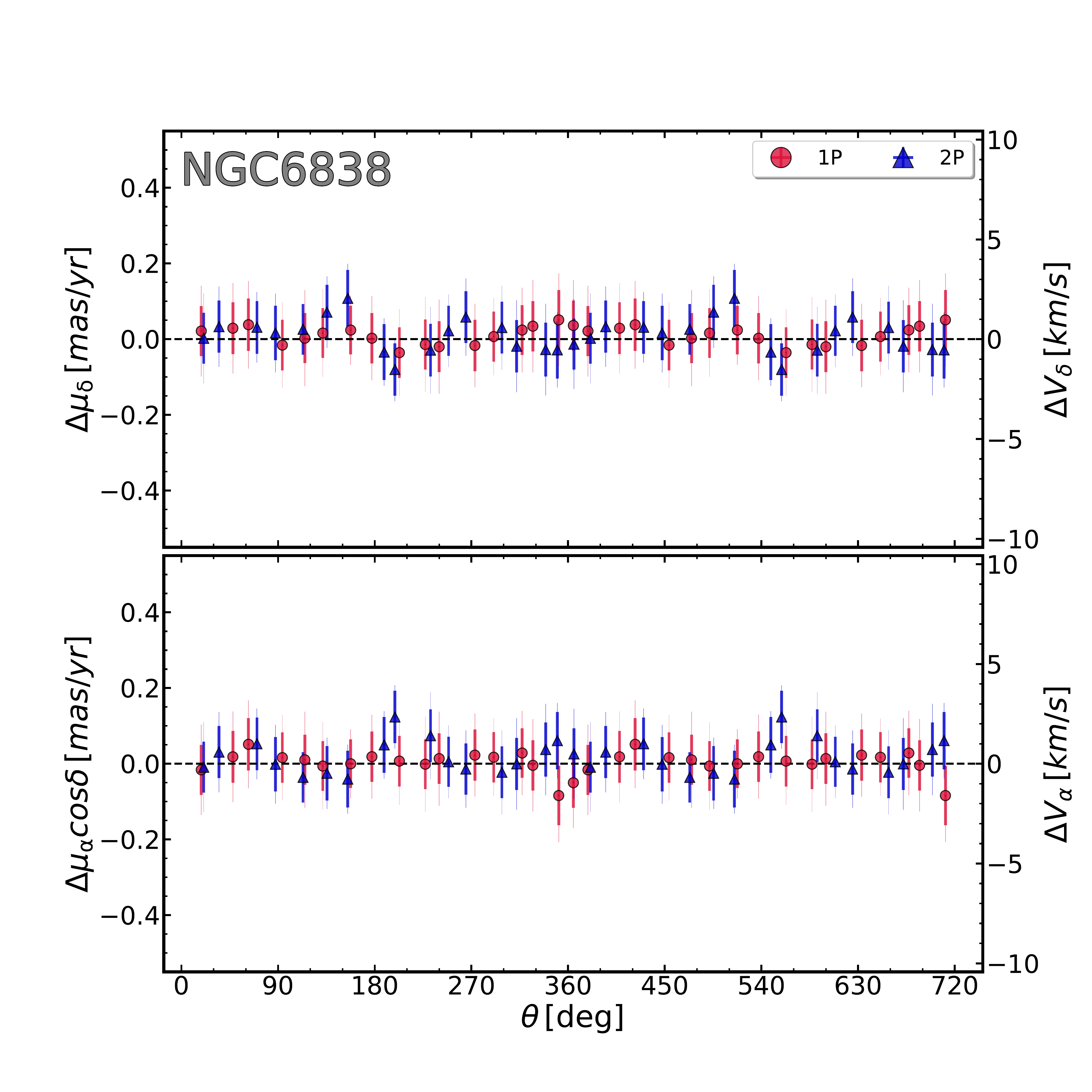}
    \caption{Same as Figure \ref{fig:AllCLrot} for NGC\,6838. LoS velocities are not available for NGC\,6838.}
    \label{fig:rot6838}
\end{figure}

\begin{table*}
\centering
\begin{tabular}{lll|cccc}
\toprule
\toprule
 ID &  & field & motion component & $\phi$ (PA) & $A$ \\
  &  &  &  & rad & mas/yr (km/s) \\
\midrule
 NGC\,0104  & 1P & all  & $\mu_\alpha\cos\delta$ & 2.96$\pm$0.06 & 0.237$\pm$0.007    \\ 
            &    &      & $\mu_\delta$           & 1.54$\pm$0.07 & 0.257$\pm$0.008    \\ 
            &    &      & $V_{\rm LoS}$          & 5.53$\pm$0.30 & 3.35$\pm$0.35      \\ 

		    & 2P &      & $\mu_\alpha\cos\delta$ & 3.13$\pm$0.04 & 0.236$\pm$0.005    \\ 
            &    &      & $\mu_\delta$           & 1.61$\pm$0.05 & 0.253$\pm$0.006    \\ 
            &    &      & $V_{\rm LoS}$          & 5.50$\pm$0.16 & 4.23$\pm$0.18     \\

\hline
NGC\,5904   & 1P & all  & $\mu_\alpha\cos\delta$ & 2.691$\pm$0.24 & 0.078$\pm$0.010 \\ 
            &    &      & $\mu_\delta$           & 2.254$\pm$0.11 & 0.089$\pm$0.007  \\ 
            &    &      & $V_{\rm LoS}$          & 2.44$\pm$0.33 & 2.69$\pm$0.40  \\ 
		    & 2P &      & $\mu_\alpha\cos\delta$ & 2.91$\pm$0.11 & 0.099$\pm$0.007 \\ 
            &    &      & $\mu_\delta$           & 1.38$\pm$0.13 & 0.089$\pm$0.006  \\ 
            &    &      & $V_{\rm LoS}$          & 1.90$\pm$0.19 & 3.20$\pm$0.27  \\ 

\hline \hline
NGC\,0104   & 1P & inner  & $\mu_\alpha\cos\delta$ & 3.10$\pm$0.08 & 0.288$\pm$0.013 \\ 
            &    &        & $\mu_\delta$           & 1.50$\pm$0.14 & 0.265$\pm$0.021  \\ 
            &    &        & $V_{\rm LoS}$          & 4.40$\pm$0.39 & 4.29$\pm$0.57  \\ 

		    & 2P &        & $\mu_\alpha\cos\delta$ & 3.18$\pm$0.05 & 0.289$\pm$0.009 \\ 
            &    &        & $\mu_\delta$           & 1.60$\pm$0.07 & 0.282$\pm$0.008  \\ 
            &    &        & $V_{\rm LoS}$          & 5.50$\pm$0.15 & 4.70$\pm$0.24  \\ 
\hline
NGC\,0104   & 1P & outer  & $\mu_\alpha\cos\delta$ & 2.86$\pm$0.08 & 0.221$\pm$0.009 \\ 
            &    &        & $\mu_\delta$           & 1.58$\pm$0.07 & 0.285$\pm$0.009  \\
            &    &        & $V_{\rm LoS}$          & 5.84$\pm$0.45 & 2.71$\pm$0.58  \\ 

		    & 2P &        & $\mu_\alpha\cos\delta$ & 2.92$\pm$0.09 & 0.205$\pm$0.008 \\ 
            &    &        & $\mu_\delta$           & 1.69$\pm$0.08 & 0.240$\pm$0.007  \\ 
            &    &        & $V_{\rm LoS}$          & 5.34$\pm$0.16 & 3.86$\pm$0.20  \\ 
\bottomrule
\bottomrule
\end{tabular}
\caption{Amplitudes and phases (Position Angle in the case of the line-of-sight component) of the best-fit functions (Equation~\ref{eq:1}) describing the observations of 1P and 2P stars in the $\mu_{\alpha} cos{\delta}$ vs.\,$\theta$, $\mu_{\delta}$ vs.\,$\theta$ planes and $V_{\rm LoS}$ vs. $\theta$. The upper twelve lines in the Table refer to the entire sample of analyzed 1P and 2P stars of NGC\,0104 and NGC\,5904, while in the lower twelve lines we consider 1P and 2P stars in the inner and outer fields of NGC\,0104.}
\label{tab:goodness}
\end{table*}

The values of $A$ and $\phi$ that provide the best-fit to the observations of NGC\,0104  and NGC\,5904 are listed in Table~\ref{tab:goodness}.
In both GCs the zero point, $M$, is consistent with zero within 0.01 ($mas/yr$ for the proper motion components and $km/s$ for the LoS velocity) and the frequency $F$ is consistent with 1.00 within 0.01 as expected for stellar rotation in GCs. 
 
To estimate the uncertainties on the amplitude and the phase of the sine function that best reproduces the distribution of 1P (2P) stars of each cluster in both proper motions components, we adopted a procedure based on 1,000 Monte Carlo simulations. 
In each simulation, we generated a sample of $N$ stars with the same $\theta$ distribution of the observed 1P (2P) stars. Here $N$ is the number of analyzed 1P (2P) stars. 
 
We used Equation~\ref{eq:1} to calculate the value of $f(\theta_{\rm i})$ that corresponds to each simulated star, $i$, by assuming the values of $A$ and $\phi$ listed in Table~\ref{tab:goodness}. 
Then, we added to $f(\theta_{\rm i})$ the same uncertainties that we inferred from the observations, and interpolated the simulated distribution of stars in $\Delta \mu_{\delta}$ vs.\,$\theta$ ($\Delta \mu_{\alpha}\cos\delta$ vs.\,$\theta$, $\Delta V_{\rm LoS}$ vs.\,$\theta$) with Equation~\ref{eq:1} by means of least-squares, thus estimating the values of $A$ and $\phi$. 
 
We calculated the differences between the 1,000 determinations of $A$ and the true value and assumed the 68.27$^{\rm th}$ percentile of the distribution of the absolute values of these differences as the uncertainty on the determination of $A$. Similarly, we defined the error associated with the best-fit phase. 
To further compare the distributions of 1P and 2P stars in the $\Delta \mu_{\alpha}\cos\delta$ vs.\,$\theta$ and $\Delta \mu_{\rm\delta}$ vs.\,$\theta$ planes we used the k-sample Anderson-Darling test \citep{adtest}, which provides the probability of two populations to belong to the same parent distribution. In NGC\,0104, NGC\,0288, NGC\,6121, NGC\,6254, NGC\,6752 and NGC\,6838, 1P and 2P stars have probability p$\gtrsim$0.15 to come from the same parent distribution.
Hence, we conclude that there is no significant difference between the distributions of stellar populations of these clusters.
 NGC\,5904 represents a remarkable exception. Indeed the  k-sample Anderson-Darling test provides probabilities of  0.05, 0.03 and 0.16 that the distributions of 1P and 2P stars in the $\Delta \mu_{\alpha}\cos\delta$ vs.\,$\theta$, $\Delta \mu_{\rm\delta}$ vs. $\theta$  and $\Delta V_{\rm LoS}$ vs. $\theta$ planes are drawn from the same distribution.
 Noticeably, the large difference between the phases of the curves that best-fit 1P and 2P stars in the $\Delta \mu_{\delta}$ vs.\,$\theta$ plane suggests that the two populations of this cluster exhibit different rotation patterns. 
 
We finally determined the PA of the rotation axis of NGC\,0104 and NGC\,5904 from the line-of-sight velocity curves as the angle corresponding to a zero LoS velocity. The  PAs of the 1P and 2P, marked with red and blue arrows  respectively, are shown in the bottom panels of Figure~\ref{fig:AllCLrot} 
and their values are listed in Table~\ref{tab:goodness}. Our results suggest that multiple stellar populations in NGC\,5904 do not share the same rotation axis, with the PA of the 1P differing from that of the 2P by $31^\circ\pm 12^\circ$. On the other hand, NGC\,0104 does not show significant differences in the rotation curves of 1P and 2P stars. \\
 
\subsection{Comparing the rotation of first- and second-population stars  in NGC\,5904 and NGC\,0104}
\label{subsec:diff rot}

To further investigate whether the difference in the rotation patterns of 1P and 2P stars of NGC\,5904 is significant or not, we analyzed 1,000 Monte Carlo realizations, for both 1P and 2P stars.
First, we assumed that both populations follow the same proper-motion and LoS-velocity distribution, and estimate the probability that the observed phase and amplitude differences between the corresponding rotation curves is entirely due to observational errors.
We simulated two samples of stars with the same angular distribution and the same number of stars as the observed 1P and 2P stars. We associated to each star the value of $\Delta \mu_{\alpha}\cos\delta$  ($\Delta \mu_{\rm\delta}$, $\Delta$V$_{\rm LoS}$) corresponding to the sine function that provides the best fit with the observations of 2P stars, $f(\theta_i)$ (see Table~\ref{tab:goodness}). This procedure ensures that, by construction, the simulated 1P and 2P stars belong to the same parent distribution.
Finally, we added the corresponding observational errors to the simulated proper motions of each star, and fitted the resulting distributions of 1P and 2P stars with the function provided by Equation~\ref{eq:1}. 
We calculated the difference between the phases ($\Delta \phi$) and the amplitudes ($\Delta A$) derived for 2P and 1P stars and analyzed the distributions of the corresponding absolute values. Results are summarized in Table~\ref{tab:Significance}.

We find that the fraction of simulations where the value of $|\Delta \phi|$ obtained from the  $\Delta \mu_{\rm\delta}$ vs.\,$\theta$ plane is equal or larger than the observed phase difference between 2P and 1P stars is 
0.008 
 Hence, the observed phase difference between the curves of the two stellar populations of NGC\,5904 is significant at the $\sim$2.6$\sigma$ level. 
In the $\Delta$V$_{\rm LoS}$ vs.\,$\theta$ plane the phase difference has significance of $\sim$2.3 $\sigma$.
On the other hand, we did not detect any significant difference between the amplitudes of the curves of the two populations in the $\Delta \mu_{\alpha}\cos\delta$ vs.\,$\theta$ plane.

Furthermore, 1P and 2P stars in NGC\,5904 are reproduced by sine functions with the same amplitudes.
For completeness, we extended the same analysis to NGC\,0104 and find no significant difference between the rotation curves of its 1P and 2P stars.

The large number of stars that are available in this cluster allows us to investigate rotation of 1P and 2P stars at different radial distances from the cluster center. 
We selected two regions with approximately the same number of stars, 
namely an inner annulus between $\sim 0.8\,R_{\rm h}$ and $\sim 2.4\,R_{\rm h}$ (2.3 to 6.6\,arcmin), and an outer annulus that goes from $\sim 2.4\,R_{\rm h}$ to $\sim 5.0\,R_{\rm h}$ ($6.6$ to 14.0 arcmin), with $R_{\rm h}$ being the half-light radius listed in Table~\ref{tab:parameters}.
The inner and outer annulus contain respectively 397/877 and 456/750 1P/2P stars. As expected, the star counts are consistent with a more centrally-concentrated 2P (\citealp[as previously noticed by] [] {milone2012} and \citealp {cordero2014}).

\begin{figure*}
  \centering
  \includegraphics[width=7.5cm,trim={2cm 0cm 0cm 1cm},clip]{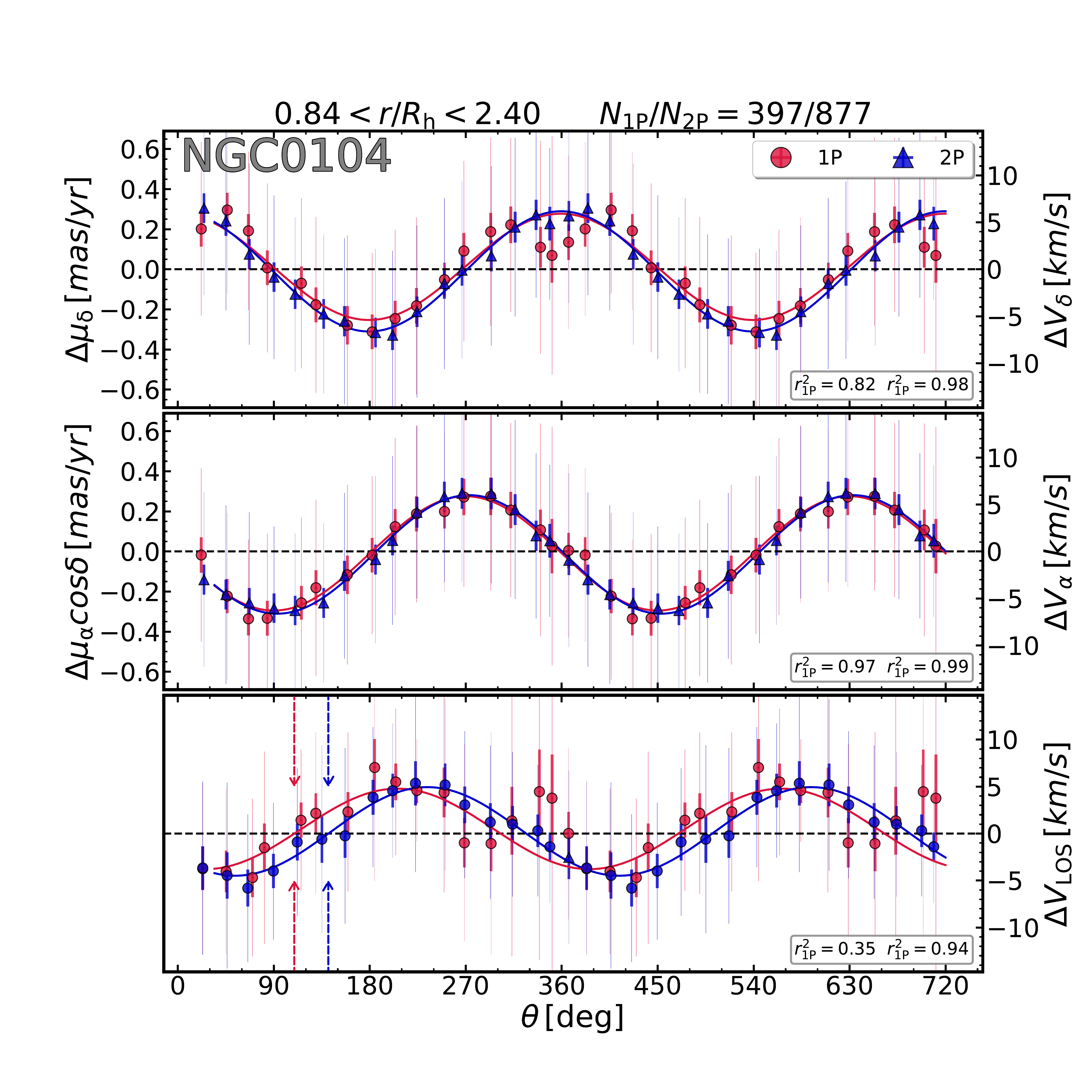}
  \includegraphics[width=7.5cm,trim={2cm 0cm 0cm 1cm},clip]{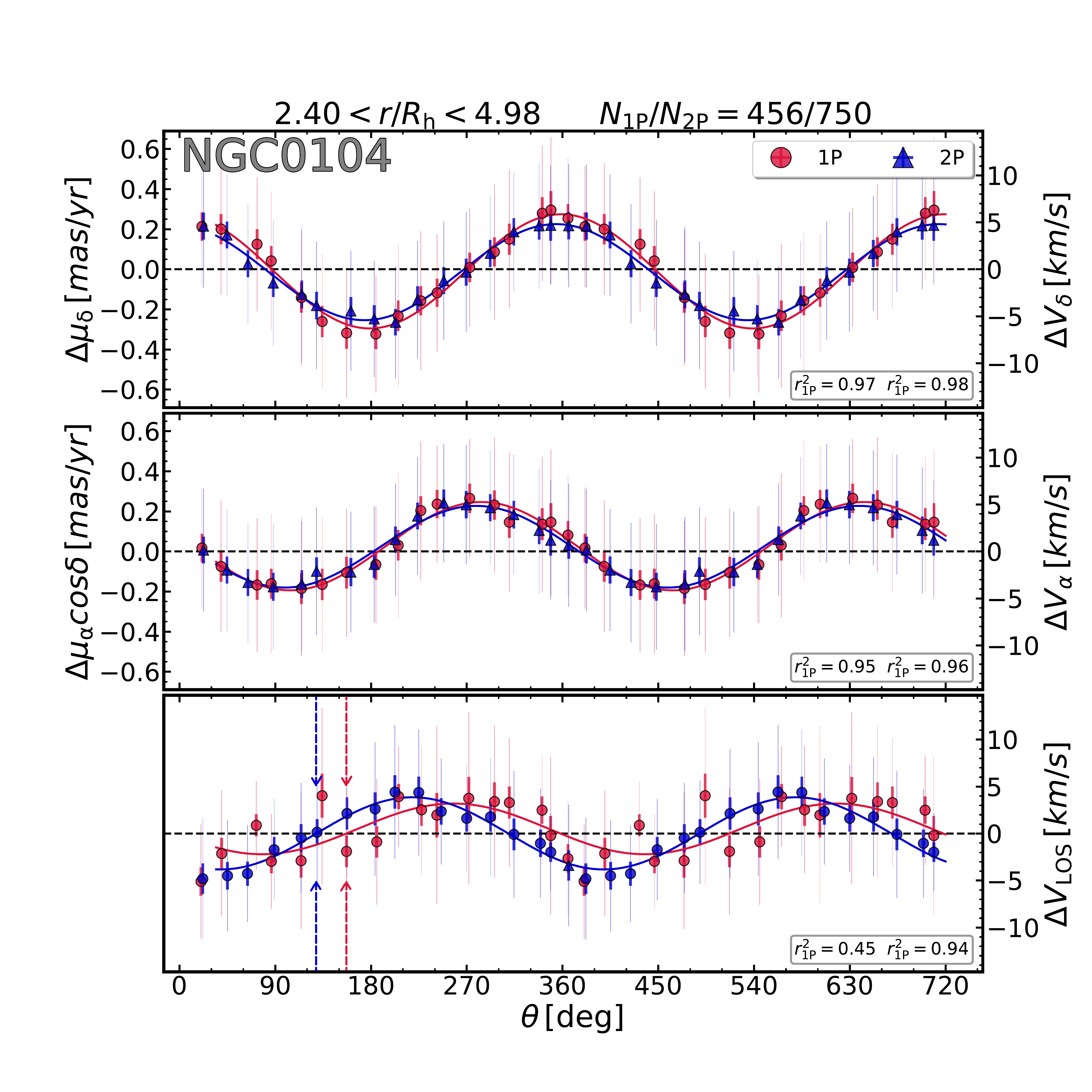}
  \caption{Same as Figure~\ref{fig:AllCLrot} for stars in the inner (left panel) and outer (right panel) regions of NGC\,0104. The curves are the best-fit sine functions. Red and blue colors refer to 1P and 2P stars, respectively. See text for details.} 
 
  \label{fig:NGC0104significance}
\end{figure*}

To investigate the rotation of 1P and 2P stars in the inner and outer region we applied the same method described in Section~\ref{subsec:diff rot} for all 1P and 2P stars. The average motions of stars in the inner and outer region are shown in Figure~\ref{fig:NGC0104significance}, while the values of $A$ and $\phi$ of the best-fit sine functions of 1P and 2P stars are listed in Table~\ref{tab:goodness}. 
We find that in the inner region the two populations are consistent with the same rotation. On the other hand, in the outer region we detect both an amplitude difference ($\Delta A=1.150$ km/s) and a phase difference ($\Delta \phi=0.500$ rad), between the curves that fit the observations of 1P and 2P stars in the $V_{\rm LoS}$ vs.\,$\theta$ plane. Only 1$\%$/4$\%$ of our simulations produce an amplitude/phase difference greater than the observed one. The observed differences are therefore significant to the $\sim 2.3/\sim 2\,\sigma$ level. \\ 
However, due to the lower number of stars with LoS velocity measurements, we obtain poor quality for the interpolation between the sine function and the observations along the LoS for 1P stars in the inner and outer regions of NCG\,0104, as shown by the values of $r^2$, listed in the bottom right insets of Figure~\ref{fig:AllCLrot}.

\begin{table*}[h]
\centering
\begin{tabular}{lll|ll|ll|cc}
\toprule
\toprule
 ID & field & motion component & A-D & p-val & $\Delta {\rm A}^{\rm obs}$ & $\Delta \phi^{\rm obs}$ 
 & P$(|\Delta {\rm A}^{\rm sim}| \geq |\Delta {\rm A}^{\rm obs}|)$ & P$(|\Delta \phi^{\rm sim}| \geq |\Delta \phi^{\rm obs}| )$ \\ 
 & & & & & mas/yr (km/s) & rad & & \\
\midrule
NGC\,0104  & all   & $\mu_{\alpha} cos{\delta}$ & 1.53 & 0.17 & $0.001\pm 0.020$ & $0.160\pm 0.090$ & 0.960 & 0.083 \\ 
           &       & $\mu_\delta$               & 1.59 & 0.16 & $0.004\pm 0.022$ & $0.070\pm 0.080$ & 0.104 & 0.460 \\
           &       & $V_{\rm LoS}$              & 1.59 & 0.16 & $0.88\pm 0.40$ & $0.030\pm 0.280$ & 0.390 & 0.281 \\
           
NGC\,5904  & all   & $\mu_{\alpha} cos{\delta}$ & 2.62 & 0.05 & $0.020\pm 0.020$ & $0.224\pm 0.195$ & 0.260 & 0.170 \\ 
           &       & $\mu_\delta$               & 3.00 & 0.03 & $0.000\pm 0.018$ & $0.870\pm 0.224$ & 0.951 & 0.008 \\ 
           &       & $V_{\rm LoS}$              & 1.59 & 0.16 & $0.51\pm 0.48$ & $0.541\pm 0.200$ & 0.453 & 0.021 \\

\hline                                                                                                                          
 NGC\,0104 & inner & $\mu_{\alpha} cos{\delta}$ & 1.44 & 0.20 & $0.001\pm 0.038$ & $0.080\pm 0.134$ & 0.980 & 0.503 \\ 
           &       & $\mu_\delta$               & 0.59 & 0.66 & $0.017\pm 0.039$ & $0.100\pm 0.125$ & 0.422 & 0.434 \\ 
           &       & $V_{\rm LoS}$              & 1.59 & 0.16 & $0.41\pm 0.50$ & $0.533\pm 0.410$ & 0.550 & 0.062 \\
           
 NGC\,0104 & outer & $\mu_{\alpha} cos{\delta}$ & 2.54 & 0.05 & $0.016\pm 0.026$ & $0.060\pm 0.121$ & 0.532  & 0.593 \\ 
           &       & $\mu_\delta$               & 1.98 & 0.10 & $0.044\pm 0.028$ & $0.110\pm 0.103$ & 0.100  & 0.402 \\ 
           &       & $V_{\rm LoS}$              & 1.59 & 0.16 & $1.15\pm 0.62$ & $0.500\pm 0.500$ & 0.010  & 0.041 \\
           
\bottomrule
\bottomrule
\end{tabular}
\caption{Comparison of the rotation curves in the $\mu_{\rm \alpha} cos{\delta}$ vs.\,$\theta$,  $\mu_{\delta}$ vs.\,$\theta$ and $V_{\rm LoS}$ vs. $\theta$ planes of 1P and 2P stars in the entire field of view of NGC\,0104 and NGC\,5904 and in the inner and outer region of NGC\,0104. We provide the A-D values from the  Anderson-Darling test and the corresponding probability that 1P and 2P stars comes from the same parent distribution (p-val). We list the amplitude ($\Delta{A^{\rm obs}}$) and phase differences ($\Delta{\phi}^{obs}$) of the curves that provide the best-fit with 2P and 1P stars and the probability that the observed difference in phase and amplitude are due to observational errors as inferred from Monte-Carlo simulations.}
\label{tab:Significance}
\end{table*}

\subsection{Comparison with theory}
Figure~\ref{fig:AllCLrot} suggests that the rotation curves of 1P and 2P stars of NGC\,5904 exhibit different phases in the $\mu_\delta$ vs.\,$\theta$ plane and  along the line of sight. On the contrary, the two populations seem to share the same rotation pattern when we consider the $\mu_\alpha\cos\delta$ component of the motion.   
To shed light on this phenomenon, we further investigate the rotation curves of 1P and 2P stars in NGC\,5904 by qualitatively comparing the observations with  mock simulated stars. Specifically, we generated two stellar populations composed of 50,000 stars each, by extracting their positions and velocities from a \citet{king1966} model with maxwellian velocity distributions. We then added to the motions of each population a specific rotation pattern characterized by the same amplitude $A$, and different inclination of the rotation axis with respect to the line-of-sight, $i$, and position angle $\theta_0$, \citep[as in][their Equation A1]{sollima2019}.

\begin{figure}
  \centering
  \includegraphics[width=10cm, trim={0cm 0cm 0cm 0cm},clip]{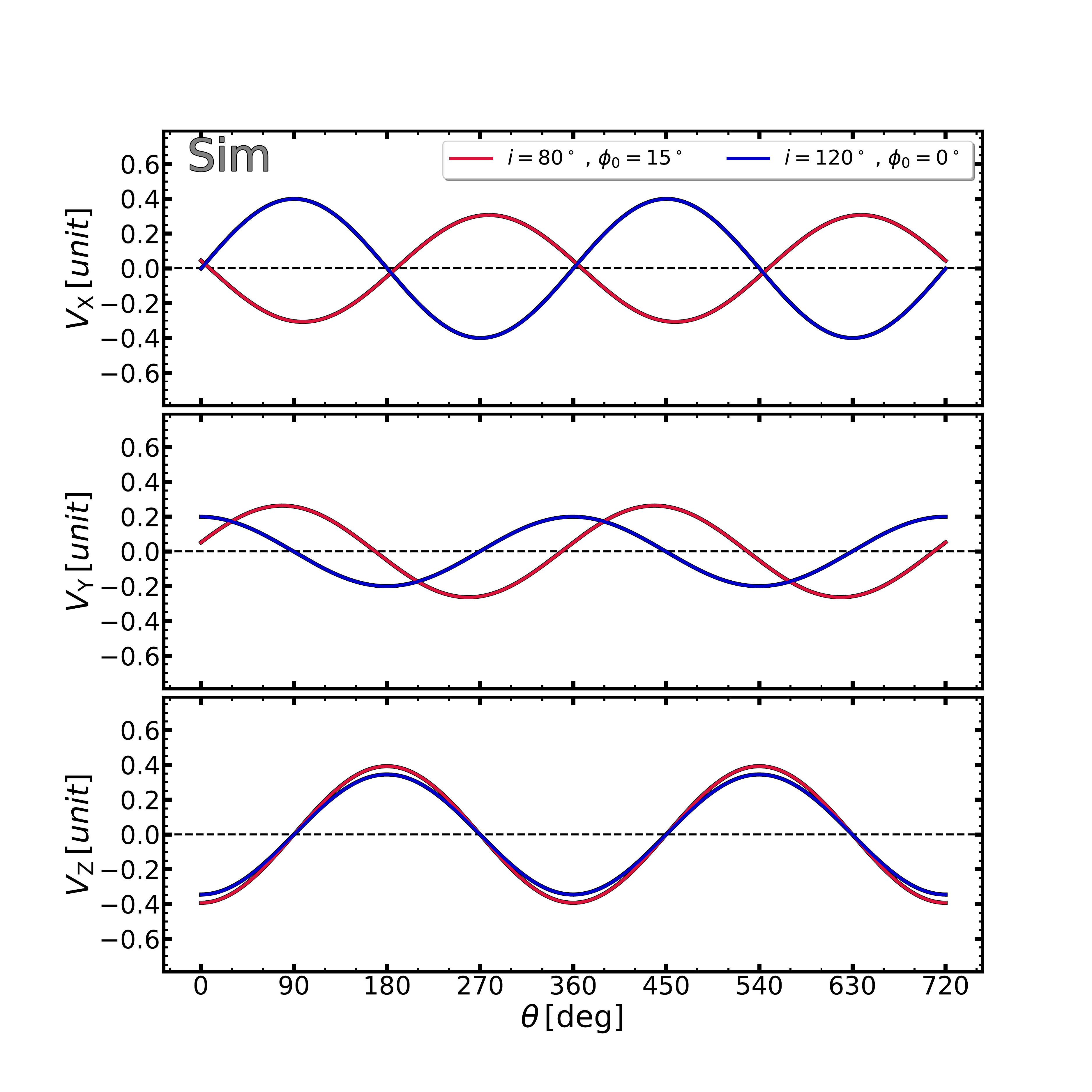}
     \caption{Projected motions along three space directions, X, Y and Z,  of two simulated stellar populations with different rotation patterns. See text for details.}
\label{fig: rotAle}
\end{figure}

We compared the rotation curves of pairs of mock stellar populations with different rotation patters along the three space directions X, Y and Z. We find that stellar populations with the same amplitudes but different inclinations and phases qualitatively reproduce the observations of NGC\,5904. 
As an example, we show in Figure~\ref{fig: rotAle} that two stellar populations with amplitude $(A=0.4)$ inclinations ($i=80^{\circ}$ and $120^{\circ}$) and phases ($\phi_{0}=15^{\circ}$ and $0^{\circ}$) qualitatively reproduce the observed pattern. Indeed the simulated rotation curves along the direction X and Y exhibit different phases, while sharing nearly the same phase along the Z direction. 

\subsection{Line of sight velocity map}

\begin{figure*}
  \centering
  \includegraphics[width=8.35cm, height=7.3cm, trim={4.1cm 2.6cm 4cm 0cm},clip]{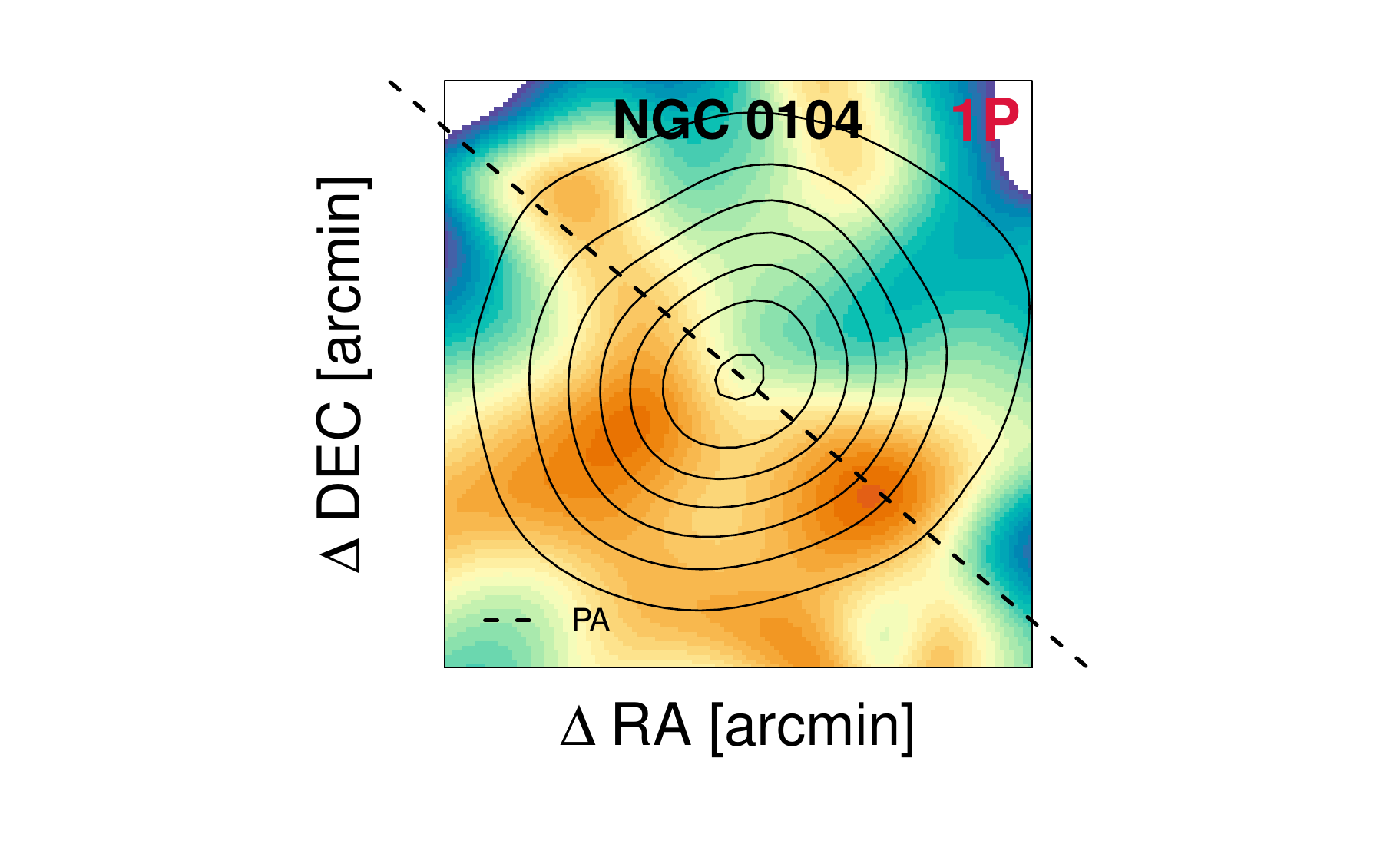}
  \includegraphics[width=9.05cm, height=7.1cm, trim={5.6cm 2.8cm 2cm 0.5cm},clip]{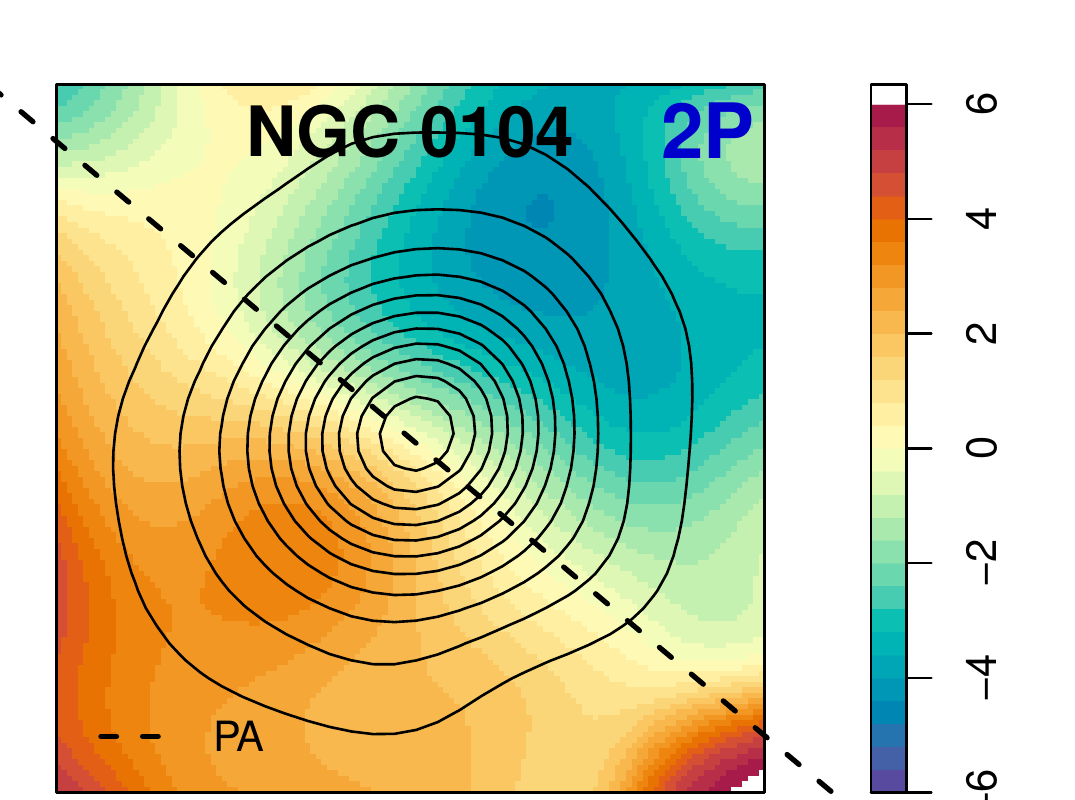}
  \includegraphics[width=8.5cm, height=8.9cm, trim={4cm 0cm 4cm 0cm},clip]{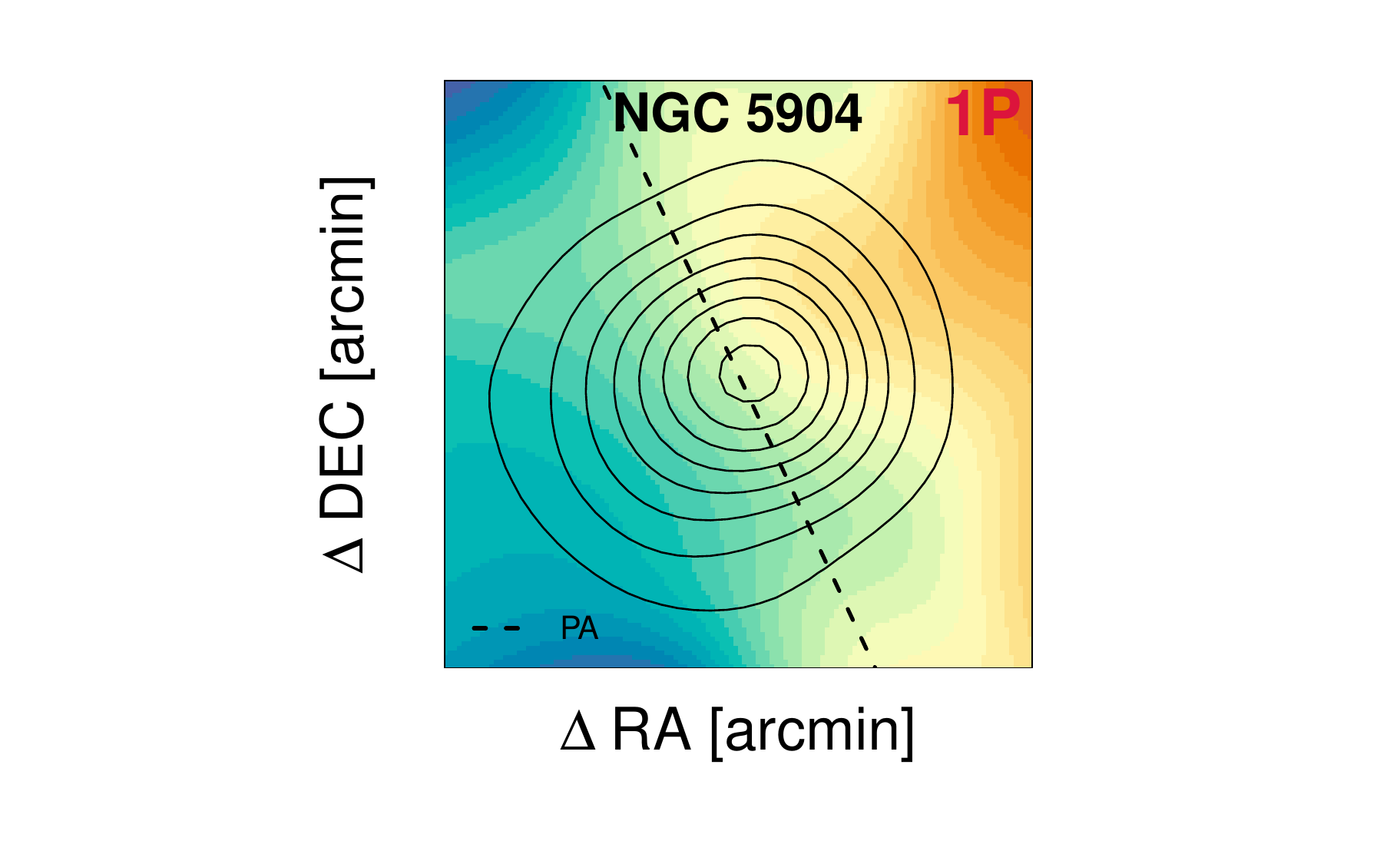}
  \includegraphics[width=9.4cm, height=8.8cm, trim={5.5cm 0.3cm 1.7cm 0.5cm},clip]{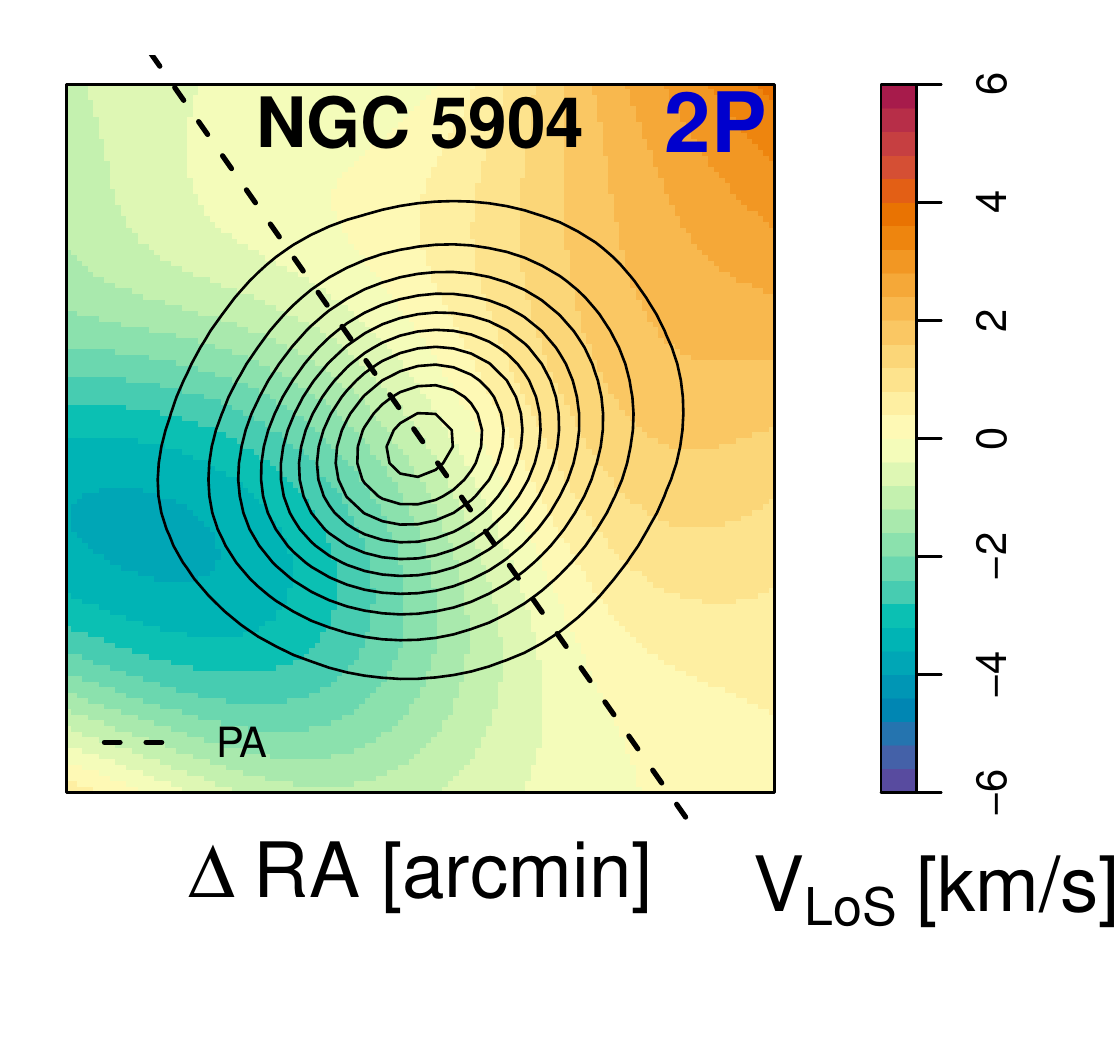}
  \caption{Line-of-sight velocity map of different stellar populations in NGC\,0104, top panels, and NGC\,5904, bottom panels. Superimposed are the isodensity contour lines, derived in Section~\ref{sec: spatial}. The thick black dashed line represents the position angle of the rotation axis, as determined in Section~\ref{sec:rot}. The colorbars on the right indicate the LoS velocity in km/s.} 
  \label{fig: losVelMap}
\end{figure*}  

For completeness we compare in Figure~\ref{fig: losVelMap} the LoS velocity map with the spatial distribution of 1P and 2P stars in NGC\,5904. The four panels show the smoothed LoS velocity map, color coded as indicated by the colorbar. The black dashed lines represent the PA of the rotation axis, derived from the rotation curves as explained in Section~\ref{sec:rot}, while the black solid lines are the same isodensity contour lines derived in Section~\ref{sec: spatial}. Clearly, the rotation axis is consistent with the minor axis of the best-fit ellipses, as expected for oblate rotators. This agreement is even more important since these quantities, i.e. the PA of the rotation axis and the major/minor axis of the best-fit ellipses, are determined with independent techniques.

\section{Velocity profiles}
{\label{sec:vprofile}}

To study the internal motion of stars as a function of the radial distance from the cluster center we divided the cluster field into different circular annuli, that are determined with the naive estimator method \citep{silverman1986}. \\
For each annulus we computed the systematics-corrected median values of the radial ($\Delta \mu_{\rm RAD}$) and tangential ($\Delta \mu_{\rm TAN}$) components of proper motions for 1P and 2P stars relative to the corresponding median proper motion components of all stars. \\ 
These proper motions have been converted into velocities, $\Delta V_{\rm RAD}$ and $\Delta V_{\rm TAN}$, by assuming for each cluster the distances listed in Table~\ref{tab:parameters}, from \cite{baumgardt2018}.

\begin{figure*}
  \centering
  \includegraphics[width=7.5cm,trim={0cm 0.2cm 0.cm 0cm},clip]{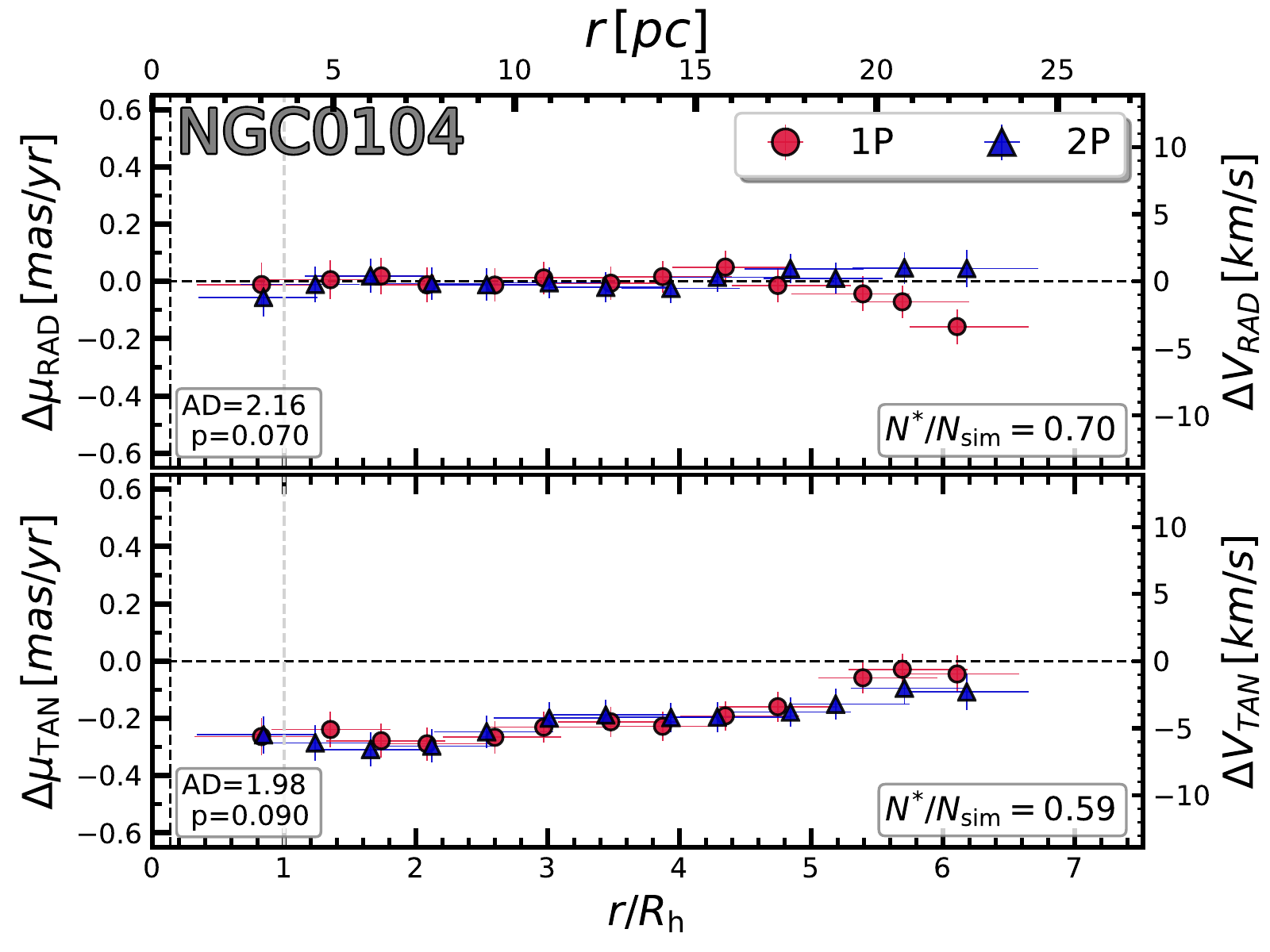}
  \includegraphics[width=7.5cm,trim={0cm 0.2cm 0.cm 0cm},clip]{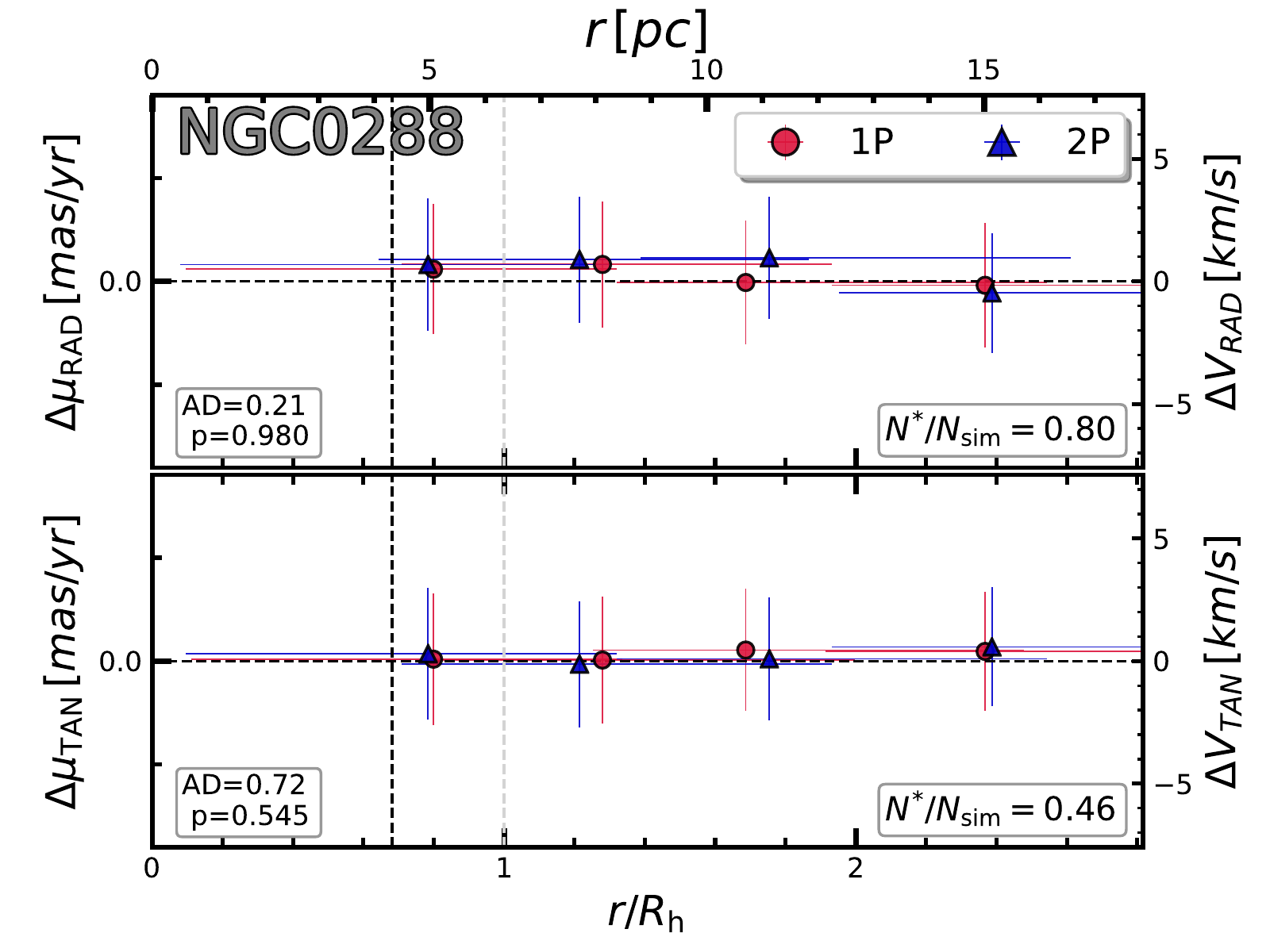}
  \includegraphics[width=7.5cm,trim={0cm 0.2cm 0.cm 0cm},clip]{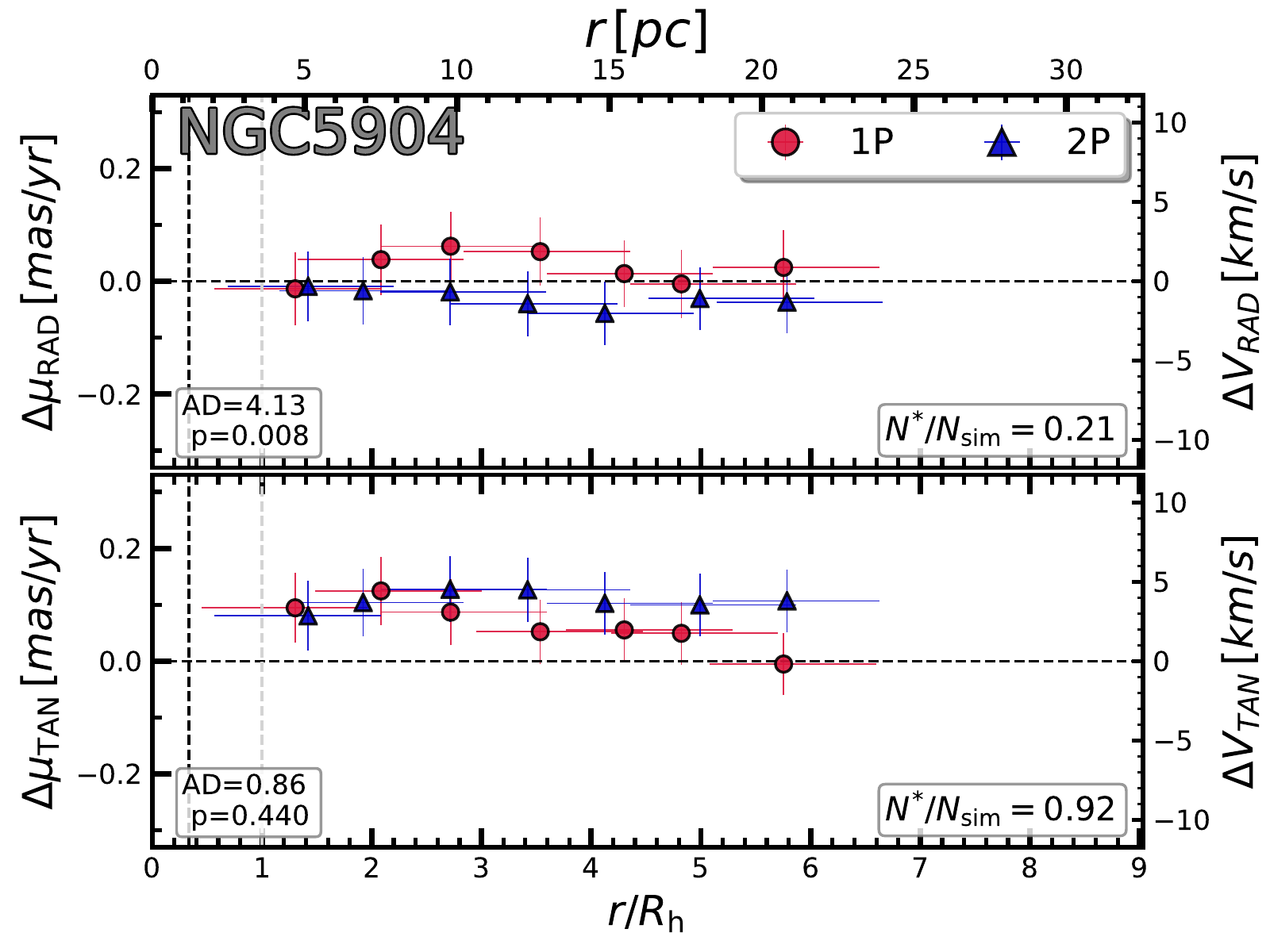}
  \includegraphics[width=7.5cm,trim={0cm 0.2cm 0.cm 0cm},clip]{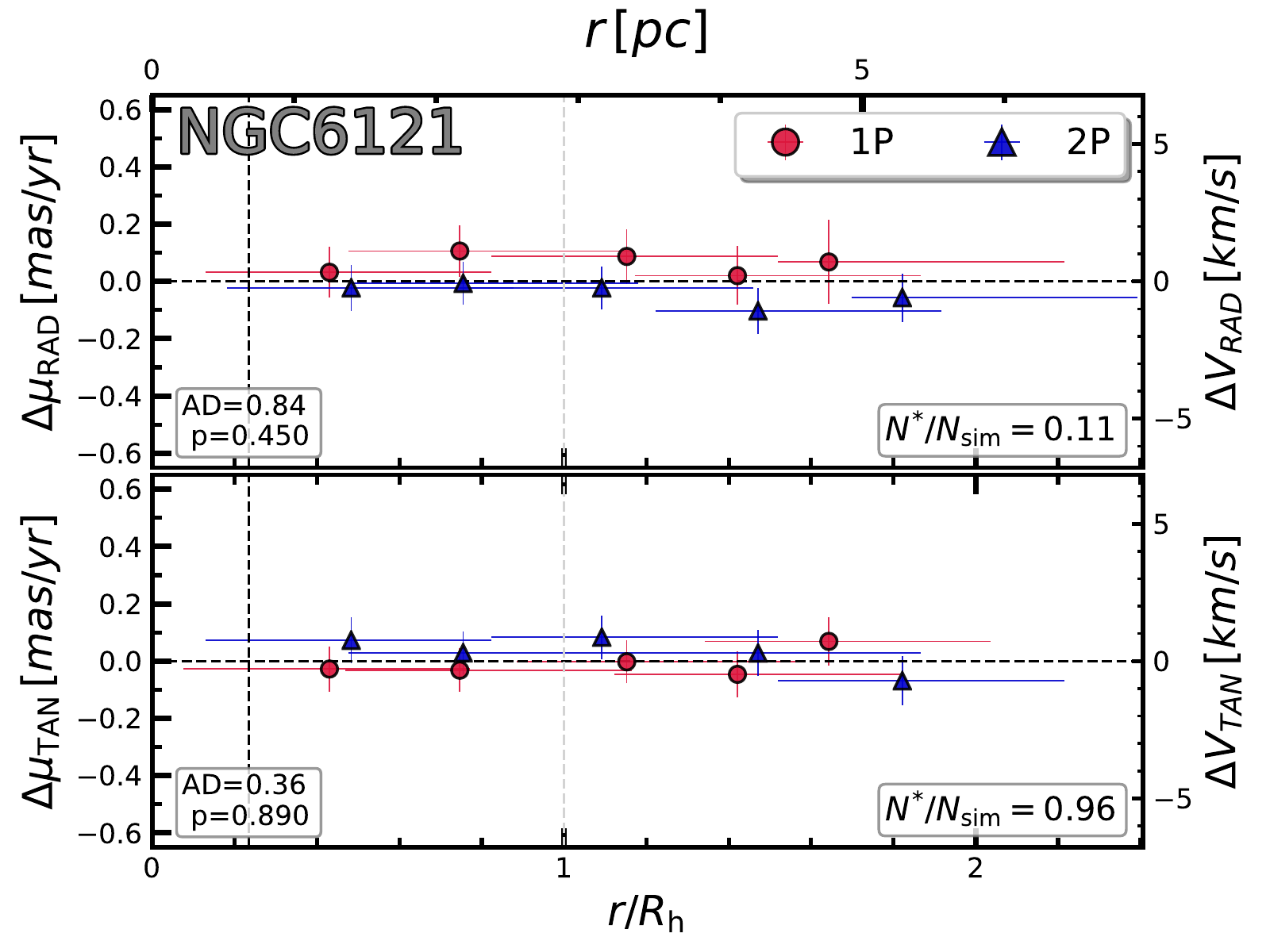}
  \includegraphics[width=7.5cm,trim={0cm 0.2cm 0.cm 0cm},clip]{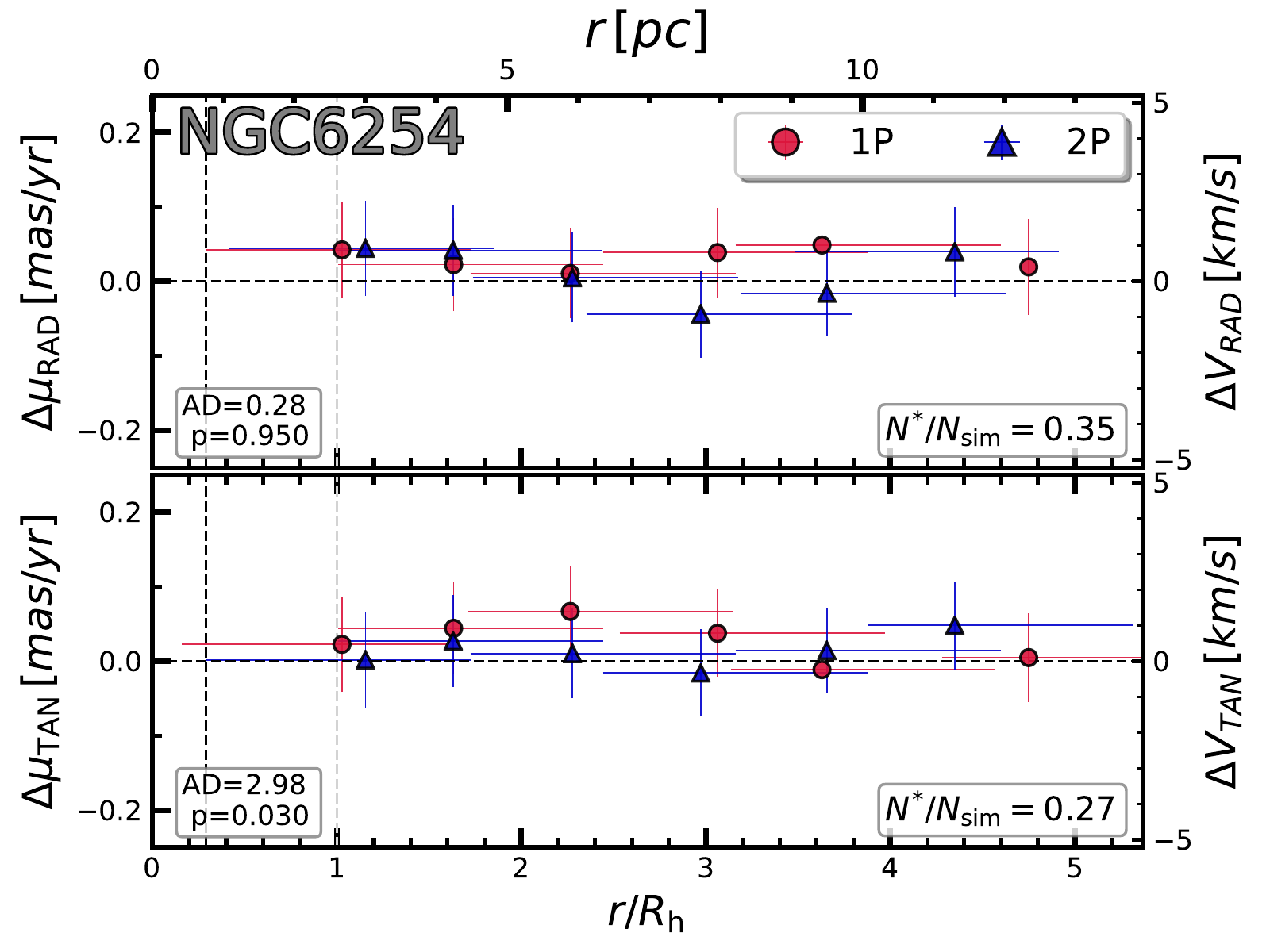}
  \includegraphics[width=7.5cm,trim={0cm 0.2cm 0.cm 0cm},clip]{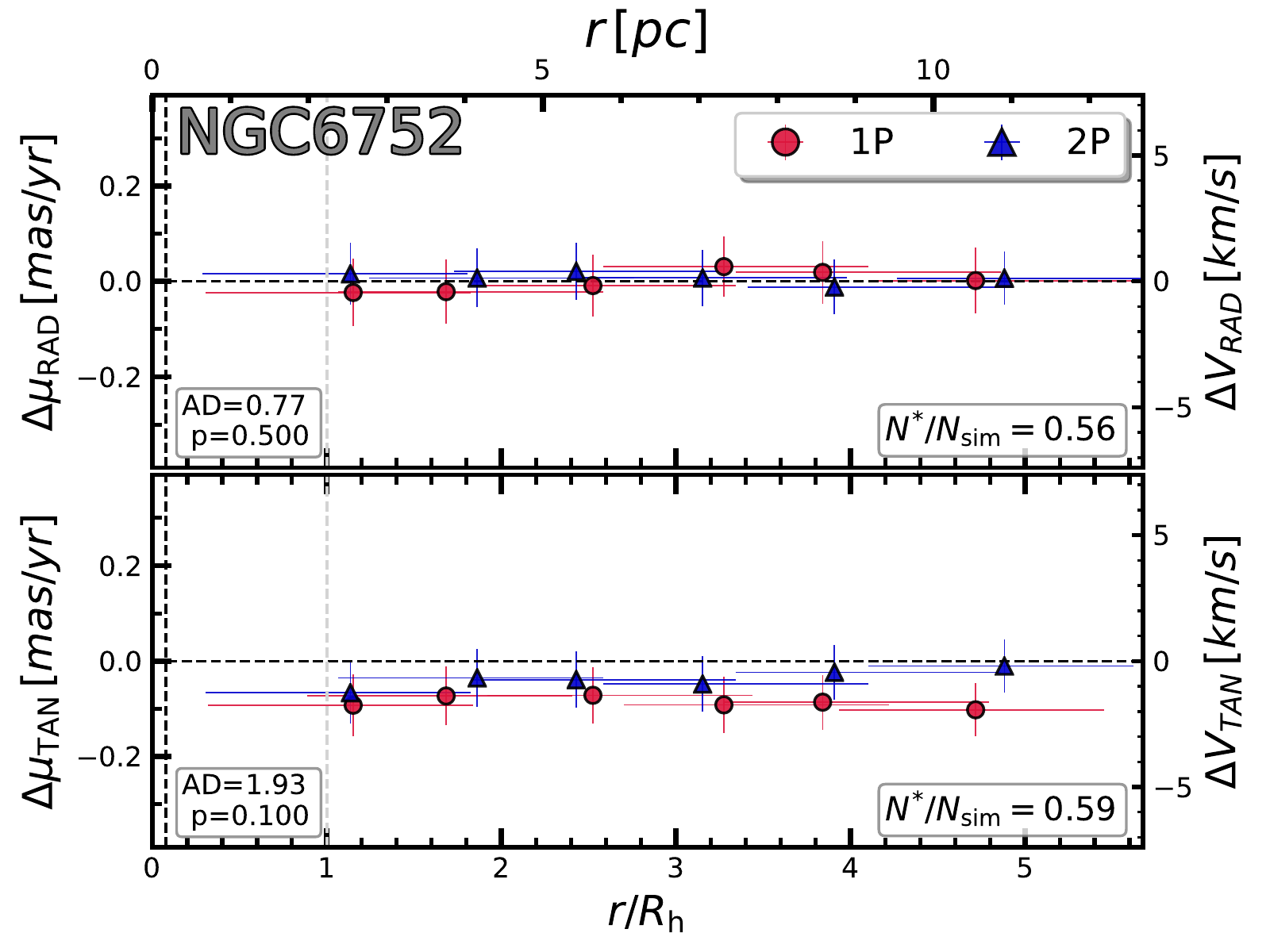}
  \includegraphics[width=7.5cm,trim={0cm 0.2cm 0.cm 0cm},clip]{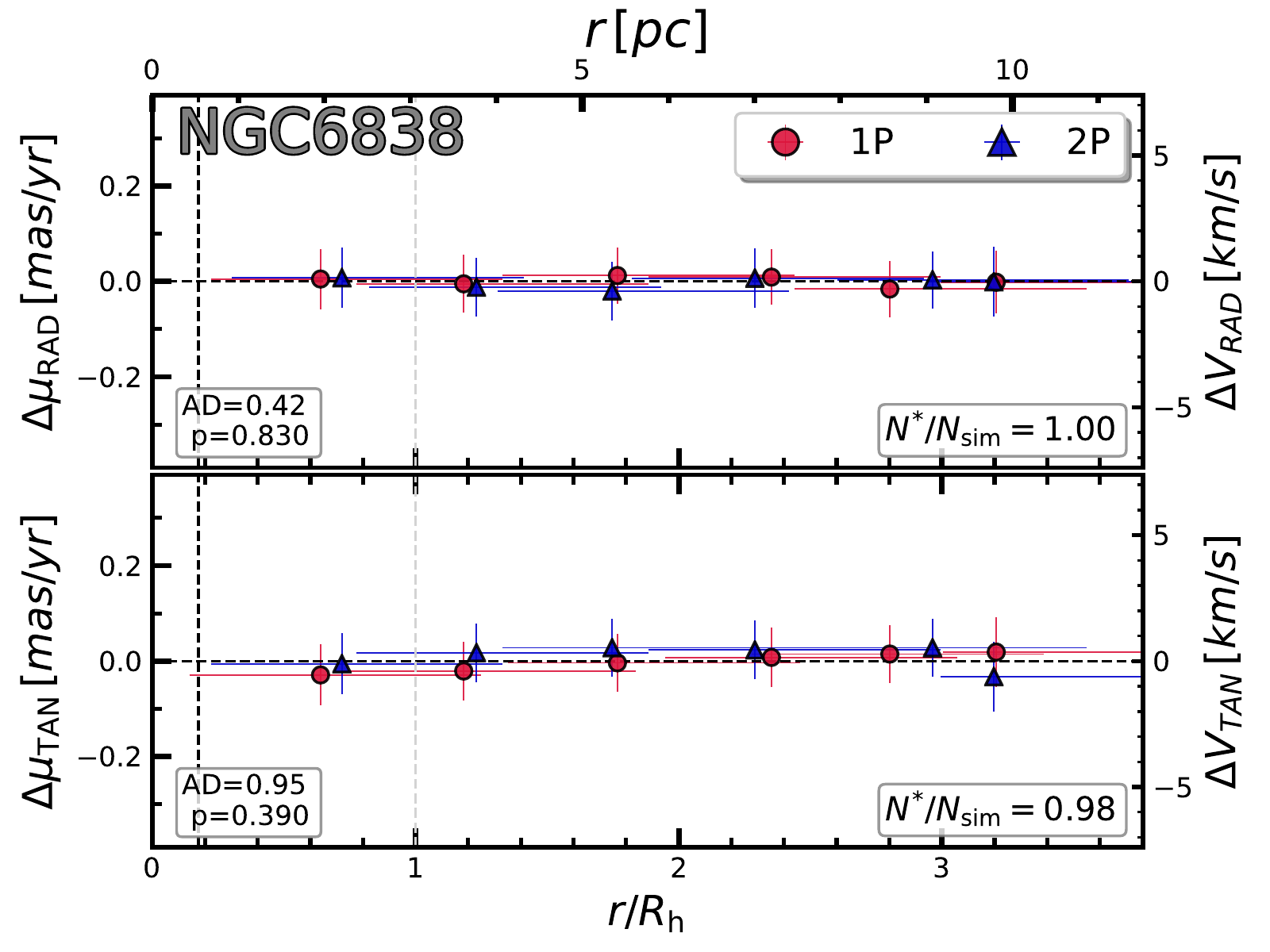}

\caption{Average tangential and radial motions for NGC\,0104, NGC\,0288, NGC\,5904, NGC\,6121, NGC\,6254, NGC\,6752 and NGC\,6838 as a function of the radial distance from the cluster center. The radial coordinate is normalized to the half-light radius from \cite{baumgardt2018}. 
  Horizontal lines mark the radial extension of the radial bins.
  The black and gray dashed lines indicate the core and the half-light radius, respectively.
  We quote for each cluster the probability, p, of the velocity distribution of 1P and 2P stars to be drawn from the same parent distribution according to the Anderson-Darling test \citep[AD,][]{adtest}. In the bottom-right corner of each panel is shown the significance of the differences between the median profile of 1P and 2P, computed as explained in the text. }
  \label{fig:AllCLmean}
\end{figure*}

Figure~\ref{fig:AllCLmean} shows the velocity profiles of the analyzed clusters as a function of the radial distance from the cluster center.  To better compare the various clusters we normalized the radial distance from the cluster center to the value of its half-light radius provided by \cite{baumgardt2018} and converted the radial distances from angular to physical units by means of the distances provided in Table~\ref{tab:parameters}.

The two populations of most GCs share similar velocity profiles and any difference between the velocities of 1P and 2P stars is smaller than $\sim$1 km/s. 
These conclusions are corroborated by the Anderson-Darling test, which provides the probabilities for 1P and 2P stars to be drawn from the same parent distribution that are quoted in the insets of Figure~\ref{fig:AllCLmean}.
As a further determination of the statistical significance of the differences between the velocity profiles of 2P and 1P stars we used the following procedure. We computed the $\chi^2_{\rm obs}$ between the observed profiles of the 1P and 2P. We then simulated 1,000 profiles for 1P and 2P, where we assumed that the two populations have the same distribution, and we scattered each star according to its observed uncertainty. For each simulation we computed the $\chi^2_{\rm sim}$ between 1P and 2P profiles, and we counted the number of realizations for which $\chi^2_{\rm sim} \gtrsim \chi^2_{\rm obs}$, $(N^*)$. The ratio between $(N^*)$ and the total number of realizations, $(N_{\rm sim})$ is indicative of the significance, and it is quoted in the bottom-right corner of each panel in Figure~\ref{fig:AllCLmean}.

NGC\,5904 seems a remarkable exception. Indeed, in the radial interval between $\sim$2 to $\sim$5 half-light radii from the center, 1P stars exhibit higher radial motions than 2P stars. 

However, such difference would be attributed to systematics as suggested by the high ratio  $N^{*}/N_{\rm sim}=0.21$. Improved proper motions, as those from next GAIA data releases, are mandatory to understand whether the observed difference is real or not.

\subsection{Velocity dispersions of 1P and 2P stars}
\begin{figure*}
  \centering
  \includegraphics[width=8cm,trim={0cm 0.2cm 0cm 0cm},clip]{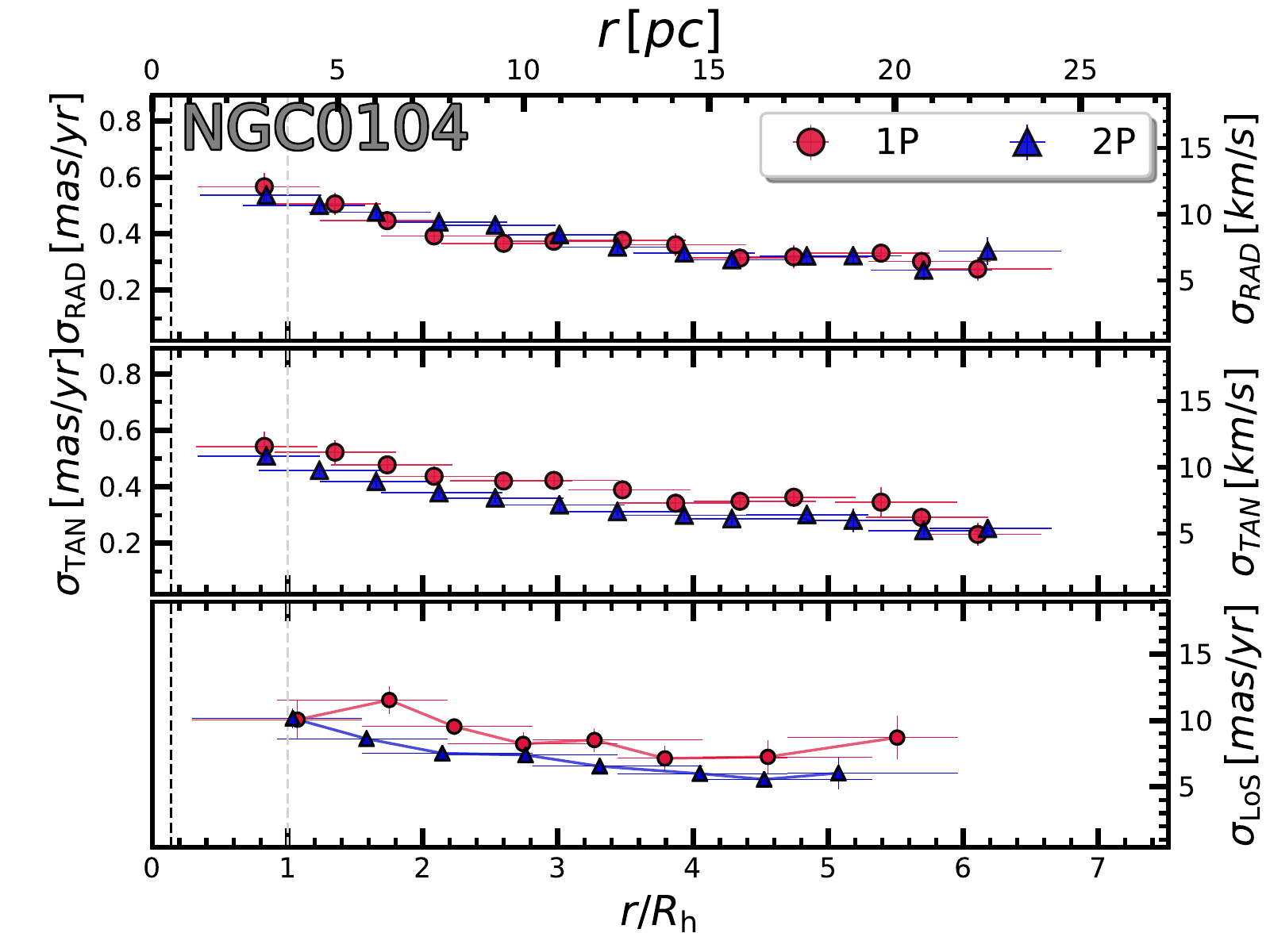}
  \includegraphics[width=8cm,trim={0cm 0.2cm 0cm 0cm},clip]{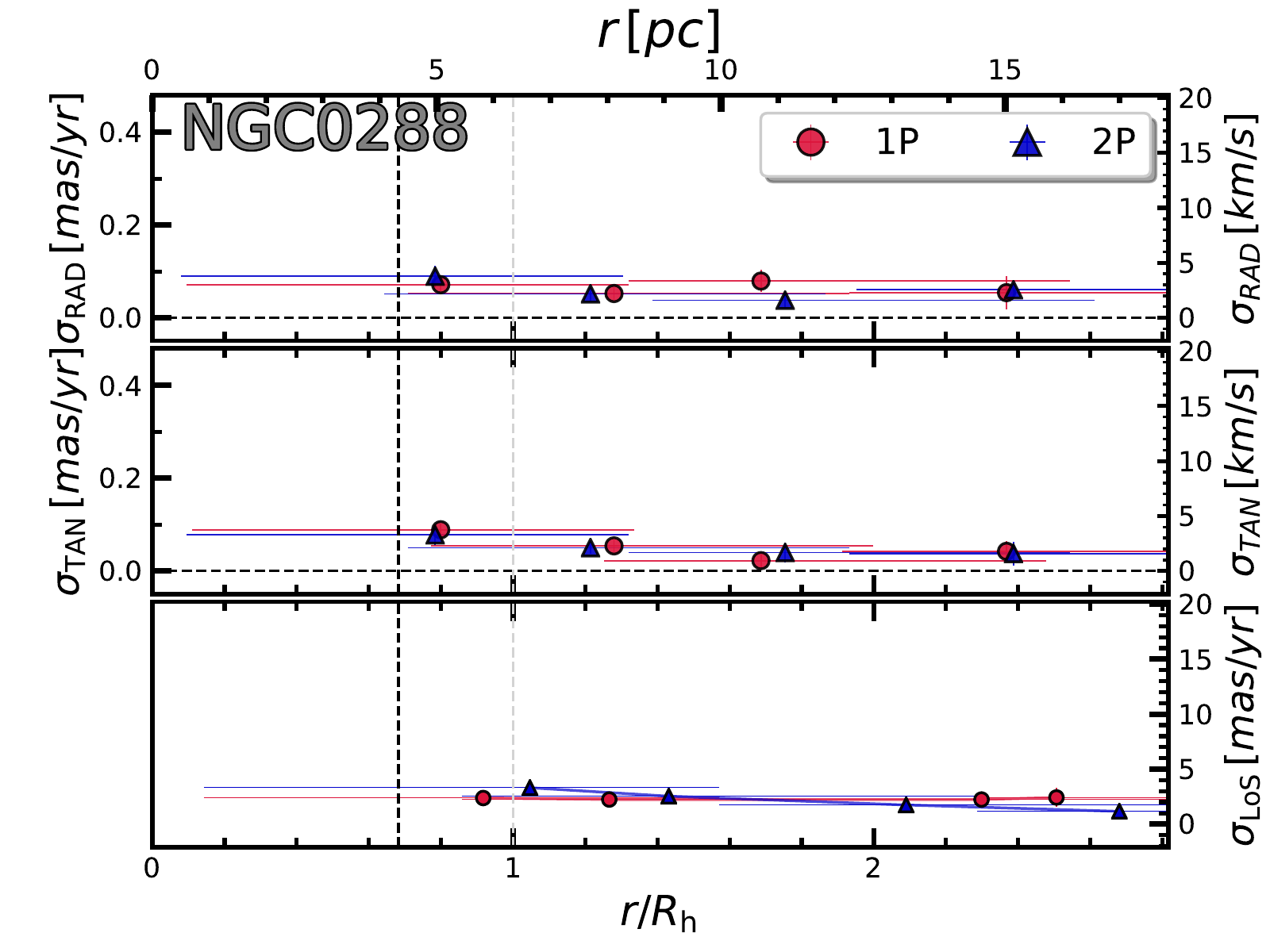}
  \includegraphics[width=8cm,trim={0cm 0.2cm 0cm 0cm},clip]{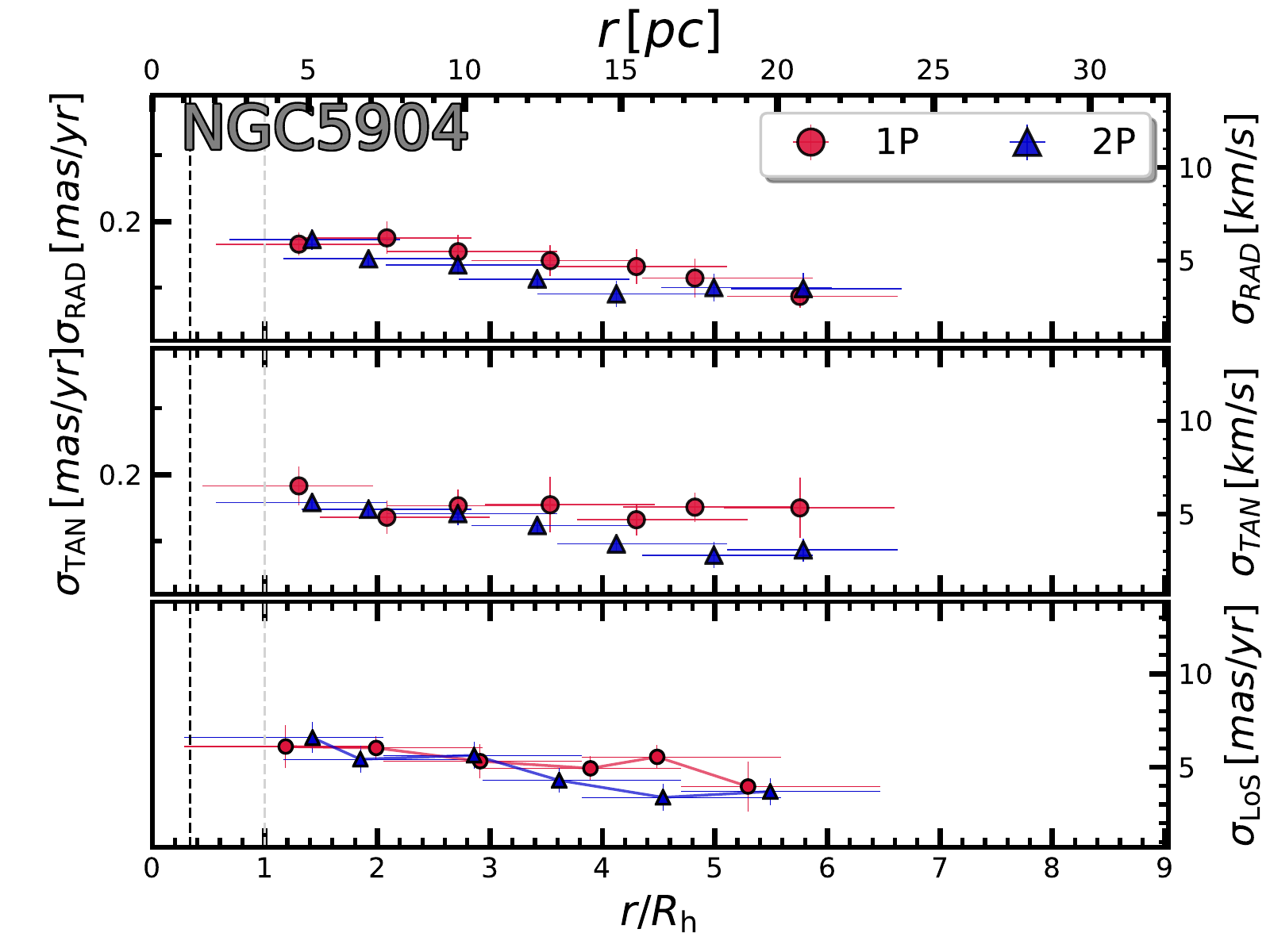}
  \includegraphics[width=8cm,trim={0cm 0.2cm 0cm 0cm},clip]{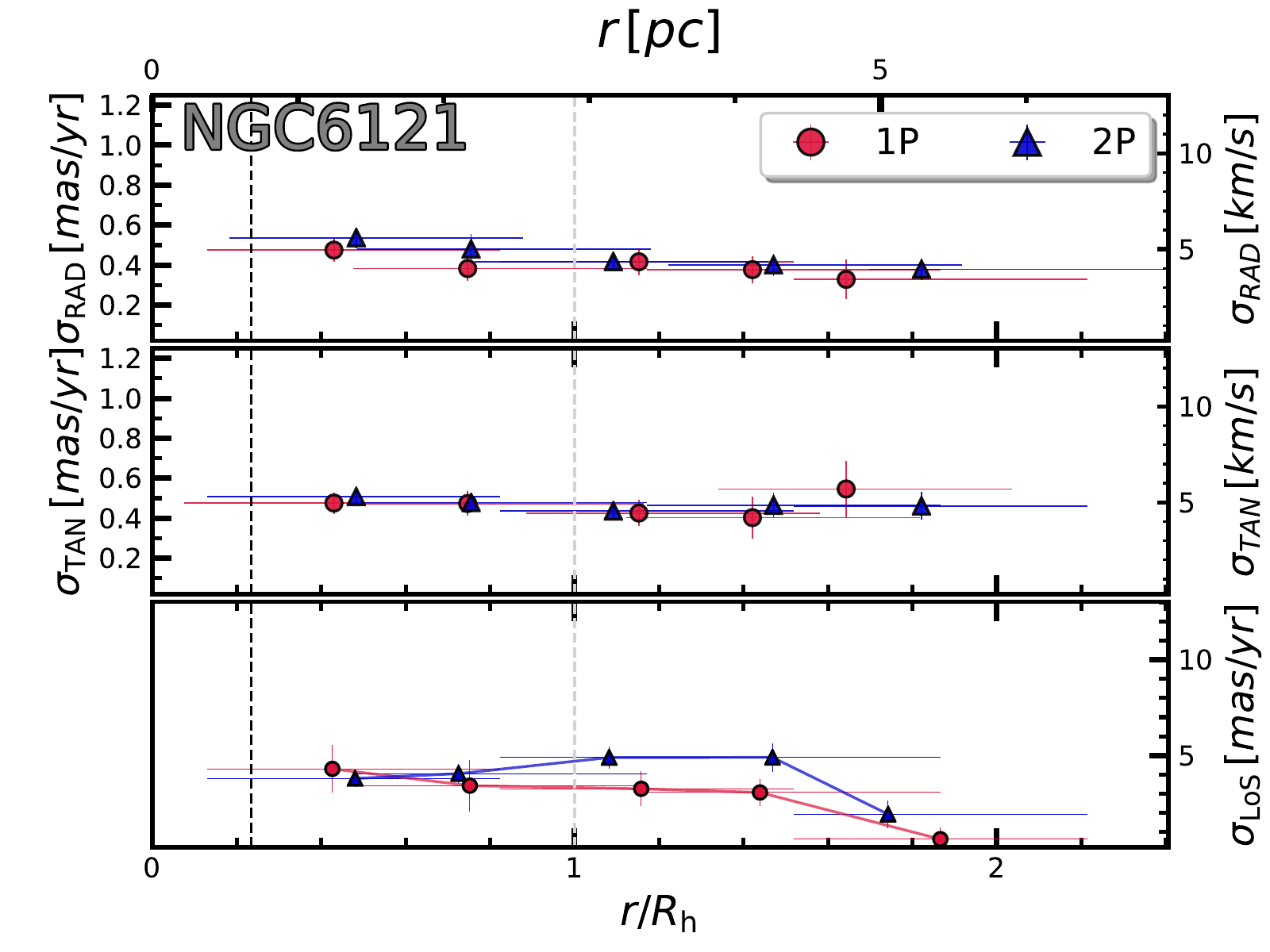}
  \includegraphics[width=8cm,trim={0cm 0.2cm 0cm 0cm},clip]{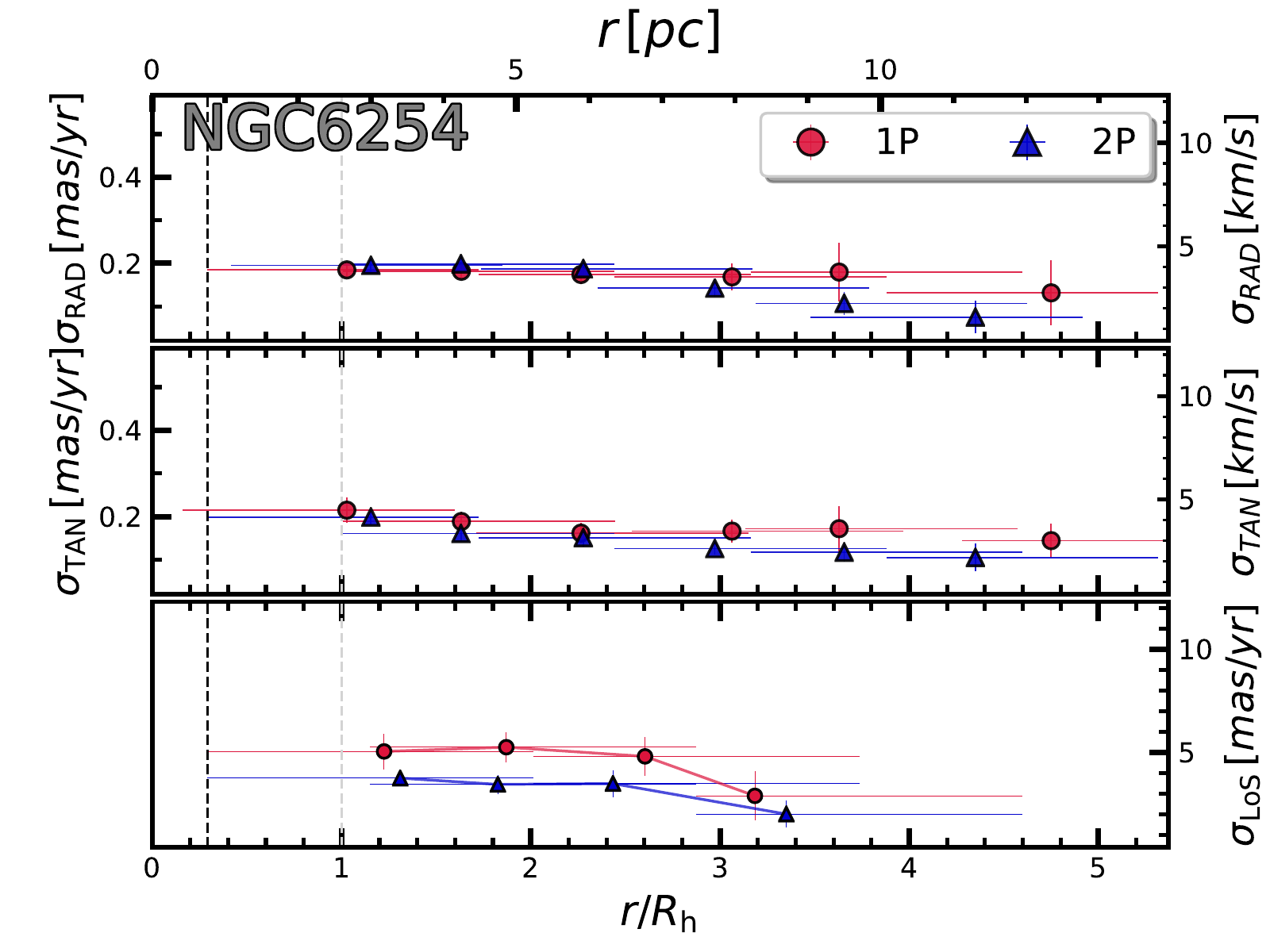}
  \includegraphics[width=8cm,trim={0cm 0.2cm 0cm 0cm},clip]{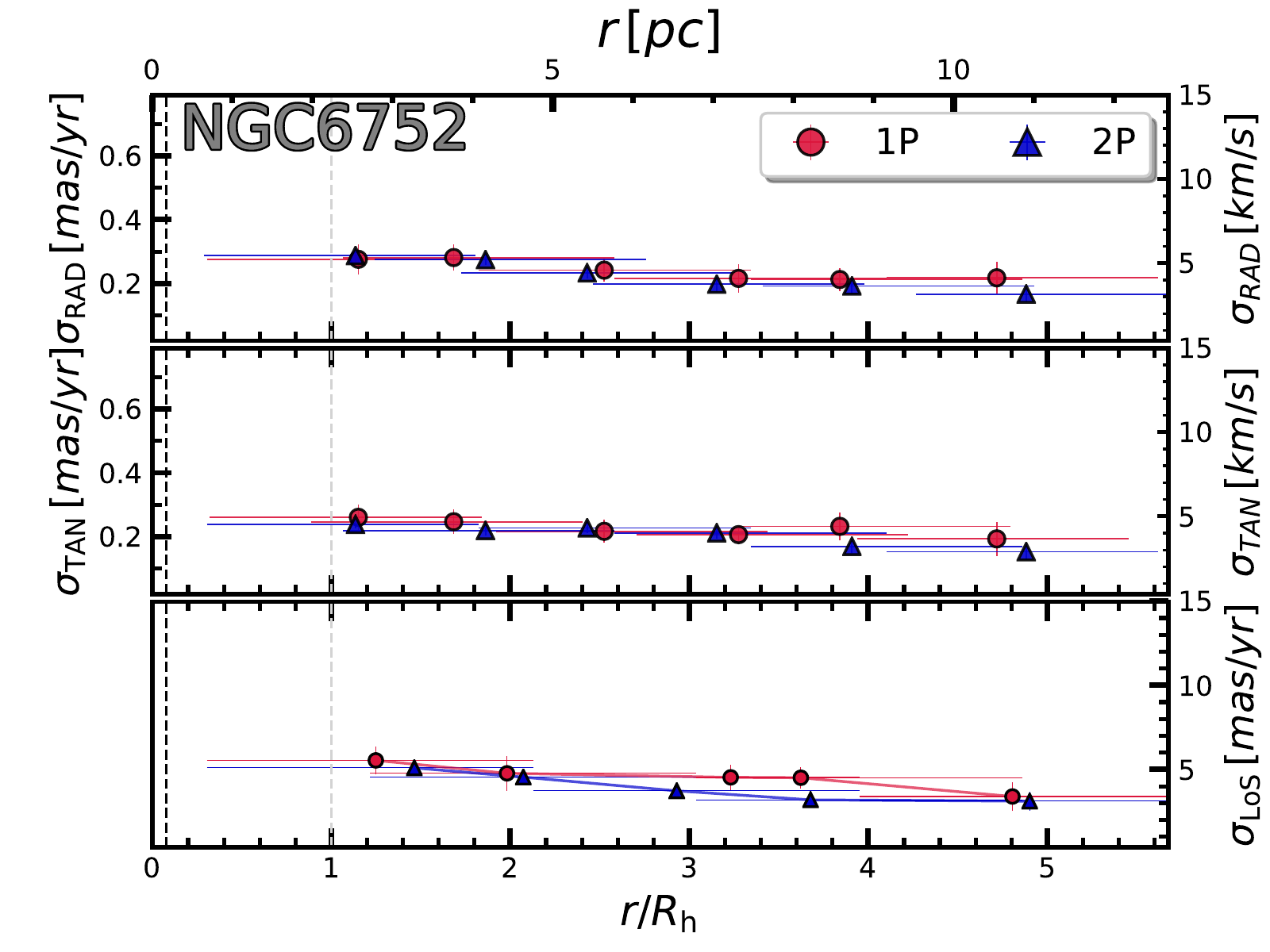}
  \includegraphics[width=8cm,trim={0cm 0.2cm 0cm 0cm},clip]{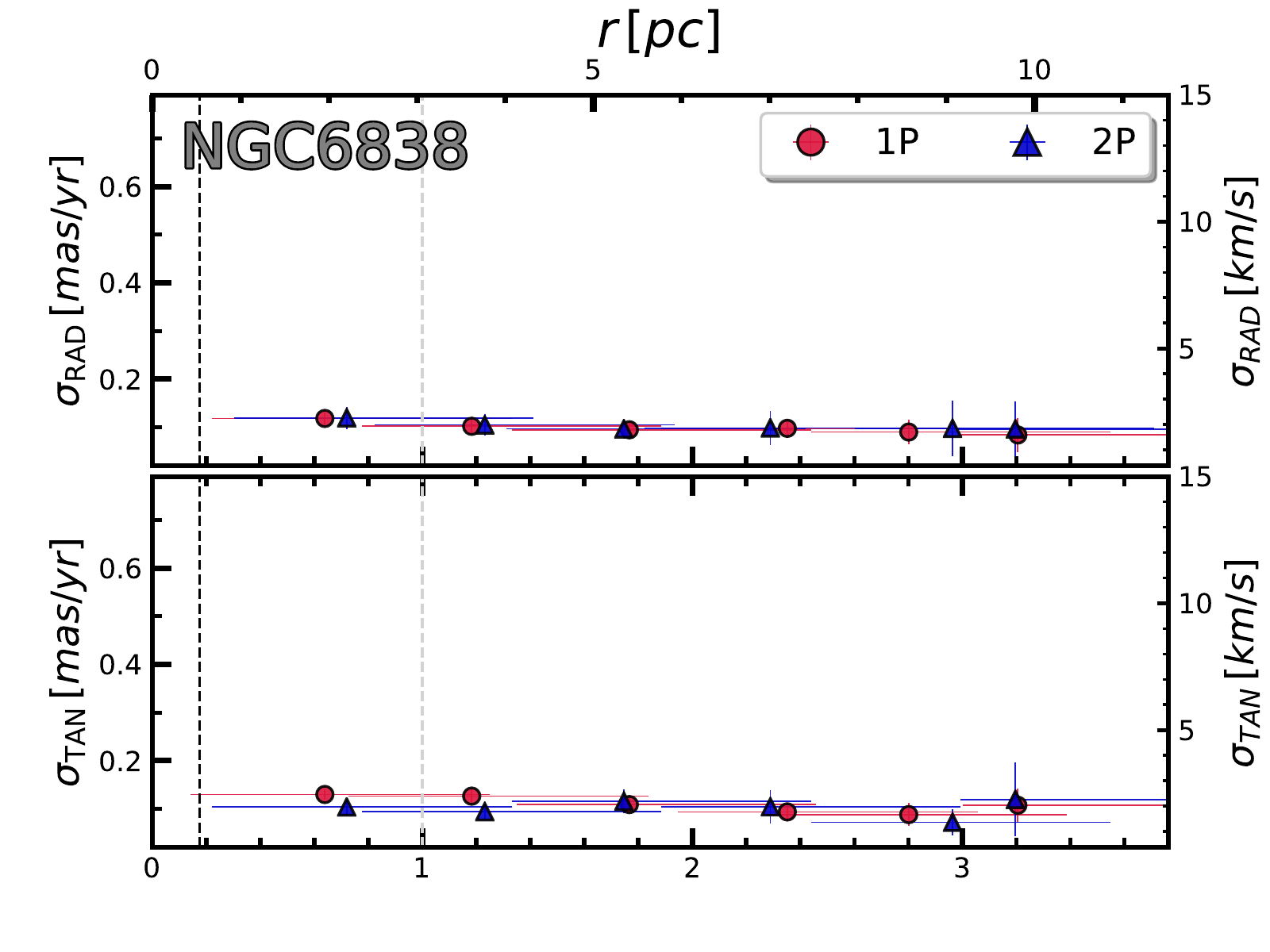}
  \caption{Velocity dispersion profiles for the radial, tangential and LoS velocity, for the analyzed GCs, except NGC\,6838 for which LoS velocities are not available.
   As in Figure~\ref{fig:AllCLmean}, the radial coordinates have been normalized to the half-light radius. Black and gray dashed lines mark the core and the half-light radius, respectively. }
  \label{fig:AllCLdisp}
\end{figure*}

To derive the velocity dispersion of 1P and 2P stars in each annulus we followed the procedure described in \cite{mackey2013} and \cite{marino2014}. Briefly we considered the negative log-likelihood function: 
$$
\lambda=\prod_{i=1}^Np(v_i,\epsilon_i)
$$
 with the probability of finding a star with velocity $v_i$ and uncertainty $\epsilon_i$ given by:
$${}
p(v_i,\epsilon)=\frac{1}{2\pi\sqrt{(\sigma^2+\epsilon_i^2)}}\exp \left(-\frac{(v_i-v)^2}{2(\sigma^2+\epsilon^2_i)}\right)
$$

and we found the intrinsic dispersion by maximizing the likelihood.  Again, the uncertainties associated with each point are determined by bootstrapping with replacements performed 1,000 times.
Figure \ref{fig:AllCLdisp} shows the velocity dispersion profile for the studied clusters, where the radial coordinated has been normalized to the half-light radius from \cite{baumgardt2018}. 

\begin{figure*}
  \centering
  \includegraphics[width=8.4cm,trim={0cm 0.2cm 0cm 0.2cm},clip]{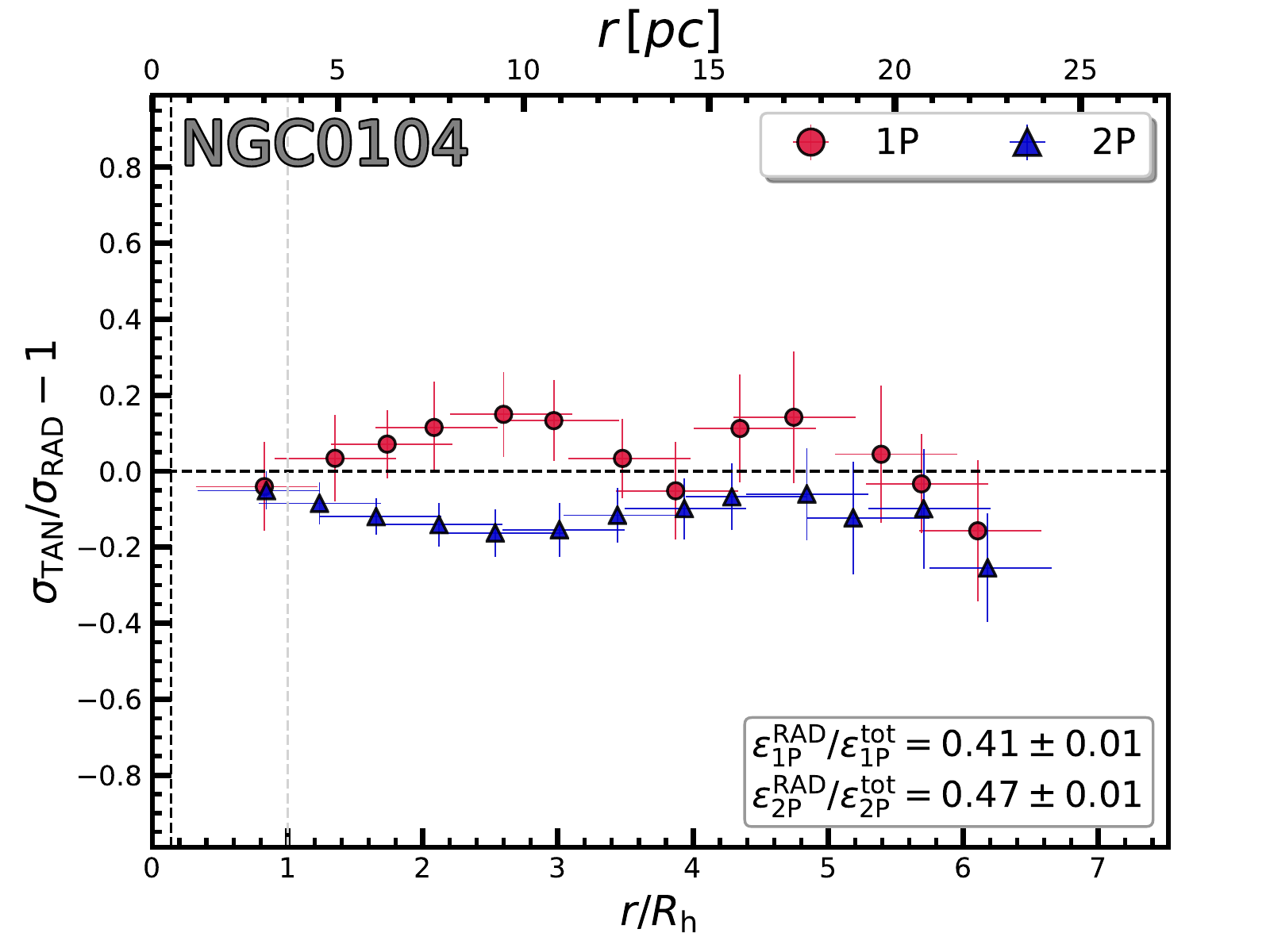}
  \includegraphics[width=8.4cm,trim={0cm 0.2cm 0cm 0.2cm},clip]{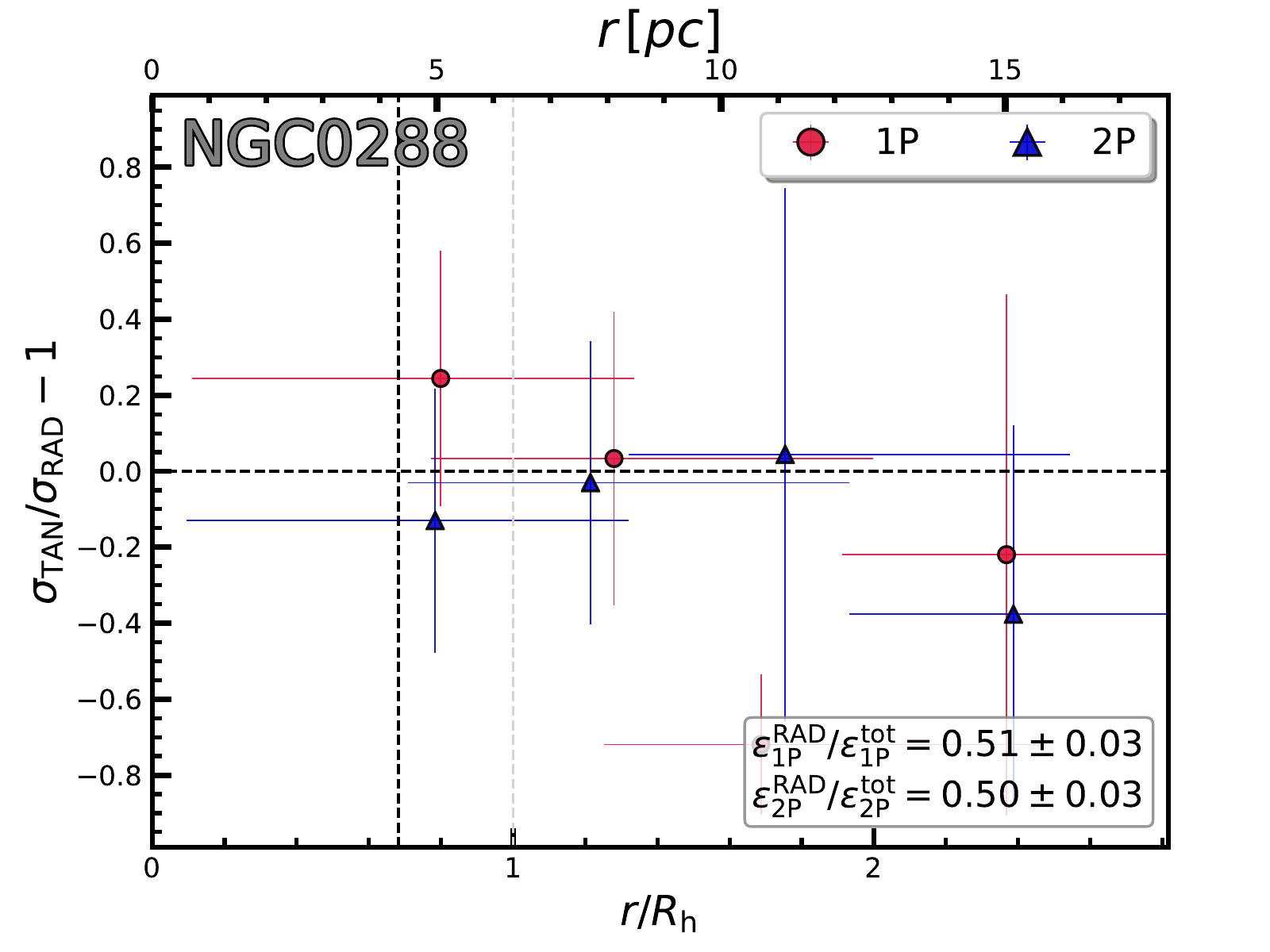}
  \includegraphics[width=8.4cm,trim={0cm 0.2cm 0cm 0.2cm},clip]{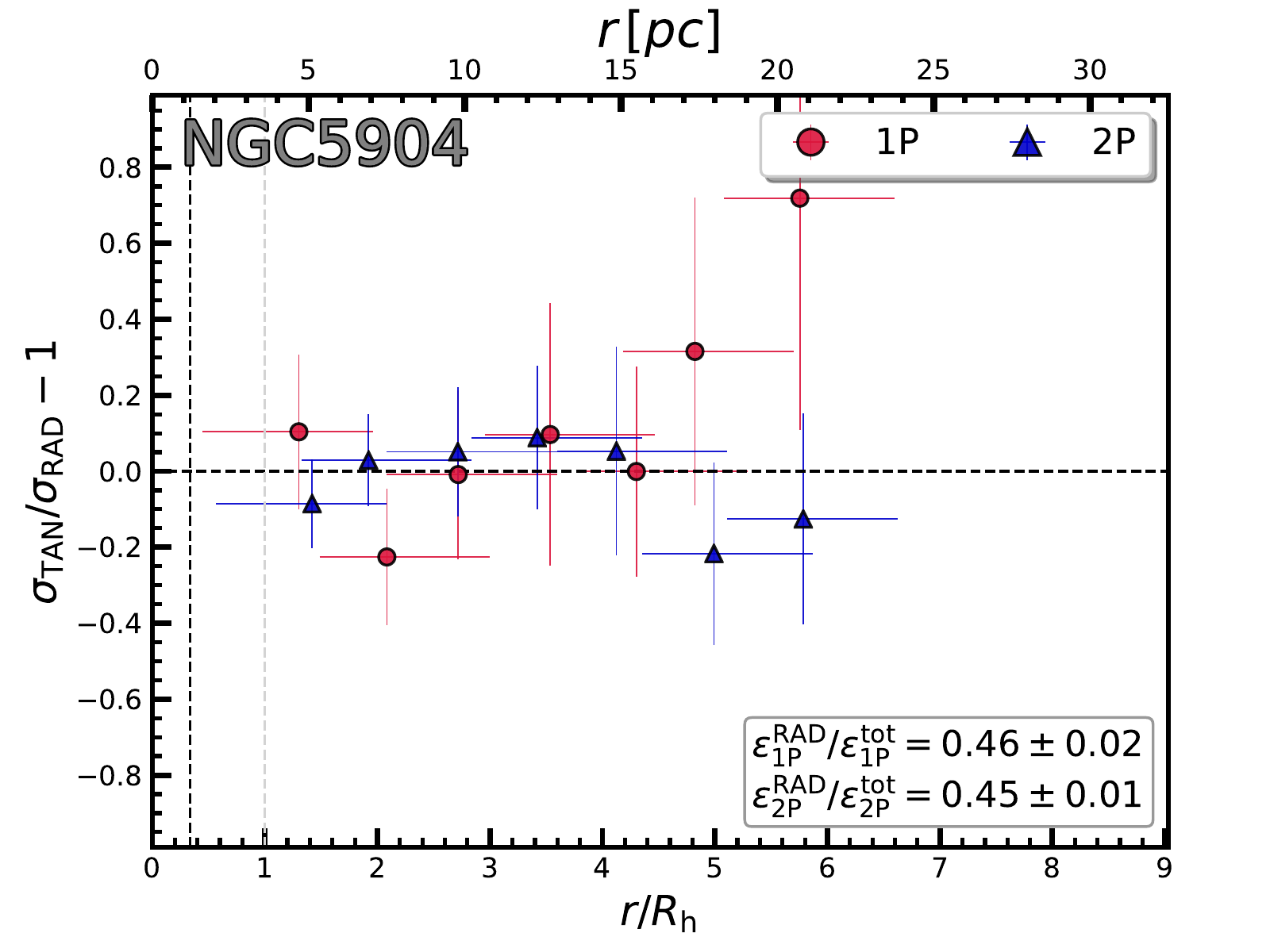}
  \includegraphics[width=8.4cm,trim={0cm 0.2cm 0cm 0.2cm},clip]{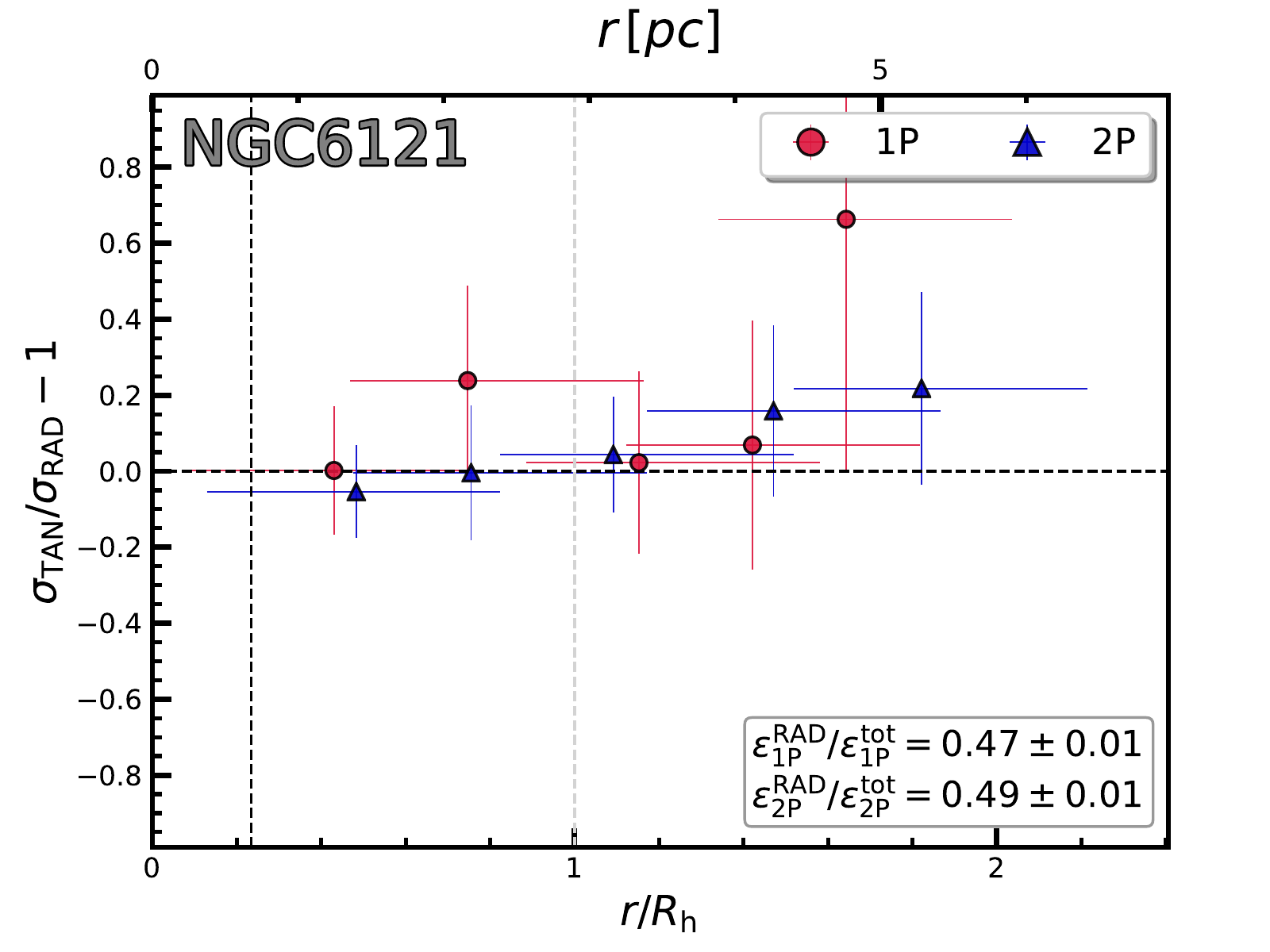}
  \includegraphics[width=8.4cm,trim={0cm 0.2cm 0cm 0.2cm},clip]{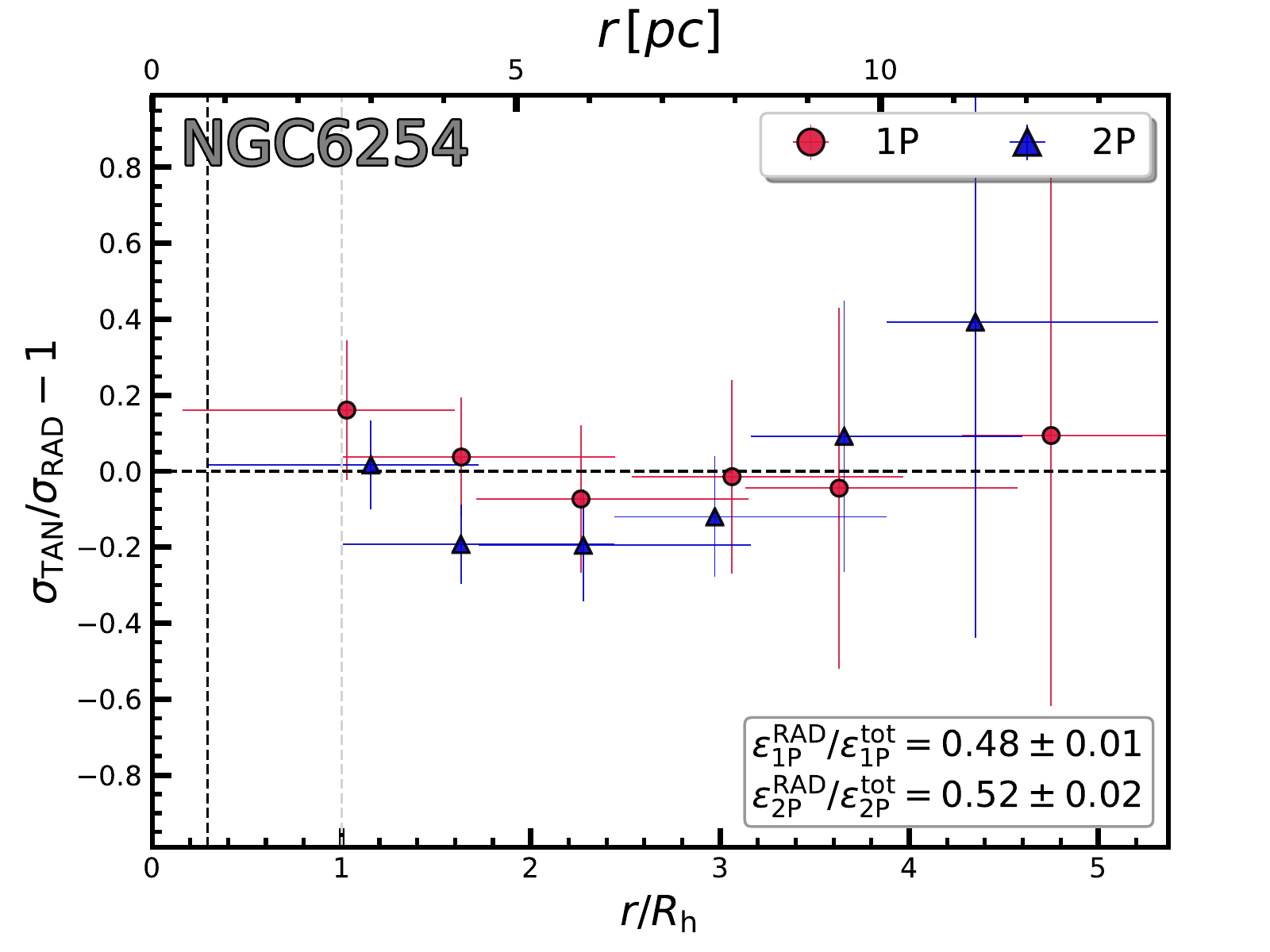}
  \includegraphics[width=8.4cm,trim={0cm 0.2cm 0cm 0.2cm},clip]{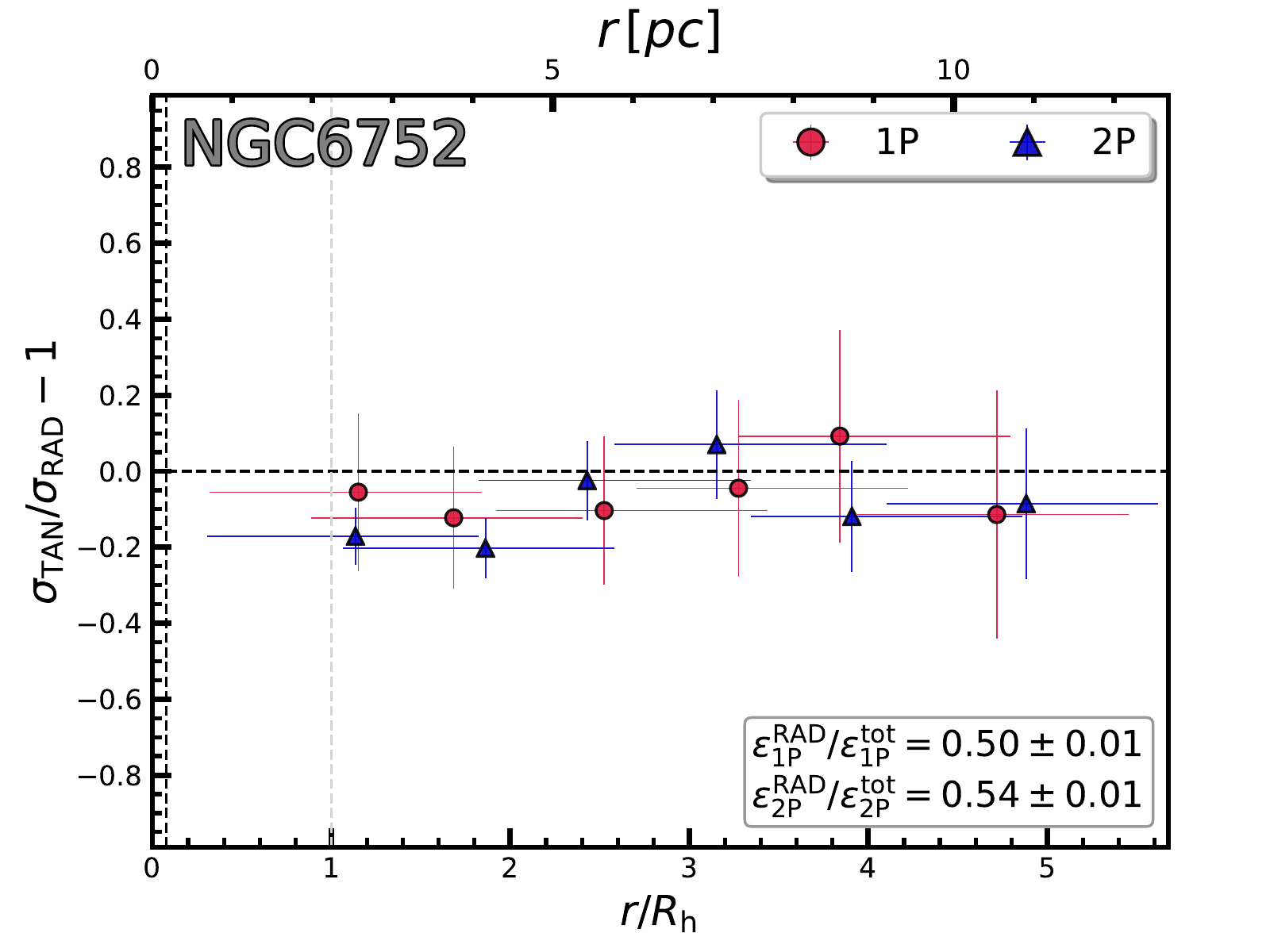}
  \includegraphics[width=8.4cm,trim={0cm 0.2cm 0cm 0.2cm},clip]{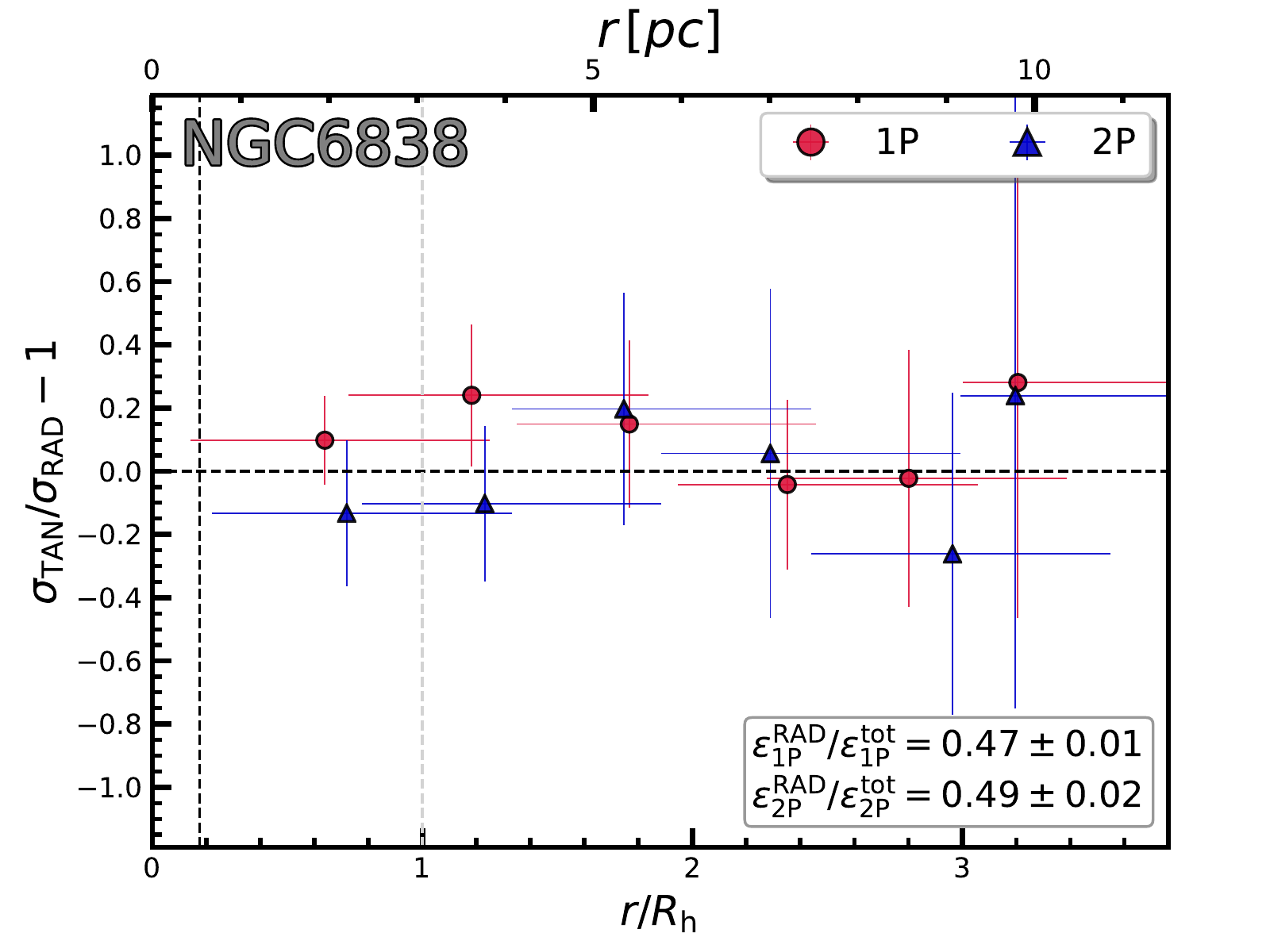}
  \caption{Anisotropy profiles for the analyzed clusters.
   The radial coordinate is normalized to the half-light radius from \citet[][]{baumgardt2018}. Black and gray dashed lines represent the core- and the half-light radius.}
  \label{fig:AllCLbeta}
\end{figure*}

We computed the quantity $\sigma_{\rm TAN}/\sigma_{\rm RAD}-1$, which is indicative of the anisotropy of the internal motion,   
and show its radial profile in Figure~\ref{fig:AllCLbeta}. The horizontal lines in the plots correspond to isotropic stellar systems. 
As a global and independent measure of the degree of anisotropy, we determined, for each population, the ratio between the radial kinetic energy and the total kinetic energy, $k = \epsilon_{\rm RAD}/(\epsilon_{\rm TAN}+\epsilon_{\rm RAD})$. The results are listed in bottom-right corners of Figure~\ref{fig:AllCLbeta}. 
As expected, non rotating clusters are characterized by a value of $k$ close to $k=0.5$, expected for isotropic stellar systems. On the other hand, NGC\,0104 and NGC\,5904 show a higher degree of tangential anisotrpy, as a consequence of the non-zero tangential velocity.
We confirm that NGC\,0104 exhibits strong differences in the degree of anisotropy of the two populations, with the 2P being more radially anisotropic than the 1P. The external region of NGC\,5904 shows hints of a more radially anisotropic 2P, but the large uncertainties prevent us from any further discussion. \\ 
The remaining clusters are consistent with being isotropic stellar systems. Concerning the LoS velocity dispersion profile we find some differences in NGC\,0104, in the outermost part of NGC\,5904 and also in NGC\,6254. 

\section{Summary and discussion} \label{sec:summary}
We exploited Gaia DR2 proper motions and parallaxes of stars in the field of views of seven GCs, namely NGC\,0104, NGC\,0288, NGC\,5904, NGC\,6121, NGC\,6254, NGC\,6752 and NGC\,6838 to separate cluster members from field stars. We analyzed the $V$ vs.\,$C_{\rm U,B,I}$ diagrams corrected for differential reddening of clusters members to identify 1P and 2P stars along the RGB and study their spatial distributions and internal kinematics by using Gaia DR2 stellar positions and proper motions and ESO/VLT and Keck LoS velocities.
To our knowledge, this is the first homogeneous study of the three velocity component internal kinematics of distinct stellar populations in a large sample of GCs over a wide field of view.

We find that 1P and 2P stars of NGC\,0104, NGC\,5904 and NGC\,6254 exhibit different spatial distributions. Specifically, in NGC\,5904 and NGC\,6254 2P stars exhibit higher ellipticities than the 1P, while NGC\,0104 seems consistent with a more-elliptical 1P. The two populations of the other clusters share the same spatial distribution. 
The entire sample of analyzed 1P and 2P stars of NGC\,0104 share similar rotation patterns and that 2P stars show stronger anisotropies than the 1P stars thus corroborating previous findings from our group \citep{milone2018}. 
When we divide stars of NGC\,0104 into two annuli with different radial distances, we find that the sine functions that best reproduce the rotation curves 1P and 2P exhibit different phases and amplitudes in the LoS component. However, such difference is significant at 2-$\sigma$ level only.

We confirm that NGC\,0104 and NGC\,5904 exhibit strong rotation both in the plane of the sky \citep{anderson2003, bianchini2018, milone2018, sollima2019} and along the line-of-sight \citep{kamann2018, lanzoni2018}.

\cite{lee2017} studied multiple populations in NGC\,5904 by using ground-based on  Ca-CN photometry. He separated 1P and 2P along the RGB by using the $V$ vs.\,cn$_{\rm JWL}$ diagram, which is a powerful tool to identify stellar populations with different nitrogen abundances along the RGB.   
Lee used the radial velocities of 100 stars by \cite{carretta2009} to investigate the projected rotations of the two populations identified photometrically. He found that 2P has a substantial net projected rotation whereas there is no evidence for any net projected rotation of 1P stars. 

Our results, based on Gaia DR2 proper motions of 263 1P and 535 2P stars and Eso/VLT LoS velocities of 106 and 238 1P/2P stars, show that both populations exhibit significant rotation along the plane of the sky and the line of sight. The sine functions that describe rotation of 2P and 1P stars exhibit different phases in the $\Delta \mu_{\delta}$ vs.\,$\theta$ and in the $V_{\rm LoS}$ vs.\,$\theta$ planes and such differences are significant at the $\sim$2.6-$\sigma$ and $\sim$2.3-$\sigma$ level, respectively. The two populations exhibit the same phase when we consider the rotation in the $\Delta \mu_{\alpha}cos{\delta}$ vs.\,$\theta$ plane. Such rotation pattern is qualitatively consistent with different position angles and inclinations of the rotation axis.

Our analysis confirms no evidence of rotation in NGC\,0288, NGC\,6121, NGC\,6254 and NGC\,6838 \citep [e.g.] []{bianchini2018, sollima2019}. On the other hand, our results are in apparent disagreement with the conclusion by Bianchini and collaborators who detected a significant rotation of  NGC\,6752 stars in the plane of the sky. We attribute the discrepancy to the small sample of 1P and 2P NGC\,6752 stars studied in our paper. We verified that, when we extend our analysis to all the stars of NGC\,6752 as done by \cite{bianchini2018} and \cite{sollima2019} we confirm previous evidence of rotation.

There is no significant difference between the tangential and radial motions of 1P and 2P stars in the analyzed clusters. 1P stars of NGC\,5904 seem to exhibit, on average, larger motions in the radial direction than 2P stars in the region between $\sim$2 and 5 half-light radii from the cluster center but such difference is not statistically significant when we account for systematic errors in Gaia DR2 proper motions.

We investigate the velocity-dispersion profile of multiple populations in all the GCs and confirm that 2P stars of NGC\,0104 show significant anisotropy with respect to the 1P. In the other clusters there is no evidence for strong anisotropy among 1P and 2P stars, with NGC\,6121 being a possible exception.\\
To summarize our results, we find significant kinematical differences in NGC\,0104 and NGC\,5904, while the remaining clusters are consistent with the presence of multiple stellar populations sharing the same internal dynamic. It is worth mentioning that these two clusters have the highest values for the half-mass relaxation time in our sample (listed in Table~\ref{tab:parameters}), with the exception of NGC\,0288.
Finally, these results are consistent with the criterion in \citet{henault2015}. According to the authors, multiple stellar populations are not expected to be fully mixed if the relation in Equation~\ref{eqn: critical mass} is satisfied.
\begin{equation}
M \gtrsim 10^5 M_\odot\cdot \left(\frac{4\,kpc}{R_{\rm G}} \right)
\label{eqn: critical mass}    
\end{equation}
where $R_{\rm G}$ is the Galactocentric radius, listed in Table~\ref{tab:parameters}. Among the clusters in our sample, only NGC\,6121 and NGC\,6838 do not fulfill Equation~\ref{eqn: critical mass}, and indeed we do not find significant dynamical differences between the 1P and 2P in these two clusters.

All our findings constitute strong constraints for existing and future multiple population scenarios. Self-enrichment scenarios, and in particular the AGB scenario, seem to be able to produce different spatial distributions and kinematics between the first and second generation. This scenario, which is the one that has been studied more in detail in terms of dynamics, predicts a higher central concentration for the 2P with respect to 1P stars. 1P stars have a higher velocity dispersion compared to 2P stars and they show a smaller amount of radial anisotropy. If the 1P cluster is initially rotating, the 2P will form in a centrally concentrated disc and will initially rotate faster than 1P stars. All these signatures are washed out but the two-body relaxation of the clusters. Rotational difference could therefore be absent due to the relaxation process in the velocity space. 

Further tests and dynamical models exploring a larger phase-space of the parameters are necessary to understand if the AGB scenario, or any of the other proposed 2P formation mechanisms, are able to reproduce simultaneously all the observed features.

\section*{acknowledgments} 
\small
We thank the anonymous referee for his/her helpful suggestions that improved the quality of our work. 
We would also like to thank Paolo Bianchini, Vincent H{\'e}nault-Brunet, Antonio Sollima, Maria Tiongco, Eugene Vasiliev and Enrico Vesperini for their helpful comments and suggestions.
This work has received funding from the European Research Council (ERC) 
under the European Union's Horizon 2020 research innovation programme 
(Grant Agreement ERC-StG 2016, No 716082 'GALFOR', PI: Milone, \url{http://progetti.dfa.unipd.it/GALFOR}), 
and the European Union's Horizon 2020 research and innovation programme 
under the Marie Sk\l odowska-Curie (Grant Agreement No 797100, beneficiary: 
Marino). APM and MT acknowledge support from MIUR through the FARE project 
R164RM93XW SEMPLICE (PI: Milone).  AMB acknowledges support by Sonderforschungsbereich (SFB) 881 `The Milky Way System' of the German Research Foundation (DFG).

\bibliographystyle{aa}

\end{document}